\documentclass[aps,prx,twocolumn, nofootinbib,superscriptaddress,floatfix,footinbib, bibnotes]{revtex4-2}
\usepackage{amsmath,amssymb,amsfonts}
\usepackage{bm}
\usepackage{booktabs}
\usepackage[colorlinks=true,citecolor=blue,linkcolor=blue,urlcolor=blue]{hyperref}
\usepackage{color}
\usepackage[dvipsnames]{xcolor}
\usepackage{graphicx}
\graphicspath{{figures/}}
\usepackage{subfigure}

\newcommand{\dd}{\mathrm d}
\newcommand{\Zbb}{\mathbb Z}

\newcommand{\eps}{\epsilon}

\newcommand{\rO}{\mathrm{O}}
\newcommand{\Tr}{\operatorname{Tr}}

\makeatletter
\def\fnum@figure{Fig.~\thefigure}
\makeatother

\begin{document}

\title{Revisiting monodromy defects in $\mathrm{O}(N)$ conformal field theories}
\date{\today}
\author{Jiechao Feng}
\email{jiechao\_feng@berkeley.edu}
\affiliation{Graduate Group in Applied Science \& Technology, University of California, Berkeley, California 94720, USA}
\author{Nayan Myerson-Jain}
\email{nayan.myerson-jain@yale.edu}
\author{G. Shankar}
\email{shankar.ganesh@yale.edu}
\affiliation{Department of Physics, Yale University, New Haven, Connecticut 06511, USA}
\author{Taige Wang}
\email{taige\_wang@fas.harvard.edu}
\affiliation{Materials Research Laboratory, Massachusetts Institute of Technology, Cambridge, MA 02139, USA}
\affiliation{Department of Physics, Harvard University, Cambridge, MA 02138, USA}
\author{Meng Cheng}
\email{m.cheng@yale.edu}
\affiliation{Department of Physics, Yale University, New Haven, Connecticut 06511, USA}
\begin{abstract}
Monodromy defects are fundamental objects in extended quantum systems with global symmetries. We study such defects in $\mathrm{O}(N)$ conformal field theories (CFT), which describe the universality class of $\mathrm{O}(N)$ spontaneous symmetry breaking transitions. In particular, we consider partial-monodromy defects, whose corresponding symmetry transformations only act on a subset of the fundamental $\mathrm{O}(N)$ vector field. We develop an analytical theory of the defect phases, using both renormalization group and large $N$ expansion. In general (spacetime) dimensions $3\leq d<4$, multiple defect CFTs are identified, which are described by appropriate boundary conditions for the vector field near the defect. Near $d=3$, we show that in the large $N$ limit, a stable defect CFT always exists, characterized by the so-called (``standard") ``alternate" boundary conditions for the (twisted) untwisted fields, whose defect correlations are (suppressed) enhanced. We then perform large-scale classical Monte Carlo simulations of monodromy defects in the O(3) CFT in an improved lattice model and numerically extract the defect scaling dimensions. The numerical results are in qualitative agreement with the stable defect CFT identified in the field theory.
\end{abstract}
\maketitle

\section{Introduction}
Defects play crucial roles in quantum and classical many-body systems and field theories: they elucidate the structure of the underlying theories and determine physical properties.
Any quantum system with global (zero-form) internal symmetry comes with a distinguished type of codimension-two defects, known as monodromy (or twist) defects. Upon encircling the defects, the fundamental fields are transformed by the corresponding symmetry action. In other words, the boundary conditions of the fields on the two-dimensional transverse plane are twisted by a symmetry of the system. 

Physically, a monodromy defect can be created by threading a localized ``magnetic" flux of the symmetry, analogous to an Aharonov-Bohm flux, producing a defect extending in the time direction. Alternatively, such defects can be generated by acting on the system with a disorder operator, i.e., a partial symmetry transformation~\cite{Zhao:2020vdn, Wang:2021lmb, WuSP2021, Chen:2022upe, Wang:2021yaf}. In a system with an exact or emergent space-time symmetry, these two constructions are equivalent and are related by a space-time rotation. Such defects, or extended operators, often provide probes of the underlying phase of matter beyond conventional local measurements. It is also worth noting that certain entanglement measures, such as R\'enyi entropies, can be formulated as the free energy of the system with replica symmetry defects \cite{Calabrese:2004eu, Metlitski:2009iyg, Bianchi:2015liz}.

The presence of a defect modifies the local correlations in its vicinity. This effect is particularly pronounced in a gapless system, where the correlation functions on the defect exhibit asymptotic long-distance behaviors distinct from those in the bulk. In particular, when the bulk is described by a conformal field theory (CFT), e.g. when the system is tuned to a critical point, the defect often preserves the residual conformal symmetry and is therefore called a conformal defect. In this case, the properties of local observables near the defect can be organized as a defect CFT (DCFT) \cite{MeineriDCFT, GaddeDCFT, Herzog:2020bqw}, which essentially describes the universality class, or the ``phase," of the defect theory.

In this work, we study monodromy defects in the $\rO(N)$ Wilson-Fisher theories in space-time dimension $d\geq 3$. These are prototypical interacting CFTs in higher dimensions, whose bulk properties are well-understood. Monodromy defects therein have been studied previously \cite{Billo:2013jda, GaiottoMazacPaulos,  Soderberg, GiombiHyperbolic, BianchiFree, XiangMonodromy, monodromypinning}, which we briefly review below. 

We will denote the fundamental $\rO(N)$ vector field by $\bm{\Phi}=(\Phi_1, \Phi_2, \dots, \Phi_N)^\mathsf{T}$. A general element of $\rO(N)$ can always be brought into block-diagonal form of $2\times 2$ SO(2) rotations labeled by a set of monodromies, and remaining $\pm 1$ blocks. We refer to the fields transforming nontrivially under the symmetry as ``twisted," and the others as ``untwisted". 

We mostly consider the following type of monodromy. Group the first $M$ real components into $M/2$ complex vector $\bm{\psi}$ with $\psi_j = \Phi_{2j-1} + i \Phi_{2j}$ for $j=1, \dots, M/2$, where $M\leq N$ is an even integer, and denote the remaining $N-M$ untwisted real fields  by $\bm{\phi}$. The monodromy is defined by the following twisted boundary condition:
\begin{equation}
   \bm{\psi}(\vec{x},r,\theta+2\pi ) = e^{2 \pi i v} \bm{\psi}(\vec{x},r,\theta). 
   \label{eq:boundary_condition}
\end{equation}
Here $r$ and $\theta$ are polar coordinates in the plane transverse to the defect, and $\vec{x}$ is the coordinate on the $(d-2)$-dimensional defect.
Throughout this paper we assume $0\leq v\leq \frac12$, as defects with monodromy $v$ and $1-v$ are related by the charge-conjugation symmetry.

  When $M=N$, i.e. all fields are twisted, the problem has been extensively studied through both numerical simulations for Ising defects in the Ising CFT~\cite{Billo:2013jda}, and analytical approaches for general $N$~\cite{Metlitski2008, GaiottoMazacPaulos, Soderberg, GiombiHyperbolic}. An infrared (IR) stable DCFT fixed point (FP) is found in all these studies, which can be accessed perturbatively in the $\epsilon$ expansion \cite{GaiottoMazacPaulos, Yamaguchi:2016pbj, Liendo:2019jpu, Soderberg}, or through the large $N$ approximation \cite{Metlitski2008, GiombiHyperbolic}.

 In this work, we study monodromy defects with $M<N$, which we will call partial-monodromy defects. Having untwisted fields substantially changes the defect dynamics, resulting in a richer IR phase diagram. Using the $\epsilon$ expansion, we identify multiple perturbative FPs, characterized by having ``standard" or ``alternate" boundary conditions of the $\rO(N)$ vector fields near the defect. Equivalently, twisted or untwisted modes may have positive or negative mass pinning to suppress or enhance correlations. 
 
 We then study a large $N$ limit where $M, N\to\infty$ but $\gamma=M/N$ is fixed. In addition to the perturbative FPs at ``weak-pinning'' of the defect mass, the large $N$ approximation also finds non-perturbative ``strong-pinning" saddle point solutions, which we identify as the endpoint of the runaway RG flows in the $\epsilon$ expansion. For these solutions, the scaling dimension of the order parameter becomes parametrically small. We analyze how these solutions evolve as $d$ is varied between 3 and 4, and argue that for small $\gamma$ the stable perturbative solution continues to $d=3$. For all $\gamma$ and $v$, we show that there is always a \emph{stable} saddle point solution at $d=3$, featuring alternate (standard) boundary conditions for the untwisted (twisted) fields, implying that their correlations are enhanced (suppressed) compared to their bulk counterparts. We also show that near $d=3$ additional saddle points exist for $v<v_c^{\text{std},\text{alt}}(\gamma)$, with standard (alternate) boundary condition for the untwisted (twisted) fields.  We support these results with analytic solutions of the large $N$ gap equations in the limit where $v$ or $\gamma$ becomes small. We summarize the FP structures from large $N$ in Table~\ref{tab:large-N-fixed-points}.
 
 Lastly, in $d = 3$, we study the problem for $N=3$ with classical Monte Carlo simulations.  Two types of defects are considered: $\mathbb{Z}_2$ monodromy defect where the symmetry acts as $\Phi_3\rightarrow -\Phi_3$, and ${\rm SO}(2)$ defects with the symmetry rotating $\Phi_1$ and $\Phi_2$. In the latter case we consider $v=1/2$ and $v=1/4$. Motivated by the analytical theory, we add a negative mass pinning ($\sim \tilde{c}\Phi^2$, with $\tilde{c}<0$) for the untwisted field on the defect. We then measure the two-point functions of twisted and untwisted fields to extract their scaling dimensions. For the ${\rm SO}(2)$ defects, when the pinning strength $|\tilde{c}|$ is large enough, we observe strong evidence for $\Zbb_2$ spontaneous symmetry breaking (SSB) on the defect, which significantly impacts the behavior of the correlation functions. The numerically extracted scaling dimensions show substantial drift as $|\tilde{c}|$ increases. For the $v=1/2$ defect, the scaling dimension of the untwisted field $\Delta_\phi$ decreases from $0.409\pm 0.011$ for $\tilde{c}=0$ to $0.194\pm 0.006$ at $\tilde{c}_*=-1.8$. Similar drifting behaviors are observed at $v=1/4$. On the other hand, for the $\Zbb_2$ (reflection) defect, the dependence of the measured scaling dimensions on the pinning strength is weaker. With this caveat in mind, the extracted scaling dimensions are in qualitative agreement with the DCFT with alternate/standard boundary conditions imposed on untwisted/twisted fields. The numerical results for defect scaling dimensions are summarized in Table \ref{tab:defect-dimensions}.
 
\begin{table}[t]
\centering
\caption{Evolution of defect FPs in the large $N$ limit as $d:4^-\to3$,
at fixed $(v,\gamma)$. Boundary conditions are ordered as $(\phi,\psi)$.
$W$ and $S$ label weak- and strong-pinning branches of solutions as identified near $d=4$.}
\label{tab:large-N-fixed-points}
\small
\setlength{\tabcolsep}{3pt}
\setlength{\arrayrulewidth}{0.4pt}
\setlength{\doublerulesep}{1.5pt}
\setlength{\lightrulewidth}{0.4pt}
\renewcommand{\arraystretch}{1.2}
\begin{tabular*}{\columnwidth}{@{\extracolsep{\fill}}ccc@{}}
\hline\hline
\addlinespace[2pt]
$(\phi,\psi)$ & Parameter region & FPs $d:4^-\to3$ \\
\midrule
$(\mathrm{alt},\mathrm{std})$ & $\gamma < \frac{1}{2}$ & $S\to(2W+S)\to W$ \\
& $\gamma > \frac{1}{2}$ & $S$ \\
\midrule
$(\mathrm{std},\mathrm{alt})$
& $v<v_c^{\text{std},\text{alt}}(d,\gamma)$ & $W+S$ \\
\addlinespace[2pt]
& Otherwise & $(W+S)\to0$ \\
\midrule
$(\mathrm{alt},\mathrm{alt})$ & All $v,\gamma$ & $(W+S)\to0$ \\
\midrule
$(\mathrm{std},\mathrm{std})$ & All $v,\gamma$ & $0$ \\
\addlinespace[2pt]
\hline\hline
\end{tabular*}
\end{table}


\section{Renormalization group analysis}

In this section we study the defect dynamics using the renormalization group analysis. 

It is helpful to review the case of the free theory for our choice of monodromy. The fields can be decomposed according to the angular momentum around the defect in Eq.~\eqref{eq:boundary_condition}, with the spin $\ell$ of each mode of the twisted complex fields being shifted by $v$ such that $
\bm{\psi}(\vec{x},r,\theta)=  \sum_{\ell \in \Zbb +v} \bm{\psi}_\ell (\vec{x},r) e^{i \ell \theta} $, while for untwisted fields $\bm{\phi}(\vec{x},r,\theta)=  \sum_{\ell \in \Zbb } \bm{\phi}_\ell (\vec{x},r) e^{i \ell \theta}$. 

Due to conformal invariance, each mode obeys a bulk-to-defect operator product expansion (OPE) in terms of defect primaries (denoted with a hat), as $r \rightarrow 0$
\begin{equation}
    \bm{\psi}_\ell(\vec{x},r) \sim r^{\Delta_\ell - \Delta} \hat{\bm{\psi}}_\ell(\vec{x}), \text{ }  \bm{\phi}_\ell(\vec{x},r) \sim r^{\Delta_\ell - \Delta} \hat{\bm{\phi}}_\ell(\vec{x}).
\end{equation}
The bulk scaling dimension for these fields is the usual value, $\Delta = \frac{d}{2}-1$, while defect operators $\hat{\bm{\psi}}_\ell$ and $\hat{\bm{\phi}}_\ell$ have defect scaling dimensions $\Delta_\ell = \frac{d}{2}-1 + |\ell|$. For the twisted fields, since $\ell \in \Zbb +v$, the defect operator with lowest dimension is $\hat{\bm{\psi}}_{\ell =v}$ with $\Delta_{\ell=v} = \frac{d}{2}-1+v$. For the untwisted fields, there is instead a zero mode defect operator $\hat{\bm{\phi}}_{\ell = 0}$ with $\Delta_{\ell=0} = \frac{d}{2}-1$. The mass operator $\hat{\bm{\phi}}^2_{\ell=0}$ has dimension $2\Delta_{\ell=0} = d-2$ and is classically marginal on the defect. The lowest spin modes will dominate the long-distance behavior of the defect two-point correlation of $\bm{\psi}$ and $\bm{\phi}$ respectively. 

In $d \geq 3$, there is another DCFT with $\Delta_{\ell=v} = \frac{d}{2}-1-v$. Following AdS/CFT terminology, for $\Delta_\ell > \Delta$, we refer to the mode as having standard boundary conditions; if $\Delta_\ell < \Delta$ instead, the mode has alternate boundary conditions. In the free theory, only the former is stable in the IR. We focus on the physics when $\bm{\psi}_{\ell=v}$ has standard boundary conditions and defer discussion of alternate boundary conditions for this mode to the Supplemental Material (SM). 

We now turn to the interacting theory. The action of the Wilson-Fisher theory in $d$ space-time dimensions is given by
\begin{equation}
\begin{aligned}
    S &= S_\Phi + S_{\rm defect},\\
    S_\Phi &= \int \dd^d{x}  \text{ } \left[ \frac{1}{2}(\partial \bm{\Phi})^2 + \frac{\lambda}{4!}(\bm{\Phi}^2)^2 \right], \\
S_{\rm defect} &=  \int_{\text{defect}} \dd^{d-2}{x} \text{ } \frac12\Big(c\hat{\bm{\phi}}_{\ell= 0}^2 + h|\hat{\bm{\psi}}_{\ell=v}|^2 \Big). \label{eq:WF_action}
\end{aligned}
\end{equation}
Here we include the classically marginal mass term for the untwisted fields on the defect, as well as the mass for the twisted field with the lowest spin $\ell=v$, which is necessary when $v \lesssim \epsilon$. A similar defect perturbation has previously been considered in the context of a Rényi conical defect in Ref.~\cite{Metlitski:2009iyg}.   For $d\!=\!4\!-\!\epsilon$ with $\epsilon\!\ll\!1$, the bulk FP is at $\lambda_*\!=\!\frac{48 \pi^2 \epsilon}{N+8}$.  

We find that the $\beta$-functions for the defect couplings $c$ and $h$ at one-loop are given by 
\begin{align}
    \beta_c &= \frac{c^2}{2\pi} + \frac{M \kappa \lambda}{24 \pi} + \frac{(N-M+2) \lambda c }{48 \pi^2} + \frac{ M \lambda h}{48 \pi^2}, \label{eq:betafunc} \\
    \beta_h &= \frac{h^2}{2\pi} + \frac{ (M+2) \kappa \lambda}{24 \pi} + \frac{(N-M)\lambda c}{48 \pi^2} + \frac{ (M+2)\lambda h}{48\pi^2} \nonumber\\ &\quad+ 2vh \nonumber.
\end{align}
Here we use the convention $\beta_c = \frac{\dd c}{\dd \log \mu}$, $\beta_h = \frac{\dd h}{\dd \log \mu}$, and define $\kappa=v(1-v)$. We refer the reader to Section \ref{sm:RG} in the SM for the derivation of $\beta_c$ and $\beta_h$, and additional results. 

To better understand the landscape of FPs of the defect RG flow, we analyze the $1 \gg v \gg \epsilon$ and $v \lesssim \epsilon$ regimes separately. The scaling dimensions of defect operators $\hat{\bm{\phi}}_{\ell = 0}$ and $\hat{\bm{\psi}}_{\ell = v}$ are given to leading order by
\begin{equation}
\begin{aligned}
   \hat{\bm{\phi}}_{\ell = 0}&:  \Delta_{\ell = 0} = \frac{d}{2}-1+\alpha_{\ell=0}, \text{ } \alpha_{\ell=0} = \frac{c_*^\pm}{2\pi};\\
    \hat{\bm{\psi}}_{\ell = v}&:  \Delta_{\ell = v} = \frac{d}{2}-1+v+\alpha_{\ell = v}, \text{ } \alpha_{\ell=v} = \frac{h_*^\pm}{2\pi}.
\end{aligned}
\end{equation}
Accordingly, $\bm{\phi}_{\ell = 0}$ has standard (alternate) boundary conditions when $c_* > 0$ ($c_* < 0$). Similarly, $\bm{\psi}_{\ell = v}$ has standard (alternate) boundary conditions when $h_*+2 \pi v >0$ $(h_*+2\pi v < 0)$.

\subsection*{Defect FPs in the $1 \gg v \gg \epsilon$ regime}

In this regime, the RG analysis can be simplified. To leading order $\beta_h = \frac{h^2}{2 \pi} + 2v h$, there are two simple FPs for the twisted field: $h_*^- = 0$ (stable) and $h_*^+ = - 4 \pi v$ (unstable). These solutions correspond, respectively, to standard or alternate boundary conditions for the twisted field in the bare theory. This analysis is equivalent to dropping the defect mass perturbation for the twisted fields in Eq.~\eqref{eq:WF_action} and choosing their boundary condition by hand. For standard boundary conditions,
\begin{equation}
    \beta_c^{1 \gg v \gg \epsilon} = \frac{c^2}{2 \pi} + \frac{M \kappa \lambda}{24 \pi} + \frac{(N-M+2) \lambda c}{48 \pi^2} \label{eq:betafunc1}.
\end{equation}
We define the critical value $\kappa_c$ by
\begin{equation}
    \kappa_c=\frac{(N-M+2)^2\lambda}{192\pi^2 M}= \frac{(N-M+2)^2}{4M(N+8)}\epsilon \label{eq:kappac}.
\end{equation}
Because $v(1-v)\leq 1/4$, the actual threshold is $\tilde{\kappa}_c = \min(\kappa_c, \frac14)$.
There are two real FPs when $\kappa<\tilde{\kappa}_c$, given by
\begin{equation}
  c_*^\pm
  =- \pi \frac{N-M+2}{N+8} \epsilon \bigg ( 1 \pm \sqrt{1 - \frac{\kappa}{\kappa_c} }\bigg ).
  \label{eq:cstar}
\end{equation}
As a result, when $\kappa < \kappa_c$ and $1 \gg v \gg \epsilon$ the existence of two real FPs requires $\gamma = \frac{M}{N} \lesssim \frac{\epsilon}{4 \kappa} \frac{(N+2)^2}{N(N+8)} \ll 1$. For both FPs, with $c_*^\pm < 0$, {$\bm{\phi}_{\ell=0}$ has alternate boundary conditions such that} the anomalous dimension $\alpha_{\ell=0}$ is negative leading to an enhancement of defect correlations of the untwisted field. Linearizing the RG equations, we find that $c^-_*$ is stable while $c^+_*$ is unstable as the anomalous dimension of $\hat{\bm{\phi}}_{\ell=0}^2$ is positive for $c_*^-$ and negative for $c_*^+$. Increasing monodromy further enhances correlations at the stable FP $c_*^-$, but weakens the enhancement at the unstable FP $c_*^+$.

\begin{figure}
    \centering    \includegraphics[width=0.85\columnwidth]{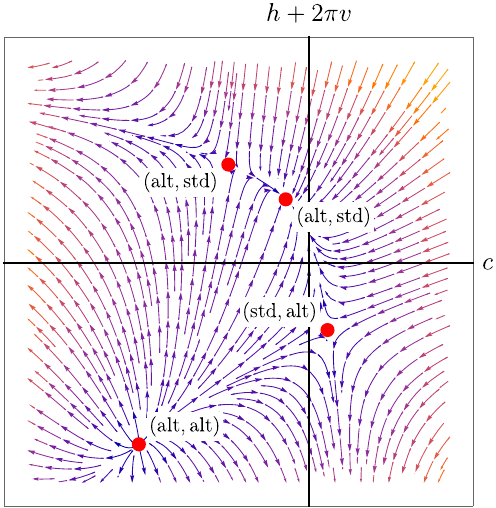}
    \caption{Prototypical RG flow (shown for $v \lesssim \epsilon$) for $v < v_c^{\text{alt},\text{std}}$ when both FPs with $\bm{\phi}_{\ell = 0}$ with alternate and $\bm{\psi}_{\ell=v}$ with standard boundary conditions exist as seen in the upper-left quadrant. In this plot, sinks correspond to {\it IR stable} FPs.} 
    \label{fig:fpgridsolo}
\end{figure}

When $\kappa > \kappa_c$, which occurs at sufficiently large $\gamma > \frac{\epsilon}{4 \kappa} \frac{(N+2)^2}{N(N+8)}$, there is no FP at finite coupling. The  runaway flow $c \rightarrow - \infty$ should end at a new FP, which we will refer to as the ``runaway'' FP below.  Only for small $N$ (e.g. $N = 3$) is this generically the only possible case if $1 \gg v \gg \epsilon$, and the two perturbative FPs will not be seen. Note that when $\tilde{\kappa}_c = 1/4$, i.e. at extremely small $\gamma$, the two FPs exist for all values of $v$ \footnote{The minimal value for the monodromy defect is $\gamma_\text{min} = 2/N$. For small $N$, $\gamma_\text{min} \gg (\epsilon/4 \kappa) (N+2)^2/N(N+8)$ and these two FPs do not appear. For larger $N>N_c \sim \frac{2}{\epsilon}$, at $\gamma_\text{min}$ which is now sufficiently small, both perturbative FPs survive for all values of $0<v \leq 1/2$.}.

We have thus seen that for $\kappa_c<1/4$, there is a critical $v_c^{\text{alt},\text{std}}(\gamma)$. The superscript refers to the alternate (standard) boundary condition for untwisted (twisted) field. For $v<v_c^{\text{alt},\text{std}}$, the RG gives two real FPs. For $v>v_c^{\text{alt},\text{std}}$, there is only the runaway FP. Hence a phase transition occurs on the monodromy defect as a function of $v$. 
 
 As already mentioned, the runaway flow should lead to a non-perturbative FP where $c$ becomes $\mathcal{O}(1)$. Physically, the negative mass pinning drives the order parameter $\bm{\phi}$ towards condensation, which would spontaneously break the remaining $\rO(N-M)$ symmetry when $N-M>1$. However, general argument for an interacting bulk shows that spontaneous breaking of a continuous symmetry is impossible on a dimension-$(2-\epsilon)$ defect~\cite{Cuomo:2023qvp}. A possible endpoint of the runaway RG flow is a new DCFT when $c_*$ becomes ${\cal O}(1)$. Below we will identify this ``runaway'' FP in the large $N$ limit. Note that when $N-M=1$, the ${\rm O}(1)=\mathbb{Z}_2$ symmetry can be broken spontaneously.

The leading correction to the defect scaling dimension for the twisted field can be extracted as well. From expansion of the solution to $\beta_h$ from Eq.~\eqref{eq:betafunc} in $\epsilon$, one finds the anomalous dimension for $\hat{\bm{\psi}}_{\ell = v}$ at the Wilson-Fisher FP to leading order is 
\begin{equation}
\begin{aligned}
\alpha_{\ell=v} &= \frac{h^-_*}{2 \pi} \simeq -\frac{(M+2)\epsilon}{2(N+8)v} \left( \kappa+  \frac{N-M}{M+2}\frac{c_*^\pm }{2 \pi }\right).
    \label{eq:dimtwisted}
\end{aligned}
\end{equation}
As $h_* + 2 \pi v \geq 0$, the scaling dimension of $\hat{\bm{\psi}}_{\ell = v}$ remains greater than that of the bulk field. This expression is consistent with known results when all fields are twisted, i.e. $M = N$, as expected \cite{Soderberg, GiombiHyperbolic}. It also agrees with results in the large $N$ limit. Defect anomalous dimensions for other $\ell\neq 0$ modes are computed in SM \ref{sm:RG}.

\begin{figure}
    \centering    \includegraphics[width=0.9\columnwidth]{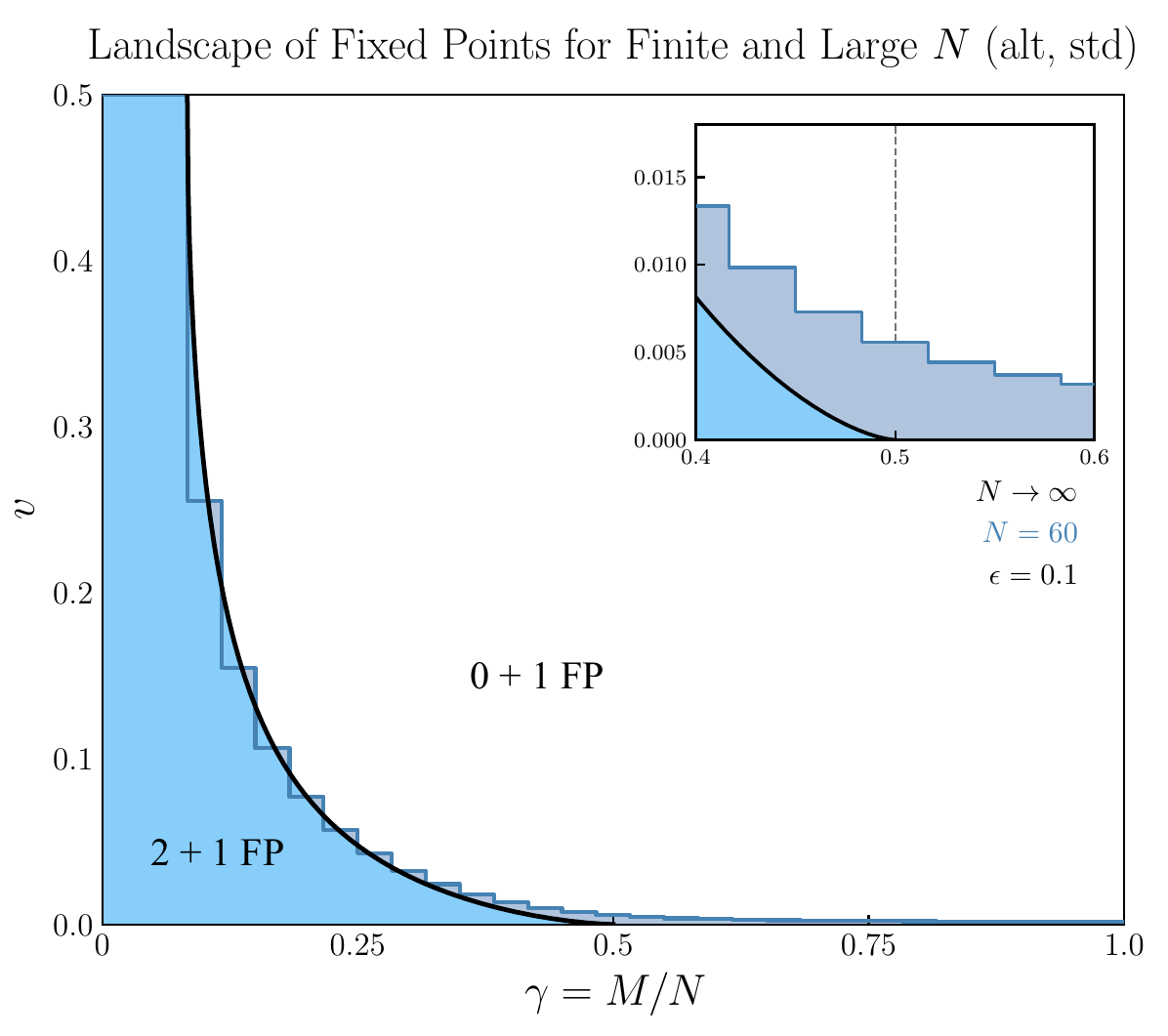}
    \caption{Landscape of FPs for $\epsilon \ll 1$ when $\bm{\psi}_{\ell=v}$ has standard boundary conditions ($h_* + 2 \pi v > 0$) from RG (and the large $N$ approximation). There are two perturbative FPs with alternate boundary conditions $(c_* < 0)$ for $\bm{\phi}_{\ell=0}$. For finite $N$ (large $N$) these FPs exist in the shaded grey (blue) region and annihilate at the boundary which is $v_c^{\text{alt},\text{std}}(\gamma)$. The difference in behavior at $\gamma = 1/2$ is displayed by the inset.  Additionally, there is a ($+1$) ``runaway'' FP.} 
    \label{fig:LargeNSmallepsfp}
\end{figure}

\subsection*{Defect FPs in the $v \lesssim \epsilon$ regime}

To accurately capture the physics in this regime with RG, it is imperative to keep the defect mass perturbation for the twisted fields in the action. The FPs are complicated, but we may comment on some new qualitative features of the solutions: 
\begin{enumerate}
    \item The same two perturbative FPs with alternate boundary conditions for $\bm{\phi}_{\ell=0}$ and standard boundary conditions for $\bm{\psi}_{\ell=v}$ exist for $v < v_c^{\text{alt},\text{std}}$ but now can be seen even up to larger $\gamma \sim \mathcal{O}(1)$, inclusive of when $N$ is small. That is because with increasing $\gamma$, $v_c^{\text{alt},\text{std}}$ decreases from $v_c^{\text{alt},\text{std}} \gg \epsilon$ as seen previously to $v_c^{\text{alt},\text{std}} \lesssim \epsilon$. These FPs can be seen in the upper-left quadrant of Fig.~\ref{fig:fpgridsolo}. 
    \item For finite $N$, even though $v_c^{\text{alt},\text{std}}$ decreases with $\gamma$, it remains nonzero as $\gamma \rightarrow 1$. That is, there exists a sufficiently small $v$ such that both FPs exist for any $\gamma$. For sufficiently large $\gamma$, by tuning $v$ there is always a phase transition on the defect. 
    \item For $N \rightarrow \infty$ when $\gamma \geq 1/2$, $v_c^{\text{alt},\text{std}} \rightarrow  0$ instead, i.e., these two FPs never exist in this range. For a fixed $v < v_c^{\text{alt},\text{std}}$, these two FPs must always annihilate as $\gamma \rightarrow 1/2$. This sharp behavior is special to this limit. 
\end{enumerate}
The regions of parameters in which these FPs exist for finite $N$ and large $N \rightarrow \infty$ are shown in Fig.~\ref{fig:LargeNSmallepsfp}. We leave discussion of when the twisted field has alternate boundary conditions to the SM \ref{sm:RG}.

When $\gamma = 1$ ($N=M$), such that all the fields are twisted, only $\beta_h$ in Eq.~\eqref{eq:betafunc} need be kept. In this case, one always finds a stable FP with standard boundary conditions for the twisted fields for any $v$ as expected. As this is not true for $\gamma < 1$, the physics is not smooth as $\gamma \rightarrow 1$. An explicit expression for $h_*$ when all the fields are twisted can be found in Section \ref{sm:RG}.

\section{Large $N$ limit}
Next we analyze the problem in the limit of large $N$. In order to have a nontrivial large $N$ limit, we assume both $M, N\rightarrow \infty$, but $\gamma=M/N$ is kept fixed. Following standard procedure, we decouple the quartic $(\bm{\Phi}^2)^2$ term using a Hubbard-Stratonovich field $\sigma(x)=\frac{\sigma}{r^2}$. The saddle-point $\sigma_*$ is determined self-consistently by a gap equation~\cite{Metlitski:2009iyg, GiombiHyperbolic}. Details can be found in the SM \ref{app:largeN}.

At this point, one has to specify the boundary conditions of the fields near the defect. As reviewed above, for each mode there are potentially two options, standard or alternate.  The defect scaling dimension is given by 
\begin{equation}
  \Delta_{\ell}=\frac{d}{2}-1\pm \sqrt{\sigma_*+\ell^2}+ {\cal O}(N^{-1}),  
\end{equation}
where $+/-$ denotes standard/alternate boundary condition for the mode $\ell$.  $\sigma_*\geq 0$ is required in order for the formula to be well-defined for $\ell=0$.  For a given mode $\ell$, the alternate boundary condition is only allowed if $\frac{d}{2}-1-\sqrt{\sigma_*+\ell^2}>0$.

One advantage of the large $N$ analysis is that it can be done for generic $d$. Hence it allows us to track how  saddle-point solutions evolve from $d$ close to $4$ all the way to $d=3$. Below we focus on the solutions with alternate (standard) boundary condition on the $\ell=0$ ($\ell\neq 0$) mode, in accordance with the RG analysis before. As a result $\sigma_*$ must satisfy
\begin{equation}
    0\leq \sigma_*<  \Big(\frac{d}{2}-1\Big)^2.
\end{equation}
Details of the derivation and more discussions on other choices of boundary conditions can be found in the SM \ref{app:largeN}. Note that when $\gamma = 1$ and all the fields are twisted and have standard boundary conditions, the unitarity bound is shifted to $-v^2 \leq \sigma_*$. Like in the RG, the physics does not change smoothly as $\gamma \rightarrow 1$.

\begin{figure}
    \centering
    \includegraphics[width=1\linewidth]{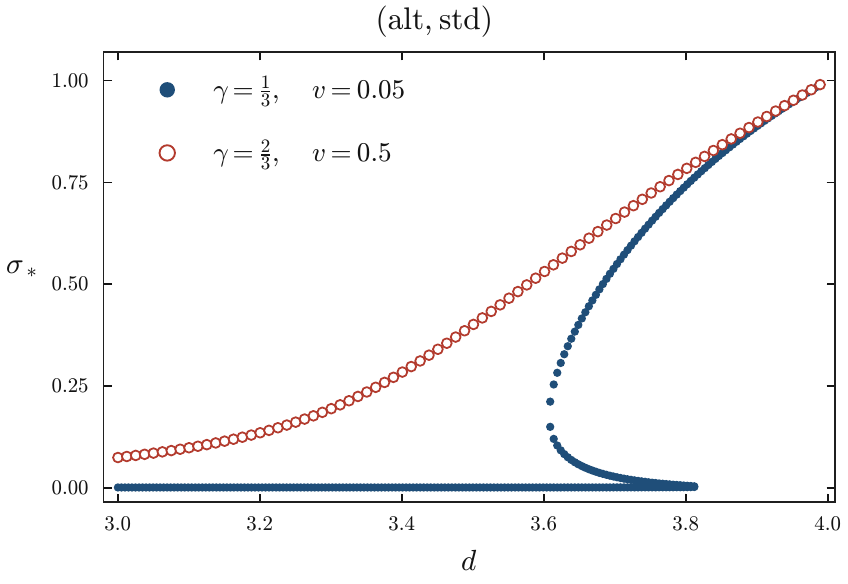}
    \caption{Solutions to the large $N$ gap equation vs $d$ for $v=0.05, \gamma=1/3$ and $v=0.5, \gamma=2/3$. Note that the solution for $\gamma = 1/3, v=0.05$ is small but nonzero for $d$ close to $3$.}
    \label{fig:sigma_d_alt_std}
\end{figure}

When $\epsilon=4-d$ is small, the gap equation can be solved analytically in two limits. First, if we also assume $\sigma_*$ is small, the gap equation reduces to the same equation for the RG FP $c_*$, once we identify $\sqrt{\sigma_*}=|c_*|/2\pi$. Thus the $\epsilon$ expansion and large $N$ results agree exactly in this limit, as expected. The equivalence holds for other choices of boundary conditions as well. 

In addition, one can also find a ``strong-pinning" solution with 
\begin{equation}
 \sqrt{\sigma_*}=1-\frac{\epsilon}{2} - (1-\gamma)\epsilon^2+ {\cal O}(\epsilon^3).   
\end{equation}
Notably, $\sigma_*$ is close to the unitarity (upper) bound, so the scaling dimension for the untwisted fields becomes parametrically small: $\Delta_{\ell=0}=(1-\gamma)\epsilon^2$. We interpret this as the large $N$ incarnation of the runaway FP: $|c|\propto \sqrt{\sigma_*}$ reaches a ${\cal O}(1)$ value at the end of the defect RG flow. From here we refer to the saddle-point solutions for $\sigma_*$ that, at small $\epsilon$, are identified with the runaway RG FPs as ``strong-pinning'' FPs, as opposed to the perturbative FPs with ``weak-pinning''. The parametrically small $\Delta_{\ell=0}$ signals the tendency towards condensation. We note that a similar saddle point was found for a conical defect in \cite{Metlitski:2009iyg}.

Having understood the small $\epsilon$ limit, we now discuss what happens when $\epsilon$ becomes ${\cal O}(1)$. For simplicity, suppose $v$ is ${\cal O}(1)$ and $\gamma$ is small, so we are in the $v\gg \epsilon$ limit. For small $\epsilon < \epsilon_c= \frac{4\gamma\kappa}{(1-\gamma)^2}$, there is only the strong-pinning FP. As $\epsilon$ increases, two perturbative weak-pinning FPs appear once $\epsilon>\epsilon_c$. Thus we expect to see three FPs in a finite range of $\epsilon$. To determine the fate of these FPs for $\epsilon\sim {\cal O}(1)$ we solve the gap equation numerically and an exemplary plot for $v=0.05, \gamma=1/3$ is shown in Fig. \ref{fig:sigma_d_alt_std}. We see that the (continuation of) strong-pinning FP annihilates with the \emph{unstable} perturbative FP, while the stable perturbative FP continues to exist all the way to $d=3$. The qualitative structure of $\sigma_*$ solutions as one varies $\epsilon$ remains largely unchanged for generic values of $v$ and $\gamma<1/2$.

\begin{figure}[t]
    \centering
    \includegraphics[width=1\linewidth]{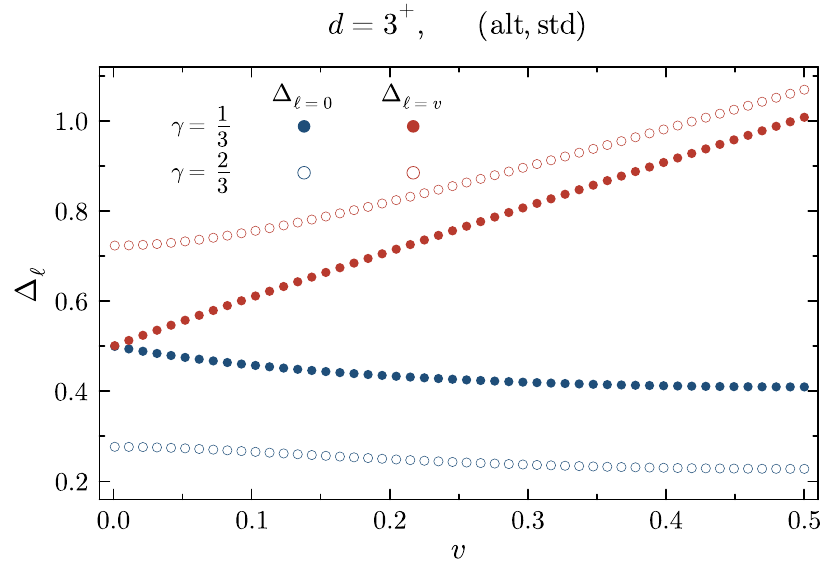}
    \caption{$\Delta_{\ell=0}$ and $\Delta_{\ell=v}$ as a function of $v$ for $\gamma=1/3$ (the saddle point obeys $\sigma_*\rightarrow 0$ as $v\rightarrow 0$) and $\gamma=2/3$ (the $v\rightarrow 0$ limit of $\sigma_*$ does not vanish). The numerical solutions are obtained at $d=3.0001$.}
    \label{fig:Delta_vs_v}
\end{figure}

To summarize, the large $N$ solution suggests that for small $\gamma$ the stable perturbative FP at small $\epsilon$ continues to $d=3$, while the other two do not.
If we further assume $\gamma\ll 1$, this perturbative solution can be calculated analytically for any $v$, and for $d=3$ the result is
\begin{equation}
    \sqrt{\sigma_*}=\frac{\gamma}{\pi}(1-2v)\tan \pi v + {\cal O}(\gamma^2).
\end{equation}
The expression for other $d$ can be found in SM.

Moreover, when $\gamma<1/2$ the stable perturbative solution can be accessed in a different limit: if we assume $v\ll 1, \sigma_*={\cal O}(v^2)$, and $\epsilon=4-d\gg v$, the gap equation can be solved to give
\begin{equation}
  \sqrt{\sigma_*}=\frac{\gamma}{\sqrt{1-2\gamma}}v+\mathcal{O}(v^2). 
  \label{eq:small-v}
\end{equation}

\begin{figure}[t]
    \centering
     \includegraphics[width=\columnwidth]{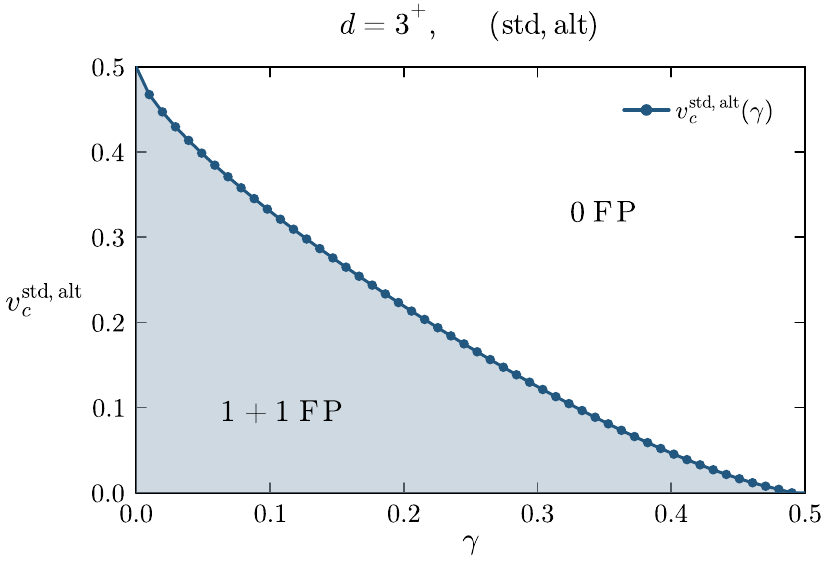}
    \caption{Landscape of solutions at large $N$ to the gap equation with (std, alt) boundary condition, calculated at $d=3.0001$. There are two solutions for $v<v_c^{\text{std},\text{alt}}(d=3, \gamma)$, and no solutions otherwise.}
    \label{fig:phase_diagram_std_alt}
\end{figure}

Now we turn to $\gamma>1/2$.  Numerically, it is found that there is a unique solution for the entire range $d\in [3,4)$, see Fig. \ref{fig:sigma_d_alt_std} for an example. The one near $d=4$ is evidently the strong-pinning solution as discussed previously. In this case, it is reasonable to interpret the $d=3$ solution as the continuation of the strong-pinning FP. Fig. \ref{fig:Delta_vs_v} shows the dependence of $\Delta_{\ell=0}, \Delta_{\ell=v}$ on the monodromy $v$, for $\gamma=1/3$ and $2/3$. Most notably, for $\gamma=2/3$, $\sigma_*$ does not vanish in the $v\rightarrow 0$ limit. In other words, the $v\rightarrow 0$ limit of the defect is not a trivial one, but a nontrivial defect where a fraction of the fields have alternate boundary conditions. This is true for any $\gamma>1/2$. On the other hand, for $\gamma<1/2$, as shown in Eq.~\eqref{eq:small-v} the $v\rightarrow 0$ limit is a trivial defect.

So far we have focused on the (alt for $\ell=0$, std for $\ell=v$) boundary condition. We now briefly discuss what happens with the other choices of boundary condition, for $d=3$ (more details can be found in the SM). For the (std, alt) boundary condition, numerically solving the gap equation finds that for  $d\in [3,4)$,  there exists a $v_c^{\text{std},\text{alt}}(d, \gamma)\geq 0$, such that  there are two physical solutions for $v<v_c^{\text{std},\text{alt}}(d,\gamma)$, and none for $v>v_c^{\text{std},\text{alt}}(d,\gamma)$. The $d=3$ case is shown in Fig. \ref{fig:phase_diagram_std_alt}. Note that in this case $v_c^{\text{std},\text{alt}}(d=3, \gamma)=0$ for $\gamma\geq 1/2$, which is a special feature of $d=3$. One of the two solutions can be solved perturbatively in either $v\ll 1$ or $\gamma\ll 1$ limit. As shown in SM, the solutions are smoothly connected to the $d\rightarrow 4$ limit, where one of them remains perturbative and agrees with the RG result, and the other is the strong-pinning FP associated with runaway flow. We also find that the strong-pinning FP at $d=3$ can be solved perturbatively when in addition to $\gamma\rightarrow 0$ we also take the limit $v\rightarrow 1/2$.  On the other hand, for (alt, alt) or (std, std) boundary conditions, no solutions exist at $d=3$.

Therefore, together with the earlier analysis of (alt, std) solutions, we find that there are three (one) physical solutions to the gap equation at $d=3$ if $v<v_c^{\text{std},\text{alt}}(\gamma)$ ($v>v_c^{\text{std},\text{alt}}(\gamma)$). The extra ones with (std, alt) boundary conditions are smoothly connected to either the perturbatively unstable FP at $d=4-\epsilon$, or to the strong-pinning FP. We believe the one smoothly connected to the strong-pinning FP should describe a stable DCFT, and will refer to it as the (std, alt) DCFTs (when it exists). Thus there are two stable DCFTs for $v<v_c^{\text{std},\text{alt}}(\gamma)$. A summary of the structure of solutions in large $N$ is given in Table~\ref{tab:large-N-fixed-points}.

\begin{figure}
    \centering
    \includegraphics[width=1.0\linewidth]{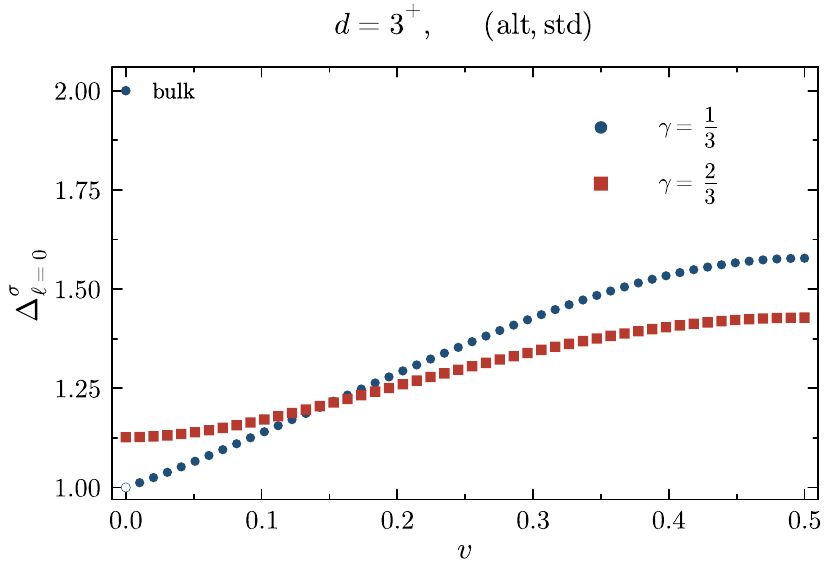}
    \caption{The defect scaling dimension $\Delta^\sigma_{\ell=0}$ of the leading singlet $\hat{\sigma}_{\ell=0}$ as a function of $v$ for $\gamma=1/3$ and $\gamma=2/3$. The large $N$ gap equations are solved at $d=3.0001$. For the branch of solutions with $\gamma = 1/3$, the $v \rightarrow 0$ solution is the trivial defect that has $\Delta_{\ell = 0}^\sigma = \Delta^\sigma= 2$. On the other hand, this is not the case for $\gamma = 2/3$, where the $v \rightarrow 0$ limit is not the trivial defect.}
    \label{fig:sigma_dimension_alt_std}
\end{figure}

Another important test for the stability of DCFTs we find in large $N$ at $d = 3$ is to compute the defect scaling dimension of $\sigma \sim \bm{\Phi}^2$. The leading nontrivial defect singlet, expected to be $\hat{\sigma}_{\ell = 0}$, is the large $N$ incarnation of the defect mass we studied in RG, and is the most relevant possible symmetric defect perturbation. By considering subleading Gaussian fluctuations around the saddle point solutions we have derived, $\sigma - \sigma_*= \frac{\delta \sigma}{\sqrt{N}}$, this dimension can be extracted from the propagator of $\delta \sigma$. The details of the calculation can be found in \ref{sm:defectsigma}. As expected, we find that for the (alt, std) solution and for the (std, alt) solution smoothly connected to the strong-pinning FP at $d = 4-\epsilon$ (when it exists), $\hat{\sigma}_{\ell = 0}$ is irrelevant on the defect, i.e. $\Delta^\sigma_{\ell = 0} > 1$. A plot of $\Delta_{\ell = 0}^\sigma$ for (alt, std) is shown in Fig.~\ref{fig:sigma_dimension_alt_std}. These nontrivial monodromy DCFTs break the $\mathrm{O}(N)$ symmetry of the bulk down to $\mathrm{O}(N-M) \times \mathrm{U}(M/2)$, and consequently, $\hat{\sigma}_{\ell = 0}$ need be a singlet of only the reduced symmetry. One such operator is a combination of symmetric-traceless tensors $T_{ij} \sim \Phi_i \Phi_j - \delta_{ij}\bm{\Phi}^2/N$ with bulk dimension $\Delta^T = 1+\mathcal{O}(N^{-1})$ \cite{BootstrapONPoland}. It is plausible this operator at the defect corresponds to $\hat{\sigma}_{\ell = 0}$, but only for these nontrivial DCFTs (e.g. if $\gamma < \frac{1}{2}$, for $v > 0$), explaining the behavior in Fig.~\ref{fig:sigma_dimension_alt_std}. For the saddle point solution that is smoothly connected to the perturbatively unstable FP at $d = 4 - \epsilon$ (when it exists), $\hat{\sigma}_{\ell = 0}$ is instead relevant, $\Delta^\sigma_{\ell = 0} < 1$. This confirms the preceding comments on the stability of each DCFT.

\section{Monte Carlo simulations}

In this section, we study monodromy line defects in the three-dimensional O($N$) Wilson–Fisher universality class, focusing on $N=3$, using a soft-spin $\phi^4$ model. Throughout this section the model is defined on an $L \times L\times L$ cubic lattice with periodic boundary condition and lattice spacing set to unity.

The dynamical variable at each site $\mathbf{x}$ is an $N$-component real vector $\boldsymbol{\Phi}_\mathbf{x}$.
The lattice Hamiltonian in the presence of a background O($N$) link field $U_{\mathbf{x},\mu}$ is
\begin{align}
    \mathcal{H} [\boldsymbol{\Phi}; U, \tilde{c}] = -\beta \sum_{\mathbf{x}} \sum_{\mu=1}^{3} \boldsymbol{\Phi}_\mathbf{x} \cdot U_{\mathbf{x},\mu}\, \boldsymbol{\Phi}_{\mathbf{x}+\hat{\mu}}  \nonumber \\ + \sum_{\mathbf{x}} \left[ \boldsymbol{\Phi}_\mathbf{x}^2 + \lambda \left( \boldsymbol{\Phi}_\mathbf{x}^2 - 1 \right)^2 \right] + H_{\rm pin}[\boldsymbol{\Phi}; \Tilde{c}],
    \label{eq:lattice_model}
\end{align}
and the partition function is $Z[U,\tilde{c}]=\int \prod_\mathbf{x} d^N \boldsymbol{\Phi}_\mathbf{x} \exp\left(- \mathcal{H} [\boldsymbol{\Phi}; U,\tilde{c}]\right)$.
For $\lambda\to \infty$, Eq.~\eqref{eq:lattice_model} reduces to the hard O($N$) spin model.
In the $(\beta,\lambda)$ plane, the bulk exhibits a second-order transition line.
Using a finite value of $\lambda$ can improve the lattice model by suppressing irrelevant perturbations in the bulk.
Our simulations are performed for the improved model at critical couplings $(\lambda,\beta)=(4.5, 0.6862385)$ for $N=3$~\cite{PhysRevB.65.144520}. 

The monodromy defect is introduced by modifying the nearest-neighbor couplings on an open, codimension-one sheet of bonds.
Specifically, the $y$-directed bonds connecting $(x,y_c,z)$ to $(x,y_c+1,z)$, for $x_1\le x < x_2$ and all $z$, are multiplied by a fixed orthogonal matrix $g\in {
\rm O}(N)$.
In terms of O($N$) link fields,
\begin{equation}
    U_{(x,y,z),y}=g, \ y=y_c, \ x_1 \le x < x_2 ,
    \label{eq:link_field}
\end{equation}
while all other link fields are unity.
The open sheet with twisted link fields has two boundaries, located on dual lattice lines $(x_1-1/2, y_c+1/2)$ and $(x_2-1/2, y_c+1/2)$.
Consequently, on the periodic lattice the implementation contains a pair of oppositely oriented defects, carrying monodromies $g$ and $g^{-1}$. 

We will study two types of monodromy defects:
\begin{enumerate}
\item For a rotational monodromy, the twist acts in the $(\Phi_1,\Phi_2)$ plane as
\begin{equation}
    g(v) = \begin{pmatrix} \cos(2\pi v) & -\sin(2\pi v) \\ \sin(2\pi v) & \cos(2\pi v) \end{pmatrix} \oplus \mathbf{1}_{N-2}.
    \label{eq:rotational}
\end{equation}
Equivalently, for the complex field $\psi=\Phi_1+i\Phi_2$, encircling one of the defect lines imposes Eq.~\eqref{eq:boundary_condition}.
\item For a reflection monodromy, we have $g=\mathbf{1}_{N-1} \oplus (-1)$.
\end{enumerate}
Only the location of the two boundary lines is physical, because the sheet connecting them can be deformed by a local ${\rm O}(N)$ change of variables without changing any gauge-invariant observable.

To represent the defect pertubation $S_{\rm defect}$ in Eq.~\eqref{eq:WF_action}, we introduce the projector $P_{\rm u}$ onto the untwisted components, and denote the four direct-lattice sites surrounding defect line $a \ (a=1,2)$ by
\begin{equation}
    \mathcal{S}_a = \{ (x_a - 1 , y_c), (x_a, y_c), (x_a - 1, y_c+1), (x_a, y_c+1)\} .
    \label{eq:plaquette}
\end{equation}
The lattice pinning term is chosen to be
\begin{equation}
    H_{\rm pin}[\boldsymbol{\Phi}; \Tilde{c}] = \frac{\Tilde{c}}{4} \sum_{z=0}^{L-1}\sum_{p\in S_{1,2}} |P_{\rm u} \boldsymbol{\Phi}_{(p,z)}|^2 .
    \label{eq:H_pin}
\end{equation}
Thus Eq.~\eqref{eq:H_pin} is a lattice representation of the continuum coupling $c$ in $S_{\rm defect}$ in Eq.~\eqref{eq:WF_action}.
For the rotation monodromy Eq.~\eqref{eq:rotational} in $N=3$, we have $P_{\rm u}\boldsymbol{\Phi}=\Phi_3$. For the reflection monodromy, we use $P_{\rm u}\boldsymbol{\Phi}=\Phi_1+i\Phi_2$.
$\Tilde{c}=0$ recovers the bare monodromy defect.

Because the defect lies on the dual lattice, defect operators are naturally represented by superposing those on the four sites on the elementary transverse plaquette (see Eq.~\eqref{eq:plaquette}).
We always measure observables around the first line $\mathcal{S}_1$.

For a rotational monodromy in $N=3$, the defect operators we measure in the simulation are
\begin{align}
    \psi^{\rm lat}(z) &= \frac{1}{4} \sum_{p\in \mathcal{S}} e^{-iv\theta_p} [\Phi_1 (p,z) + i \Phi_2(p,z)] , \label{eq:twisted} \\
    \phi^{\rm lat}(z) &=  \frac{1}{4} \sum_{p\in \mathcal{S}} \Phi_3(p,z) \label{eq:untwisted} .
\end{align}
Here $\theta_p=\arg\left[ (x_p-x_d)+i(y_p-y_d)\right]$ denotes the polar angle from the defect line to the measured site in the plaquette plane, where $(x_d,y_d)=(x_1-1/2, y_c+1/2)$.
Eqs.~\eqref{eq:twisted} and \eqref{eq:untwisted} denote the twisted and untwisted field respectively, where the phase factors are introduced to enforce the expected behavior under SO(2) rotations.

\begin{figure}[t]
    \centering    \includegraphics[width=0.99\columnwidth]{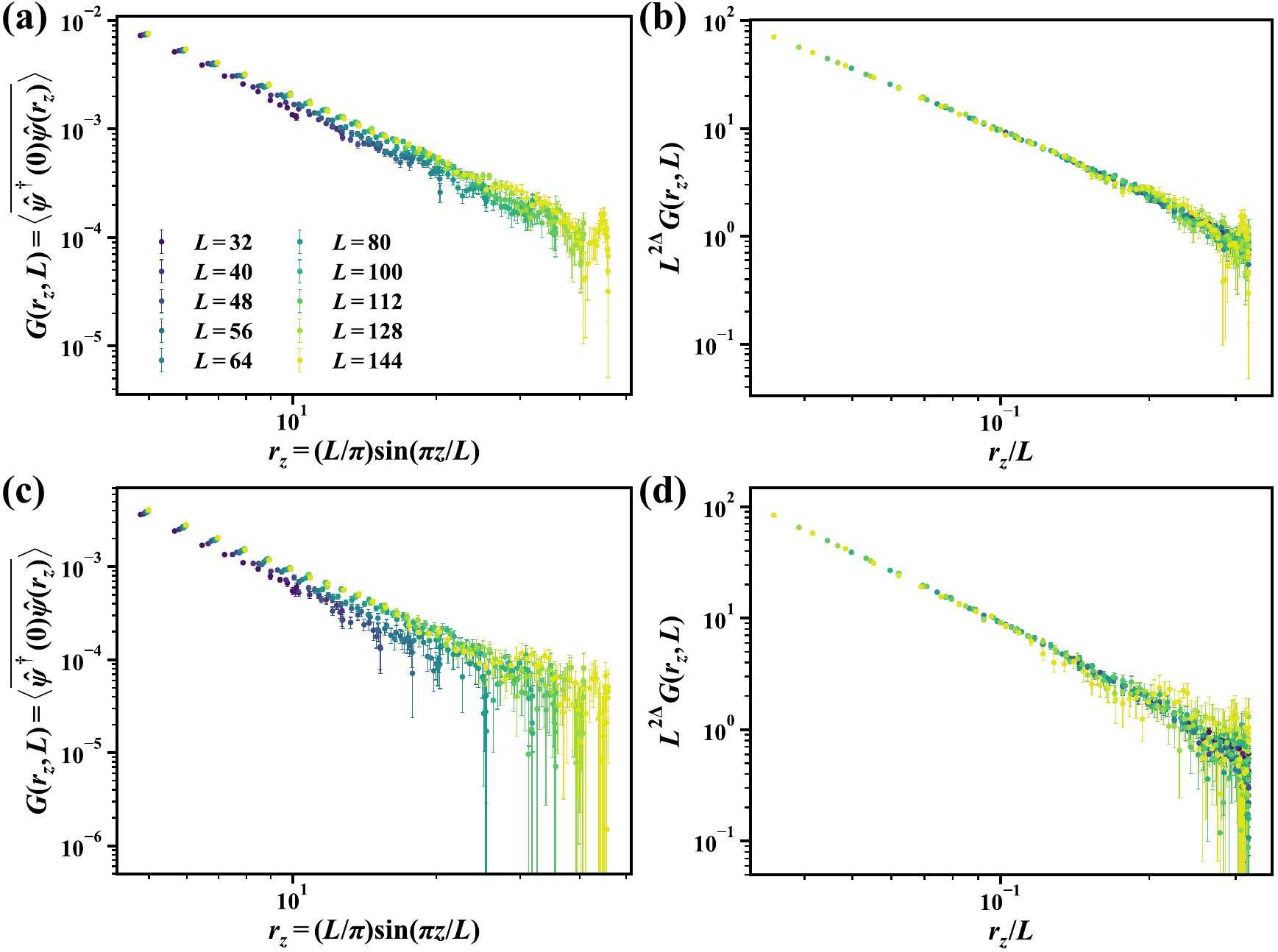}
    \caption{Determination of the scaling dimension of the twisted defect primary $\hat{\psi}$ for $v=1/2$ rotational monodromy in $N=3$ (a)(b) without explicit lattice pinning ($\Tilde{c}=0$) and (c)(d) with explicit lattice pinning ($\Tilde{c}_*=-1.8$) using finite-size scaling collapse. (a)(c) Average two-point correlator $\langle \overline{\hat{\psi}^\dagger(0) \hat{\psi}(r_z)} \rangle$ versus chord distance $r_z=(L/\pi)\sin (\pi z/L)$ for various system sizes $L$. (b)(d) Data collapse using finite-size scaling form in Eq.~\eqref{eq:scaling}, from which we get $\Delta_{\psi}=0.920 \pm 0.036$ for $\Tilde{c}=0$ and $\Delta_{\psi}=0.998 \pm 0.061$ for $\Tilde{c}_*=-1.8$. The fitting cost functions are $S=1.21$ for (b) and $1.03$ for (d).} 
    \label{Fig:twisted_collapse}
\end{figure}

For the reflection monodromy defect, we measure the defect operators
\begin{align}
    \phi^{\rm lat}(z) &= \frac{1}{4} \sum_{p\in \mathcal{S}} [\Phi_1 (p,z) + i \Phi_2(p,z)] , \label{eq:twisted_o1} \\
    \psi^{\rm lat}(z) &=  \frac{1}{4} \sum_{p\in \mathcal{S}} e^{-i\theta_p/2} \Phi_3(p,z) \label{eq:untwisted_o1} .
\end{align}
Eqs.~\eqref{eq:twisted_o1} and \eqref{eq:untwisted_o1} denote the untwisted and twisted field respectively.

In these expressions, the superscript ``lat'' indicates that these are interpolating lattice operators. 
Their long-distance correlation functions are dominated by the corresponding lowest defect primaries, although the microscopic operators may also overlap with descendants and higher-dimensional primaries.
For each run, we measure the two-point correlator for defect operator $\hat{\mathcal{O}}(z)$
\begin{equation}
G(z,L)=\langle \overline{\hat{\mathcal{O}}^\dagger(0) \hat{\mathcal{O}}(z)} \rangle  ,
\end{equation}
where $\hat{\mathcal{O}}(z)$ is defined by Eqs.~\eqref{eq:twisted}, \eqref{eq:untwisted}, \eqref{eq:twisted_o1} and \eqref{eq:untwisted_o1}, and $\langle ...\rangle$ denotes statistical average.
Moreover,
\begin{equation}
    \overline{\hat{\mathcal{O}}^\dagger(0) \hat{\mathcal{O}}(z)} = \frac{1}{L} \sum_{z_0=0}^{L-1} \mathcal{O}(z_0)^* \mathcal{O}(z_0+z), 
\end{equation}
where $0\le z\le L/2$ due to periodic boundary condition.
When fitting to the scaling form, we also replace $z$ by the chord distance $r_z=(L/\pi)\sin (\pi z/L)$.
Due to conformal symmetry, the leading-order finite-size scaling form is
\begin{equation}
\label{eq:scaling}
    G(r_z,L) = L^{-2\Delta}  \mathcal{F}\left( \frac{r_z}{L} \right) ,
\end{equation}
where $\Delta$ is the corresponding scaling dimension and $\mathcal{F}$ is a scaling function.
We determine the scaling dimensions of twisted and untwisted defect primaries by fitting to the scaling form Eq.~\eqref{eq:scaling}, and statistical errors are reported using jackknife method. In the figures we also include the cost functions $S$ for the data collapse (see Eq. \eqref{eq:fitting-cost-function} for the definition).
Moreover, the nonzero pinning strengths $\tilde c_*$ used for the finite-size scaling analysis were selected from preliminary scans with $L\leq64$, favoring good finite-size collapse and no observed signatures of SSB.
Specifically, we used $\tilde c_*=-1.8, -1.5$, and $-1.75$ for the $v=1/2$ rotational, $v=1/4$ rotational, and reflection monodromies, respectively.

\begin{figure}[t]
    \centering    \includegraphics[width=0.99\columnwidth]{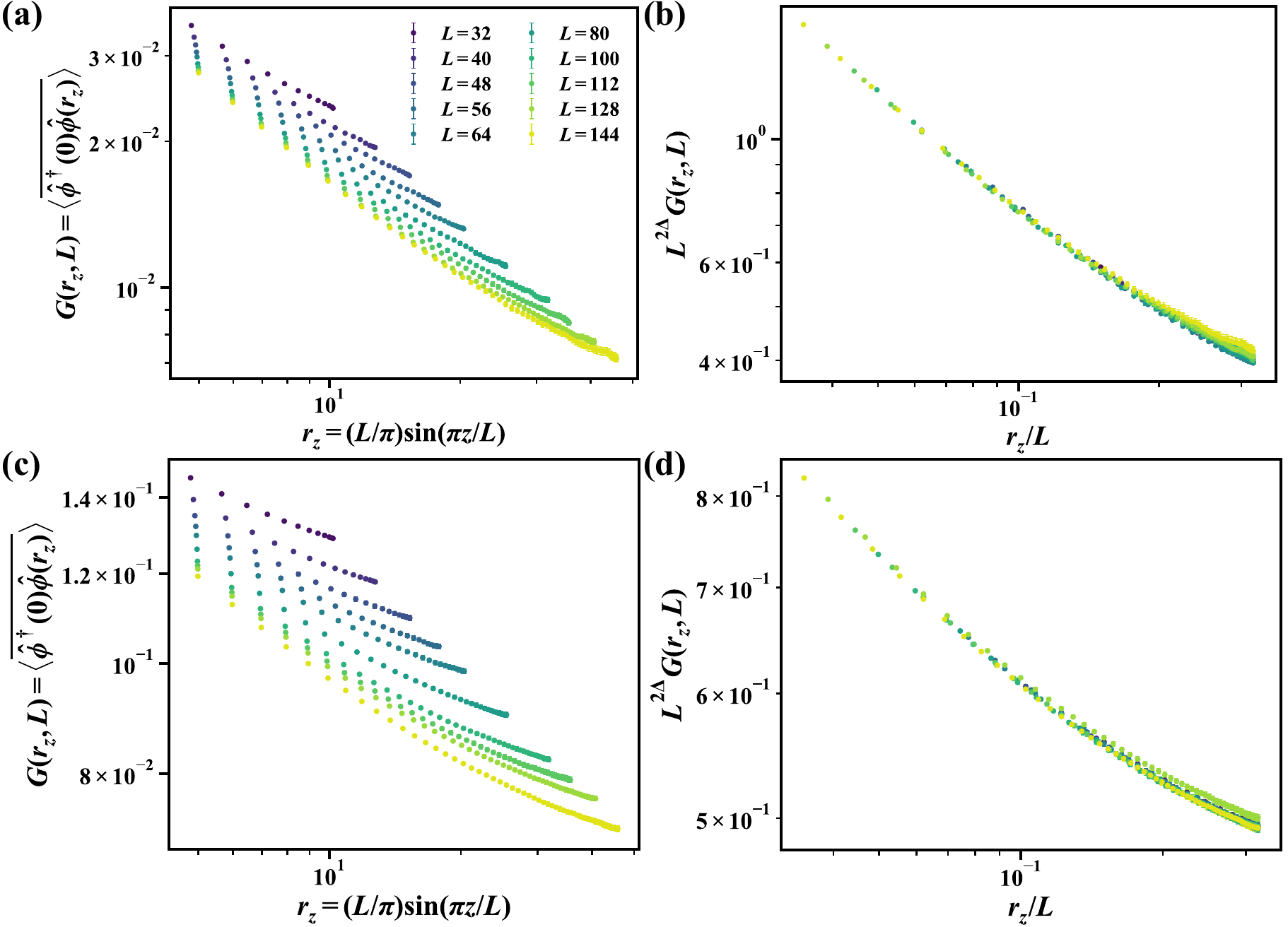}
    \caption{Determination of the scaling dimensions of the untwisted defect primary $\hat{\phi}$ for $v=1/2$ rotational monodromy in $N=3$ (a)(b) without explicit lattice pinning ($\Tilde{c}=0$) and (c)(d) with explicit lattice pinning ($\Tilde{c}_*=-1.8$) using finite-size scaling collapse. We get $\Delta_{\phi}=0.409 \pm 0.011$ when $\Tilde{c}=0$ and $\Delta_{\phi}=0.194 \pm 0.006$ when $\Tilde{c}_*=-1.8$. The fitting cost functions are $S=2.52$ for (b) and $1.85$ for (d).} 
    \label{Fig:untwisted_collapse}
\end{figure}

Figures~\ref{Fig:twisted_collapse} and \ref{Fig:untwisted_collapse} show the results for the $v=1/2$ rotational monodromy.
First, without adding the pinning term (i.e., $\Tilde{c}=0$ in Eq.~\eqref{eq:H_pin}) we  obtain $\Delta_{\psi}=0.920 \pm 0.036$ for the twisted field (Fig. \ref{Fig:twisted_collapse}), and $\Delta_{\phi}=0.409 \pm 0.011$ for the untwisted field (Fig. \ref{Fig:untwisted_collapse}). 

Next we turn on the lattice pinning. In general, we find that when $|\Tilde{c}|$ is not too large, the finite-size correlators can still be well collapsed using the scaling form in Eq. \eqref{eq:scaling} to extract scaling dimensions. We find that as $|\Tilde{c}|$ increases, the extracted $\Delta_{\phi}$ decreases and $\Delta_{\psi}$ increases. For example, for $\Tilde{c}_*=-1.8$, we have $\Delta_{\phi}=0.194 \pm 0.006$, significantly smaller than the value at $\Tilde{c}=0$, and $\Delta_\psi=0.998\pm 0.061$. For comparison, the large $N$ approximation for $\gamma=2/3, v=1/2$ finds $\Delta_{\ell=0}\approx 0.228$ and $\Delta_{\ell=\frac12}\approx 1.069$. 

\begin{figure}[t]
    \centering    \includegraphics[width=0.99\columnwidth]{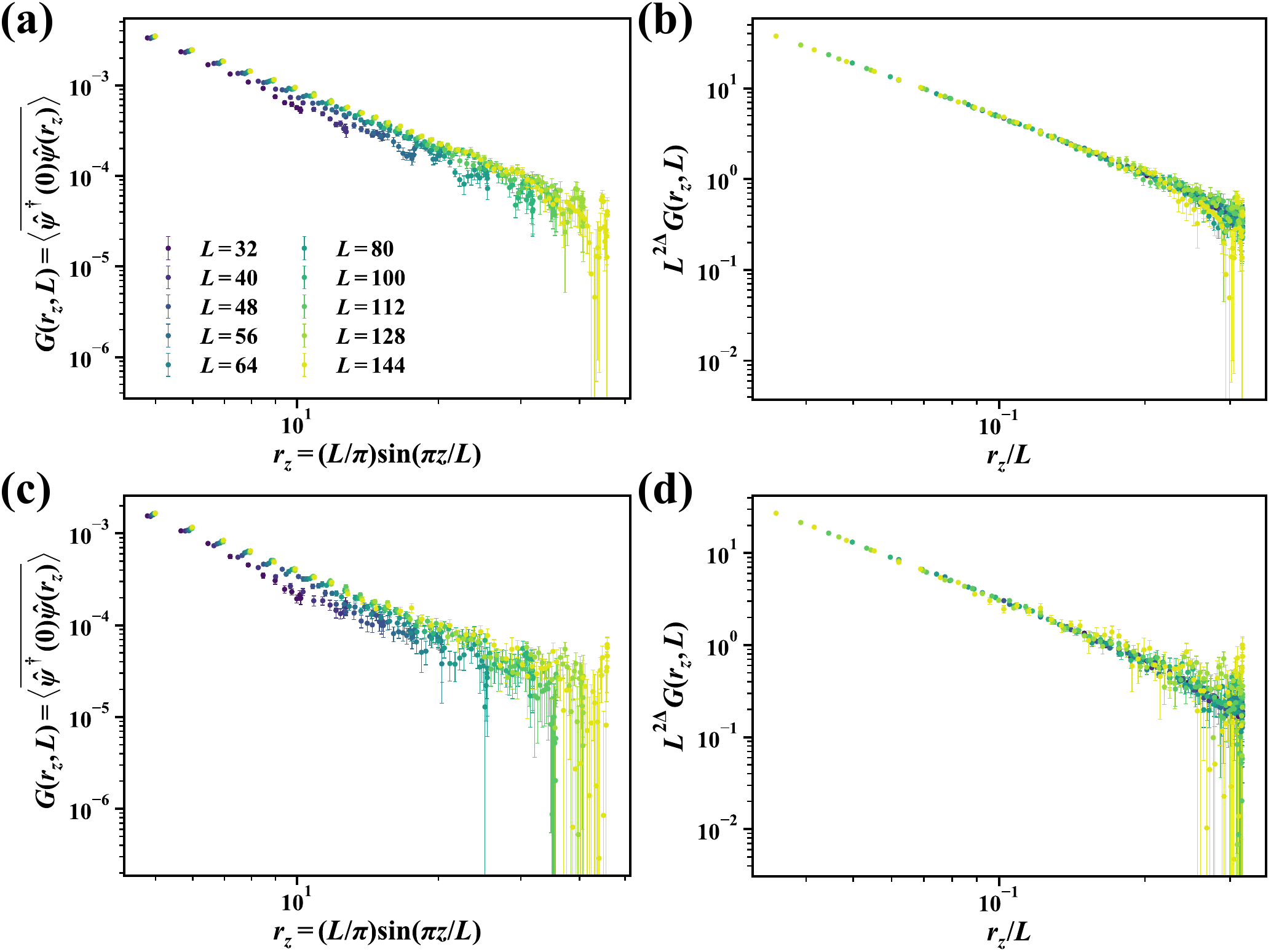}
    \caption{Determination of the scaling dimension of the twisted defect primary $\hat{\psi}$ for reflection monodromy in $N=3$ (a)(b) without explicit lattice pinning ($\Tilde{c}=0$) and (c)(d) with explicit lattice pinning ($\Tilde{c}_*=-1.75$) using finite-size scaling collapse. We get $\Delta_{\psi}=0.934 \pm 0.036$ when $\Tilde{c}=0$ and $\Delta_{\psi}=0.976 \pm 0.066$ when $\Tilde{c}_*=-1.75$. The fitting cost functions are $S=1.09$ for (b) and $S=1.71$ for (d).} 
    \label{Fig:reflection_twist_collapse}
\end{figure}

Importantly, the pinning field can not be made too strong: when $|\Tilde{c}|>2.0$, we observe signatures of the spontaneous breaking of the remaining $\Zbb_2$ symmetry, through the long-range order of $\phi$ along the defect, see Fig. \ref{fig:Z2_SSB}. Across the range of $\tilde{c}$ that shows no SSB on the line defect, including $\tilde{c}=0$, the measured scaling dimensions agree qualitatively with the (alt, std) DCFT, in which $\Delta_\phi$ ($\Delta_\psi$) is below (above) the bulk dimension $\Delta_\Phi\approx 0.519$ \cite{PhysRevB.65.144520, Chester:2020iyt}. 

We have also studied the $v=1/4$ rotational monodromy. The results are qualitatively similar to the $v=1/2$ case, so we leave the details to the SM \ref{sm:MCmore}, and the measured scaling dimensions can be found in Table \ref{tab:defect-dimensions}. 

While we do not fully understand the reasons for the drifting in the fitted scaling dimensions, there are two natural mechanisms: First, it is possible that the drifting is a crossover behavior caused by slow RG flows. Second, the close proximity to the SSB phase can also affect the behavior of the correlation function of the untwisted fields.

Finally, we consider the reflection defect corresponding to $g=\mathbf{1}_2\oplus (-1)$. In this case, the remaining internal symmetry is O(2), which can not be broken spontaneously on a line \cite{Cuomo:2023qvp}. Hence this setup should be less prone from the influence of the proximate defect SSB phase.

\begin{figure}[t]
    \centering    \includegraphics[width=0.99\columnwidth]{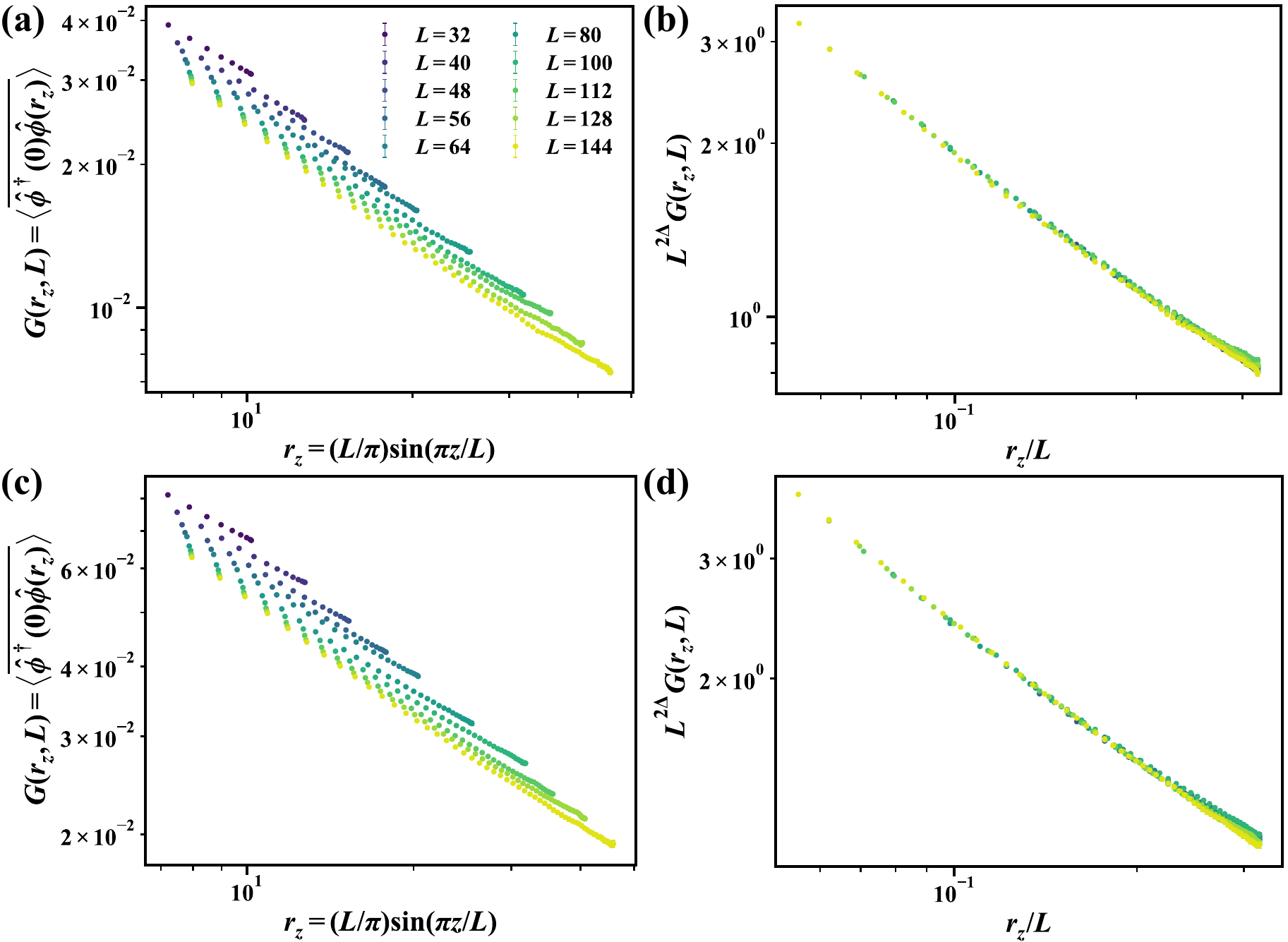}
    \caption{Determination of the scaling dimension of the untwisted defect primary $\hat{\phi}$ for reflection monodromy in $N=3$ (a)(b) without explicit lattice pinning ($\Tilde{c}=0$) and (c)(d) with explicit lattice pinning ($\Tilde{c}_*=-1.75$) using finite-size scaling collapse. We get $\Delta_{\phi}=0.472 \pm 0.006$ when $\Tilde{c}=0$ and $\Delta_{\phi}=0.411 \pm 0.006$ when $\Tilde{c}_*=-1.75$. The fitting cost functions are $S=2.47$ for (b) and $S=3.20$ for (d).} 
    \label{Fig:reflection_untwist_collapse}
\end{figure}

Indeed, in this case the drifting of the scaling dimensions with the pinning field appears to be much weaker: For the twisted field $\psi$, the scaling dimension is found to be $\Delta_\psi=0.934\pm 0.036$ for $\tilde{c}=0$ and $\Delta_\psi=0.976\pm 0.066$ for $\tilde{c}_*=-1.75$ , which agree within error bars (Fig. \ref{Fig:reflection_twist_collapse}). For the untwisted field $\phi$, we have $\Delta_{\phi}=0.472\pm 0.006$ with $\Tilde{c}=0$ and $\Delta_{\phi}=0.411\pm 0.006$ with $\Tilde{c}_*=-1.75$ (Fig. \ref{Fig:reflection_untwist_collapse}). The discrepancy is larger for the untwisted field, which we believe is likely due to finite-size effects. For comparison, the large $N$ approximation for $\gamma=1/3, v=1/2$ gives $\Delta_{\ell=0}\approx 0.410$ and $\Delta_{\ell=\frac12}\approx 1.008$, which are close to the numerical values with pinning.

We summarize the defect scaling dimensions from the Monte Carlo simulations of the three types of monodromy defects in Table. \ref{tab:defect-dimensions}.

\begin{table}[t]
\centering
\caption{Scaling dimensions of untwisted ($\phi$) and twisted ($\psi$) defect primaries
from Monte Carlo (MC) simulations at $N=3$ and the large-$N$ approximation for $d=3$.
In the order shown, the MC pinning strengths are
$\tilde c_*=-1.8,-1.5,-1.75$, and the large $N$ ratios are
$\gamma=2/3,2/3,1/3$.
The $\mathbb{Z}_2$ reflection corresponds to $v=1/2$ and $\gamma = 1/3$ in large $N$.}
\label{tab:defect-dimensions}
\small
\setlength{\tabcolsep}{4pt}
\setlength{\arrayrulewidth}{0.4pt}
\setlength{\doublerulesep}{1.5pt}
\setlength{\lightrulewidth}{0.4pt}
\renewcommand{\arraystretch}{1.15}
\begin{tabular*}{\columnwidth}{@{\extracolsep{\fill}}cccc@{}}
\hline\hline
\addlinespace[2pt]
$\mathrm{SO}(2)$, $v=1/2$ &  MC $(\tilde{c} = 0)$ & MC $(\tilde{c}_* = -1.8)$ & Large $N$ \\
\midrule
$\Delta_{\phi}$ &$0.409 \pm 0.011$& $0.194\pm0.006$ & $0.228$ \\
$\Delta_{\psi}$ &$0.920 \pm 0.036$ & $0.998\pm0.061$ & $1.069$ \\
\addlinespace[7pt]
$\mathrm{SO}(2)$, $v=1/4$ &  MC $(\tilde{c} = 0)$ & MC $(\tilde{c}_* = -1.5)$ & Large $N$ \\
\midrule
$\Delta_{\phi}$ & $0.461 \pm 0.011$ & $0.270\pm0.006$ & $0.243$ \\
$\Delta_{\psi}$ & $0.632 \pm 0.011$ & $0.725\pm0.016$ & $0.860$ \\
\addlinespace[7pt]
$\mathbb{Z}_2$ reflection & MC $(\tilde{c} = 0)$& MC $(\tilde{c}_* = -1.75)$ & Large $N$ \\
\midrule
$\Delta_{\phi}$ & $0.472 \pm 0.006$ & $0.411\pm0.006$ & $0.410$ \\
$\Delta_{\psi}$ &$0.934 \pm 0.036$ & $0.976\pm0.066$ & $1.008$ \\
\addlinespace[2pt]
\hline\hline
\end{tabular*}
\end{table}

\section{Conclusions and Discussion}
In this work, we have explored the possible DCFT FPs of partial-monodromy defects in $\text{O}(N)$ CFTs using a variety of analytical tools as well as classical Monte Carlo simulation. Such monodromy defects are natural objects to consider as they are generated by threading flux or by acting with a disorder operator of a $\mathrm{U}(1)$ symmetry. In dimensions $d$ close to $4$, we find an array of fixed points in both $\epsilon$ expansion and large $N$ labeled by so-called ``standard'' or ``alternate'' boundary conditions for the twisted field $\bm{\psi}$ and untwisted field $\bm{\phi}$. Upon continuing down to $d = 3$ in large $N$, we find at least one fixed point, expected to be stable, which has alternate (standard) boundary conditions for the untwisted (twisted) field. Monte Carlo simulations of SO(2) monodromy defects for $N = 3$ are suggestive of the existence of a DCFT, but exhibit drifting of the scaling dimensions and SSB seemingly for sufficiently large pinning field strength. On the other hand, the numerical results for the reflection defect show less drifting and no SSB, thus support the identification of a DCFT. For some range of monodromy $v$ and ratio of twisted fields $\gamma$, additional FPs with standard (alternate) boundary conditions for the untwisted (twisted) field can be found.

The drifting behavior of scaling dimensions seen in the Monte Carlo simulation for the rotational monodromy is peculiar. Considering this drift is caused by pinning of the untwisted field, a completely symmetric defect perturbation, one possible explanation is that this operator is generated in RG but the flow is extremely slow. If so, we would expect that increasing the pinning field strength on the lattice might drive us closer towards reaching the expected stable DCFT, but for sufficiently large $|\tilde{c}|$, SSB of the residual $\mathbb{Z}_2$ symmetry of the untwisted field is observed. The proximity of the SSB phase might also cause issues with extracting the scaling dimensions. The drift in scaling dimensions being much weaker for the case of reflection monodromy where there is no SSB may be suggestive of this. Likewise, we would not expect to see SSB on the defect when there is more than one untwisted field for $N > 3$.

The large $N$ analysis reveals two stable DCFTs at $d=3$, for $v<v_c^{\text{std},\text{alt}}(\gamma)$. The Monte Carlo simulations appear to find the (alt, std) DCFT. It would be interesting to realize the (std, alt) DCFT, presumably by reversing the sign of the pinning field $\tilde{c}$ to favor the standard boundary condition on the untwisted field.

We have focused on the existence of different possible partial-monodromy DCFT FPs and, in particular, their local properties such as defect scaling dimensions of the leading $\mathrm{O}(N)$ vector defect operators. In the future, it would be interesting to compute universal contributions to the free energy of these defects, such as the defect entropy and cusp anomalous dimension of monodromy line defects in $d = 3$. The cusp anomalous dimension of such a defect may be particularly relevant to studies of the disorder operator in quantum lattice models \cite{Zhao:2020vdn, Wang:2021lmb, Wang:2021yaf}.  We leave these investigations to future work. A complementary study of monodromy defects in three-dimensional $\text{O}(N)$ CFTs using fuzzy-sphere regularization will be presented in an upcoming work \cite{fuzzy3}. 

\section{Acknowledgments}
MC acknowledges enlightening conversations with Yichul Choi,  Max Metlitski, Zimo Sun, Yifan Wang and Xiao-Chuan Wu, and is grateful for Chunhao Guo and Zheng Yan for sharing unpublished results. MC would like to thank the hospitality from  the Institute for Advanced Study where part of the work was done. JF acknowledges helpful discussion with Michael P. Zaletel. We thank Barry Cimring and Yifan Wang for helpful comments on a draft, and Wei Zhu for informing us about upcoming related work and sharing unpublished results. MC is supported by NSF grant DMR-2424315. GS and NMJ acknowledge support through the Yale Prize Postdoctoral Fellowship in Condensed Matter Theory. TW acknowledges support from the Harvard Quantum Initiative Fellowship. JF acknowledges support by the U.S. Department of Energy, Office of Science, Office of Basic Energy Sciences, Materials Sciences and Engineering Division under Contract No. DE-AC02-05-CH11231 (Theory of Materials program KC2301). This research uses the Lawrencium computational cluster provided by the Lawrence Berkeley National Laboratory (supported by the U.S. Department of Energy, Office of Basic Energy Sciences under Contract No. DE-AC02-05-CH11231). Generative AI tools have been used to assist coding, checking derivations and creating figures.
\newline 

\noindent \emph{Note added:} During the late stages of finalizing the manuscript, we became aware of related works \cite{qian2026monodromylinedefectsfuzzy, ZhuMonodromy, Tang:2026xny} that study monodromy defects in ${\rm O}(N)$ CFTs on fuzzy sphere. 

\bibliography{main}

\clearpage
\onecolumngrid

\setcounter{page}{1}
\renewcommand{\thepage}{S\arabic{page}}

\setcounter{section}{0}
\setcounter{equation}{0}
\setcounter{figure}{0}
\setcounter{table}{0}

\renewcommand{\thesection}{S\arabic{section}}
\renewcommand{\theequation}{S\arabic{equation}}
\renewcommand{\thefigure}{S\arabic{figure}}
\renewcommand{\thetable}{S\arabic{table}}

\def\theHequation{supp.\arabic{equation}}
\def\theHfigure{supp.\arabic{figure}}
\def\theHtable{supp.\arabic{table}}
\def\theHsection{supp.\arabic{section}}
\def\theHsubsection{\theHsection.\arabic{subsection}}
\def\theHsubsubsection{\theHsubsection.\arabic{subsubsection}}


\begin{center}
{\large\bfseries Supplemental Material}
\end{center}
\vspace{0.5em}

\section{Renormalization group calculation and additional details}
\label{sm:RG}

We return to the Wilson-Fisher theory described in the main text. The action can be rewritten as  
\begin{equation}
\begin{aligned}
    S &= S_\Phi + S_\text{defect}\\
    S_\Phi &= \int \dd^d{x} \text{ }\left[\frac{1}{2}(\partial \bm{\Phi})^2 + \frac{\lambda}{4!}(\bm{\Phi}^2)^2 \right]=  \int \dd^d{x}\text{ } \left[\frac{1}{2}|\partial \bm{\psi}|^2 + \frac{1}{2}(\partial \bm{\phi})^2 + \frac{\lambda}{4!}(\bm{\phi}^2)^2 + \frac{\lambda}{4!}(|\bm{\psi}|^2)^2 + \frac{2\lambda}{4!}|\bm{\psi}|^2\bm{\phi}^2\right]\\
    S_\text{defect} &= \int \dd^{d-2}x \text{ } \left[\frac{c}{2}\hat{\bm{\phi}}_{\ell=0}^2 + \frac{h}{2}| \hat{\bm{\psi}}_{\ell=v}|^2\right].
\end{aligned}
\end{equation}
The free bulk-to-bulk propagator in $d=4 - \epsilon$ for the twisted fields is the following \cite{Soderberg, GiombiHyperbolic}
\begin{align}
    G^\psi(\vec{x}-\vec{y},r_x,r_y,\theta) &= \sum_{\ell \in \Zbb + v} G_{\ell}^\psi(\vec{x}-\vec{y},r_x,r_y) e^{i \ell \theta}, \label{eq:propagators}\\
    G^\psi_{\ell \in \Zbb+v}(\vec{x}-\vec{y},r_x,r_y) &= \langle \psi_{i,\ell}^\dagger(\vec{x},r_x) \psi_{i,\ell}(\vec{y},r_y) \rangle =  \frac{1}{2 \pi^{d/2} (r_x r_y)^{\Delta}} \frac{\Gamma(\Delta_\ell)}{\Gamma(\Delta_\ell+2-\frac{d}{2})} \frac{_2F_1(\Delta_\ell,\Delta_\ell + \frac{3-d}{2},2\Delta_\ell+3-d,-\zeta^{-1})}{(4\zeta)^{\Delta_\ell}}. \nonumber
\end{align}
The propagator for the untwisted fields we denote as $G_{\ell}^\phi(\vec{x}-\vec{y},r_x,r_y)$ which takes the same form but with half the normalization and $\ell \in \mathbb{Z}$. Here, $\zeta = (|\vec{x}-\vec{y}|^2+(r_x-r_y)^2)/4r_xr_y$ is the cross-ratio. At coincident points, i.e. in the $\zeta \rightarrow 0$ limit, this yields the bulk one-point function of the twisted field mass
\begin{equation}
    G^\psi(\vec{x}-\vec{x},r_x,r_x,0) = \langle |\psi_{i}(\vec{x},r_x,\theta)|^2 \rangle = \frac{C_\psi^\pm(\epsilon)}{r_x^{2\Delta}} \text{ and } C_\psi^\pm(0) = \frac{v(v \mp 1)}{4 \pi^2}
\end{equation}
with $C^+_\psi$ ($C^-_\psi$) corresponding to the normalization of twisted fields depending when specifically mode $\bm{\psi}_{\ell = v}$ has standard (alternate) boundary conditions and the rest of the modes are in standard. The contraction of a bulk field with a defect field can be written in terms of a bulk-to-defect propagator. For instance, for the untwisted fields
\begin{equation}
\begin{aligned}
    \langle \hat{\phi}^\dagger_{i,\ell}(\vec{x}) \phi_{i}(\vec{y},r_y,\theta) \rangle = e^{i \ell \theta} K^\phi_\ell(\vec{x}-\vec{y},r_y) \text{ with } K^\phi_\ell(\vec{x}-\vec{y},r_y) 
    &= \lim_{r_x \rightarrow 0} r_x^{\Delta-\Delta_\ell} G_\ell^\phi(\vec{x}-\vec{y},r_x,r_y),\\
    &= \frac{1}{4 \pi^{d/2} r_y^\Delta} \frac{\Gamma(\Delta_\ell)}{\Gamma(\Delta_\ell+2-\frac{d}{2})} \bigg ( \frac{r_y}{|\vec{x}-\vec{y}|^2+r_y^2} \bigg )^{\Delta_\ell}.
\end{aligned}
\end{equation}
Likewise the defect-to-defect propagator for the untwisted fields is given by $\hat{G}_\ell^\phi(\vec{x}-\vec{y}) = \lim_{r_y \rightarrow 0} r_y^{\Delta-\Delta_\ell} K_\ell^\phi(\vec{x}-\vec{y},r_y)$, or just
\begin{equation}
    \hat{G}_\ell^\phi(\vec{x}-\vec{y}) = \frac{1}{4 \pi^{d/2}} \frac{\Gamma(\Delta_\ell)}{\Gamma(\Delta_\ell+2 - \frac{d}{2})} \frac{1}{|\vec{x}-\vec{y}|^{2\Delta_\ell}}.
\end{equation}
The bulk-to-defect and defect-to-defect propagators for the twisted fields take the same functional form but with twice the normalization and the appropriately shifted spin.

We renormalize the Wilson-Fisher theory by rewriting bare quantities in terms of renormalized quantities and counterterms as $\bm{\phi}_0 = \sqrt{Z_\phi} \bm{\phi}, \lambda_0 = \mu^\epsilon(\lambda+\delta \lambda)/Z_\phi^2$, $c_0 = \mu^\xi (c+\delta c)/Z_\phi$ and $h_0 = \mu^{\xi - 2v}(h+\delta h)/Z_\phi$. Here, we have introduced a new regulator $\xi$ to control the dimensions of the defect fields which are modified to $\Delta_{\ell = 0} = \frac{d}{2}-1 - \frac{\xi}{2}$ and $\Delta_{\ell = v} = \frac{d}{2}-1+v - \frac{\xi}{2}$. Note that this modification changes the bulk-to-defect OPE accordingly, but the bulk dimensions of $\bm{\phi}$ and $\bm{\psi}$ remain the same. We will only work at 1-loop where $Z_\phi = 1$ and with $v \ll 1$ and $\epsilon \ll 1$. Additionally, we will be free to choose $\bm{\psi}_{\ell=v}$ having standard boundary conditions ($C_\psi^+(0) = -\frac{\kappa}{4 \pi^2}$) without loss of generality as alternate boundary conditions for this mode will be accessible perturbatively.

\begin{figure}
    \centering    \includegraphics[width=0.9\linewidth]{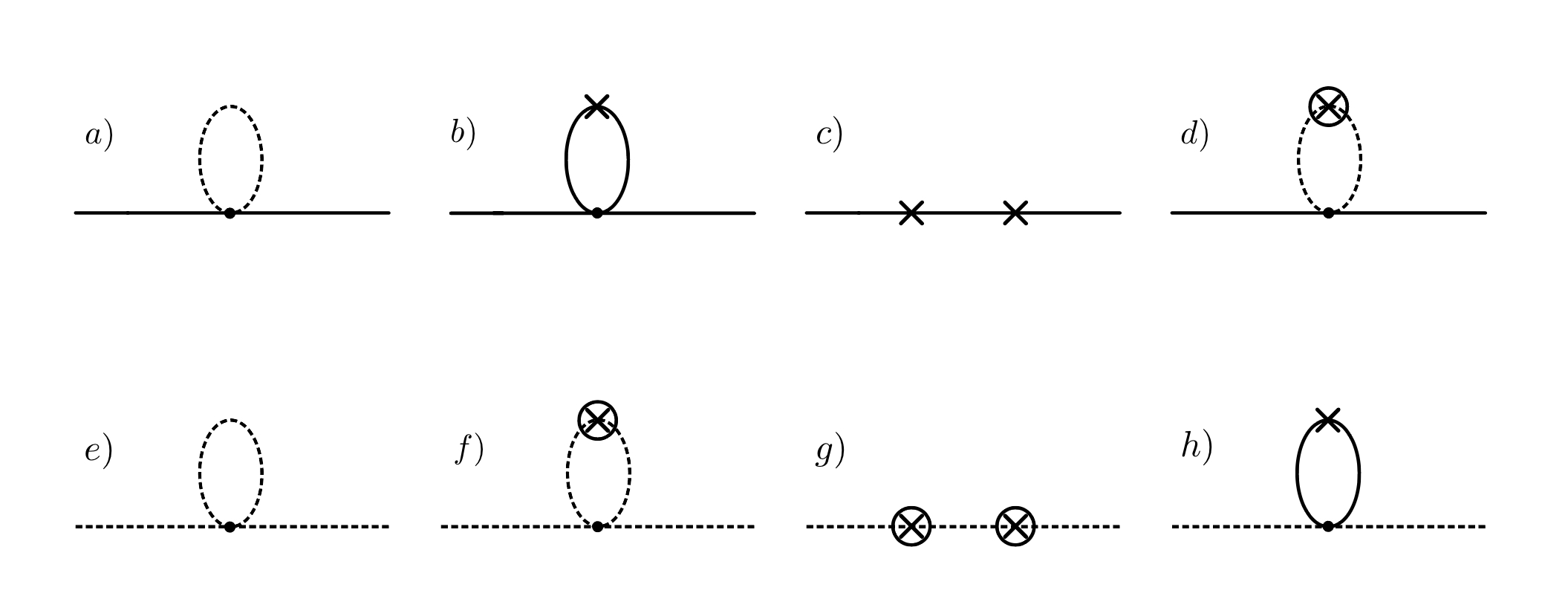}
    \caption{UV divergent diagrams for the bulk-to-bulk propagators $G_{\ell = 0}^\phi(\vec{x}-\vec{y},r_x,r_y)$ (row 1, diagrams $a$ through $d$) and $G_{\ell=v}^\psi(\vec{x}-\vec{y},r_x,r_y)$ (row 2, diagrams $e$ through $h$) that renormalize $c,h$. Vertex factors from the bulk interaction $\lambda$ are denoted by a solid dot, while those of the defect perturbations $c$ and $h$ are denoted by a cross and a circled cross respectively.} 
    \label{fig:diagrams}
\end{figure}

The $\beta$-functions for $c$ and $h$ can be derived by requiring the appropriate bulk-to-bulk propagators remain finite. At leading order, for the zero mode of the untwisted field there are four UV divergent diagrams that contribute to the bulk-to-bulk propagator which can be cured by appropriate choice of counterterm $\delta c$. These four diagrams are shown in the first row of Fig.~\ref{fig:diagrams}. The form of the correction to the bulk-to-bulk propagators from the counterterms are
\begin{align}
    \delta^\text{c.t.} G_{\ell = 0}^\phi (\vec{x}-\vec{y},r_x,r_y) &= - \delta c \int \dd^{2}{\vec{z}} \text{ } K_{\ell = 0}^\phi(\vec{x}-\vec{z},r_x) K_{\ell = 0}^\phi (\vec{z}-\vec{y},r_y),\\
    \delta^{\text{c.t.}} G_{\ell=v}^\psi(\vec{x}-\vec{y},r_x,r_y) &= - \frac{ \delta h}{2} \int \dd^{2}{\vec{z}} \text{ }K_{\ell=v}^\psi(\vec{x}-\vec{z},r_x) K_{\ell=v}^\psi(\vec{z}-\vec{y},r_y).
\end{align}
In computing the UV divergent part of diagrams we evaluate any prefactors of integrals at $\epsilon = 0$, $\xi = 0$ and $v = 0$ if they are finite in order to simplify expressions. 

The first diagram $a)$ is computed as follows
\begin{align}
    \delta^1 G^{\phi}_{\ell=0}(\vec{x}-\vec{y},r_x,r_y) &= - \frac{M\lambda \mu^\eps }{2 \times 3!} \int \dd^d{z} \text{ } G^\phi_{\ell=0}(\vec{x}-\vec{z},r_x,r_z) G^\psi(0,r_z,r_z,0) G^\phi_{\ell=0}(\vec{z}-\vec{y},r_z,r_y), \nonumber\\
    &\sim -\frac{\pi M \lambda   C_\psi^+(0)}{6} \int \frac{\dd^{d-2}{\vec{z}} \dd{r_z}}{r_z^{2\Delta-1}} \text{ }G^\phi_{\ell=0}(\vec{x}-\vec{z},r_x,r_z) G^\phi_{\ell=0}(\vec{z}-\vec{y},r_z,r_y). \nonumber
\end{align}
As $r_z \rightarrow 0$, $G^\phi_{\ell = 0}(\vec{x}-\vec{z},r_x,r_z) \rightarrow r_z^{-\xi/2} K_{\ell=0}^\phi(\vec{x}-\vec{z},r_x)$. This integral is UV divergent at the defect as $\Delta = \frac{d}{2}-1 = 1 - \frac{\epsilon}{2}$. We find the divergent contribution to be, after expanding to leading order in $\epsilon$ and $\xi$,
\begin{equation}
    \delta^1 G_{\ell = 0}^\phi(\vec{x}-\vec{y},r_x,r_y) = -\frac{ \pi M \lambda  C_\psi^+(0) }{6 (\epsilon-\xi)} \int \dd^{2}{\vec{z}} \text{ } K_{\ell = 0}^\phi(\vec{x}-\vec{z},r_x) K_{\ell = 0}^\phi(\vec{z}-\vec{y},r_y)
\end{equation}
which is canceled by choosing counterterm $\delta^1 c =  -\frac{\pi M \lambda   C_\psi^+(0)}{6 (\epsilon-\xi)} =  \frac{M \lambda   \kappa}{24 \pi (\epsilon-\xi)}$. 

The second diagram $b)$ is computed as
\begin{align*}
    &\delta^2 G^{\phi}_{\ell=0}(\vec{x}-\vec{y},r_x,r_y) = \frac{(N-M+2)\lambda c \mu^{\epsilon+\xi} }{6} \int \dd^dz_1 \dd^{d-2}{\vec{z}_2} \text{ } G_{\ell = 0}^\phi(\vec{x}-\vec{z}_1,r_x,r_{z_1})K_{\ell=0}^\phi(\vec{z}_1-\vec{z}_2,r_{z_1})^2 G_{\ell = 0}^\phi(\vec{z}_1-\vec{y},r_{z_1},r_y),\\
    &\sim  \frac{\pi (N-M+2) \lambda c}{48 \pi^4} \int \dd^{2}{\vec{z}_1} \dd{r_{z_1}} \text{ }r_{z_1}^{1-\xi} \bigg [ \int \frac{ \dd^{d-2}{\vec{z}_2}}{(|\vec{z}_1-\vec{z}_2|^2+r_{z_1}^2)^{d-2-\xi}} \bigg ] G_{\ell = 0}^\phi(\vec{x}-\vec{z}_1,r_x,r_{z_1})G_{\ell = 0}^\phi(\vec{z}_1-\vec{y},r_{z_1},r_y) .
\end{align*}
The integral over $\vec{z}_2$ can be performed directly yielding a finite result: $\int \frac{ \dd^{d-2}{\vec{z}_2}}{(|\vec{z}_1-\vec{z}_2|^2+r_{z_1}^2)^{d-2-\xi}} = \frac{\pi^{d/2-1}\Gamma(\frac{d-2}{2}-\xi)}{\Gamma(d-2-\xi) r_{z_1}^{d-2-2\xi}}$. With this in mind, the correction simplifies to
\begin{equation*}
    \delta^2 G_{\ell=0}^\phi(\vec{x}-\vec{y},r_x,r_y) = \frac{(N-M+2) \lambda c  }{48 \pi^2} \int \frac{\dd^{d-2}{\vec{z}_1} \dd{r_{z_1}}}{r_{z_1}^{d-3-\xi}} \text{ } G_{\ell = 0}^\phi(\vec{x}-\vec{z}_1,r_x,r_{z_1})G_{\ell = 0}^\phi(\vec{z}_1-\vec{y},r_{z_1},r_y). 
\end{equation*}
Again because as we approach the defect, $G_{\ell=0}^\phi(\vec{x}-\vec{z},r_x,r_z) \rightarrow r_z^{-\xi/2} K_{\ell = 0}^\phi(\vec{x}-\vec{z},r_x)$ we get that the UV divergent piece is 
\begin{equation}
    \delta^2 G^{\phi}_{\ell = 0}(\vec{x}-\vec{y},r_x,r_y) = \frac{ (N-M+2) \lambda c }{48 \pi^2 \epsilon} \int \dd^{2}\vec{z}_1 \text{ }  K_{\ell=0}^\phi(\vec{x}-\vec{z}_1,r_{x}) K_{\ell=0}^\phi(\vec{z}_1-\vec{y},r_{y})
\end{equation}
which is canceled by choosing $\delta^2 c = \frac{\lambda c (N-M+2)}{48 \pi^2 \epsilon}.$

We now compute the third diagram $c)$. This correction is at leading order
\begin{equation*}
    \delta^3 G_{\ell=0}^\phi(\vec{x}-\vec{y},r_x,r_y) = c^2 \mu^{2 \xi} \int \dd^{d-2}{\vec{z}}_1 \dd^{d-2}{\vec{z}_2} \text{ } K_{\ell=0}^\phi(\vec{x}-\vec{z}_1,r_x) \hat{G}_{\ell=0}^\phi (\vec{z}_1-\vec{z}_2) K_{\ell=0}(\vec{z}_2-\vec{y},r_y).
\end{equation*}
The integral over $\vec{z}_2$, which is logarithmically divergent, has to be regularized by hand. This is done by shifting $\vec{z}_2 \rightarrow \vec{z}_2 + \vec{z}_1$ for which the UV divergent contribution occurs now at $\vec{z}_2 = 0$. After this, we have simply $\int \dd^{2}{\vec{z}_2} \text{ }\hat{G}_{\ell=0}^\phi(\vec{z}_2) = \frac{1}{4 \pi^2} \int \dd^{2}{\vec{z}_2} \text{ }|\vec{z}_2|^{-(2-\xi)}\!=\! \frac{1}{2\pi\xi}$. 
 
The final result is that the third diagram is set to be
\begin{equation}
    \delta^3 G_{\ell=0}^\phi(\vec{x}-\vec{y},r_x,r_y) = \frac{c^2  }{2 \pi \xi} \int \dd^{2}\vec{z}_1 \text{ } K_{\ell=0}^\phi(\vec{x}-\vec{z}_1,r_x) K_{\ell=0}^\phi(\vec{z}_1-\vec{y},r_y)
\end{equation}
which is canceled by choosing counterterm $\delta^3 c = \frac{c^2 }{2 \pi \xi}$. 

The fourth diagram $d)$ contributes a correction after expanding finite prefactors accordingly
\begin{align}
    & \delta^4 G^\phi_{\ell = 0}(\vec{x}-\vec{y},r_x,r_y) = \frac{M\lambda h\mu^{\epsilon-2v+\xi}  }{4!} \int \dd^dz_1 \dd^{d-2}{\vec{z}_2} \text{ } G_{\ell = 0}^\phi(\vec{x}-\vec{z}_1,r_x,r_{z_1})K_{\ell=v}^\psi(\vec{z}_1-\vec{z}_2,r_{z_1})^2 G_{\ell = 0}^\phi(\vec{z}_1-\vec{y},r_{z_1},r_y),\nonumber\\
     &\sim \frac{\pi M\lambda h   }{48  \pi^4}\!\int\!\dd^{d-2}{\vec{z}_1}\,\dd{r_{z_1}}\,r_{z_1}^{2v+1-\xi} \left[\int \frac{ \dd^{d-2}{\vec{z}_2}}{(|\vec{z}_1-\vec{z}_2|^2+r_{z_1}^2)^{d-2+2v-\xi}}\right] G_{\ell = 0}^\phi(\vec{x}-\vec{z}_1,r_x,r_{z_1}) G_{\ell = 0}^\phi(\vec{z}_1-\vec{y},r_{z_1},r_y)\nonumber.
\end{align}
This becomes the following
\begin{align*}
    \delta^4 G^\phi_{\ell = 0}(\vec{x}-\vec{y},r_x,r_y) &= \frac{M\lambda  h }{48 \pi^2}\!\int\!  \frac{\dd^{d-2}{\vec{z}_1}\,\dd{r_{z_1}}}{r_{z_1}^{d-3+2v-\xi}} \,\ G_{\ell = 0}^\phi(\vec{x}-\vec{z}_1,r_x,r_{z_1}) G_{\ell = 0}^\phi(\vec{z}_1-\vec{y},r_{z_1},r_y).
\end{align*}
The UV divergent part of this integral is therefore merely
\begin{align}
     \delta^4 G^\phi_{\ell=0}(\vec{x}-\vec{y},r_x,r_y) = \frac{M\lambda h }{48 \pi^2(\epsilon-2v)}\int \dd^{2}{\vec{z}_1}\,K^\phi_{\ell=0}(\vec{x}\!-\!\vec{z}_1,r_x) K^\phi_{\ell=0}(\vec{z}_1\!-\!\vec{y},r_y)
\end{align}
which is canceled by choosing counterterm $\delta^4 c = \frac{M \lambda h}{48 \pi^2(\epsilon - 2v)}$. Putting everything together, the total counterterms required at this order are
\begin{equation}
    \delta c = \delta^1 c+\delta^2 c + \delta^3 c + \delta^4 c =  \frac{M \kappa \lambda  }{24 \pi (\epsilon-\xi)} + \frac{  (N-M+2) \lambda c}{48 \pi^2 \epsilon} + \frac{c^2 }{2 \pi \xi} + \frac{M \lambda h }{48 \pi^2(\epsilon-2v)}.
    \label{eq:c-counter}
\end{equation}

The diagrams that correct the bulk-to-bulk propagator $G^\psi_{\ell=v}$ and determine counterterms for the coupling $h$ are shown in the second row of Fig.~\ref{fig:diagrams}. The evaluation of these diagrams, including the method of regularization, mirrors the calculation above but with different symmetry factors since the twisted field $\psi$ is complex-valued. The counterterms are determined to be
\begin{equation}
    \delta h = \delta^1 h+\delta^2 h + \delta^3 h + \delta^4 h = \frac{(M+2)\kappa\lambda}{24\pi(\epsilon+2v-\xi)} +\frac{(M+2)\lambda h}{48\pi^2\epsilon} + \frac{h^2}{2\pi (\xi-2v)}+\frac{(N-M)\lambda c}{48\pi^2 (\epsilon +2v)}.
    \label{eq:h-counter}
\end{equation}
The counterterms Eq.~\eqref{eq:c-counter} and Eq.~\eqref{eq:h-counter} determine the $\beta$-functions:  
\begin{equation}
\begin{aligned}
    \frac{\dd c}{\dd\log \mu} = \beta_c &= \frac{c^2}{2\pi} + \frac{M \kappa \lambda}{24 \pi} + \frac{(N-M+2) \lambda c }{48 \pi^2} + \frac{ M \lambda h}{48 \pi^2},\\
    \frac{\dd h}{\dd\log \mu}= \beta_h &= \frac{h^2}{2\pi} + \frac{ (M+2) \kappa \lambda}{24 \pi} + \frac{(N-M)\lambda c}{48 \pi^2} + \frac{ (M+2)\lambda h}{48\pi^2} + 2vh.
\end{aligned}
\end{equation}

Additionally, in the $N \rightarrow \infty$ limit, $h_*^\pm = -2 \pi v \mp \sqrt{c_*^2 + (2 \pi v)^2}$ and the FP equation can be cast in the form 
\begin{equation}
    \frac{c_*^2}{2\pi\epsilon}+(1-\gamma)\mathrm{sgn\,(c_*)}|c_*| + \gamma\,\left[\mathrm{sgn}(h_*\!+\!2\pi v)\sqrt{c_*^2 + (2\pi v)^2}- 2 \pi v^2\right] = 0.
    \label{eq:rgFP-largeN}
\end{equation}

The fact that the defect anomalous dimensions for $\hat{\bm{\phi}}_{\ell=0}$ and $\hat{\bm{\psi}}_{\ell = v}$ are $\alpha_{\ell=0} = \frac{c_*}{2 \pi}$ and $\alpha_{\ell = v} = \frac{h_*}{2 \pi}$ respectively can be seen by the following. At a FP, the correction to the defect-to-defect propagator for the zero mode is given to leading order by (since $c_*$ is perturbatively small, we evaluate the integral at $d = 4$) 
\begin{equation*}
    \delta \hat{G}_{\ell=0}^\phi(\vec{x}-\vec{y}) = - c_* \lim_{r \rightarrow 0}\int \dd^{2}\vec{z} \text{ }K_{\ell=0}^\phi(\vec{x}-\vec{z},r) K_{\ell=0}^\phi(\vec{z}-\vec{y},r) = - \frac{c_*}{(4\pi^2)^2} \lim_{r \rightarrow 0} \int   \frac{\dd^{2}\vec{z}}{[|\vec{x}-\vec{z}|^2+r^2][|\vec{z}-\vec{y}|^2+r^2]}.
\end{equation*}
This can be rewritten directly using Feynman parameters as follows and by shifting $\vec{z} \rightarrow \vec{z}+\vec{y}$ one finds:
\begin{equation*}
    \int  \frac{\dd^{2}\vec{z} }{[|\vec{x}-\vec{z}|^2+r^2][|\vec{z}-\vec{y}|^2+r^2]} = \pi \int_0^1 \dd{u} \text{ } \frac{1}{[u(1-u)|\vec{x}-\vec{y}|^2+r^2]}.
\end{equation*}
Evaluating the integral over the Feynman parameter and taking the small $r$ limit yields
\begin{equation}
    \delta \hat{G}_{\ell=0}^\phi(\vec{x}-\vec{y}) \sim - \frac{c_*}{2\pi} \frac{1}{4 \pi^2}\frac{ \log(|\vec{x}-\vec{y}|^2/r^2)}{|\vec{x}-\vec{y}|^2} \Longrightarrow \alpha_{\ell = 0} = \frac{c_*}{2\pi}
\end{equation}
as in the free theory, $\hat{G}^\phi_{\ell = 0}(\vec{x}-\vec{y}) = \frac{1}{4 \pi^2}\frac{1}{|\vec{x}-\vec{y}|^2}$. One finds $\alpha_{\ell = v} = \frac{h_*}{2 \pi}$ for the twisted mode by the same exercise after expanding the leading order correction in $\epsilon$ and $v$.

\subsection{Further discussion of monodromy defect fixed points}

\begin{figure}
    \centering    \includegraphics[width=0.8\linewidth]{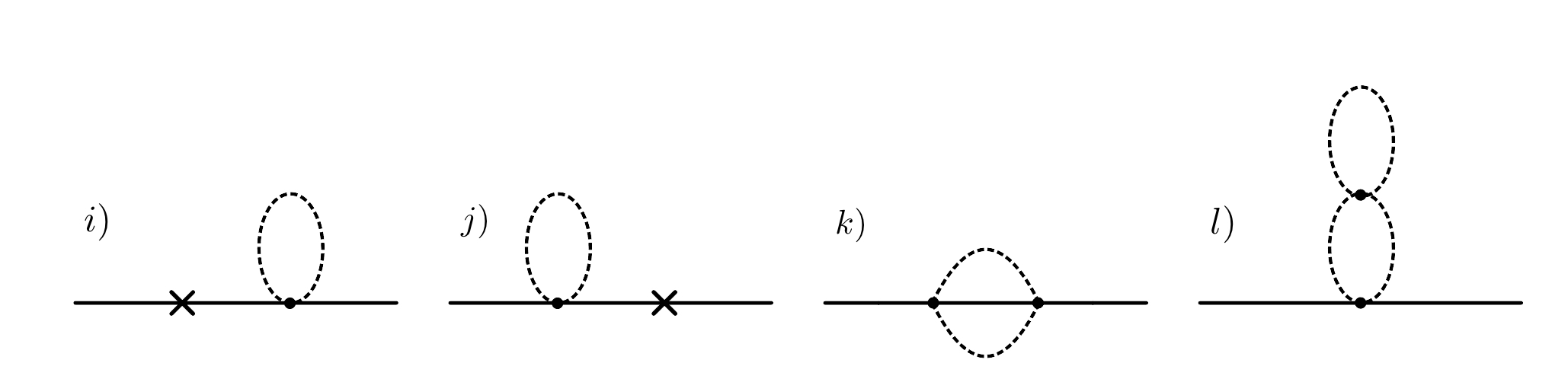}
    \caption{In our calculation, we have neglected some additional higher-order diagrams at $1$-loop and at $2$-loop. Examples of such diagrams that may lend to the renormalization of $c$ are shown here. The $1$-loop diagrams may be safely dropped as we have assumed $v \ll 1$. } 
    \label{fig:HigherOrderDiagram}
\end{figure}

\begin{figure}[h!]
    \centering    \includegraphics[width=0.8\linewidth]{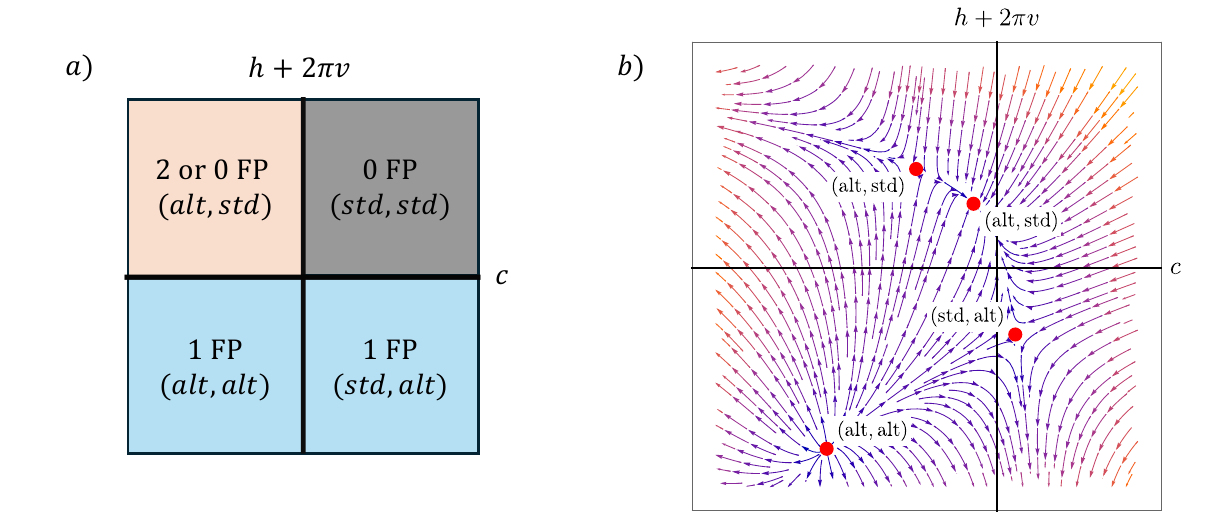}
    \caption{Landscape of monodromy defect fixed points. Figure $a)$ is a schematic picture of all four possible perturbative fixed points classified by boundary condition for $(\bm{\phi}_{\ell = 0},\bm{\psi}_{\ell = v})$. Figure $b)$ is an example of prototypical RG flow (shown for $v \lesssim \epsilon$) for $v < v_c^{\text{alt},\text{std}}$ when both fixed points with $\bm{\phi}_{\ell = 0}$ with alternate and $\bm{\psi}_{\ell=v}$ with standard boundary conditions exist as shown in the main text. In this plot, sinks correspond to IR stable fixed points. 
    } 
    \label{fig:FPgrid1}
\end{figure}

\textit{Defect fixed points in the $1 \gg v \gg \epsilon$ regime.} Here, as was stated in the main text, $\beta_h = \frac{h^2}{2 \pi} + 2vh$ for which there are two solutions $h_*^- =  0$ (stable) and $h^+_* = -4 \pi v$ (unstable) which correspond to standard and alternate boundary conditions at the level of the bare theory. For the former, there are two different solutions for $c^\pm_* < 0$ (Eq.~\eqref{eq:cstar}):
\begin{align}
    c^{\pm,\text{std.}}_* &= - \pi \bigg ( \frac{N-M+2}{N+8} \epsilon \pm \sqrt{\bigg ( \frac{N-M+2}{N+8} \epsilon \bigg )^2 - \frac{4M v(1-v) \epsilon}{N+8}} \bigg ) =- \pi \frac{N-M+2}{N+8} \epsilon \bigg ( 1 \pm \sqrt{1 - \frac{\kappa}{\kappa_c} }\bigg ).
\end{align}
with $\kappa = v(1-v)$ and $\kappa_c = (N-M+2)^2 \epsilon/4M(N+8)$. The anomalous dimension of $\hat{\bm{\phi}}^2_{\ell = 0}$ is given by the derivative of $\beta_c$ at the two fixed points: $\beta_c'(c^{\pm,\text{std}.}_*) = \frac{c^{\pm,\text{std}.}_*}{\pi} + \frac{(N-M+2) \lambda}{48 \pi^2} = \mp \frac{N-M+2}{N+8} \epsilon\sqrt{1-\kappa/\kappa_c}$. Hence $c_*^{-,\text{std}.}$ is stable while $c_*^{+,\text{std}.}$ is unstable.

As discussed, the requirement that there are perturbatively accessible two fixed-points is $\kappa < \kappa_c$. As $1 \gg v \gg \epsilon$, this essentially amounts to requiring sufficiently small $\gamma \lesssim \frac{\epsilon}{4 \kappa} \frac{(N+2)^2}{N(N+8)} \ll 1$. At the minimal value of this bound $\gamma_c = \epsilon \frac{(N+2)^2}{N(N+8)}$: when $\gamma < \gamma_c$, these two fixed points always exist for any monodromy $v$ while for $\gamma > \gamma_c$ there is a phase transition to the runaway fixed point on the defect by tuning $v$. Hence for very small $N$, due to this bound, these two fixed points cannot be seen in the $1 \gg v \gg \epsilon$ regime. These fixed points appear upon increasing $N$ and at $N > N_c \simeq \frac{2}{\epsilon}+{\cal O}(1)$, there is a range $\gamma < \gamma_c$ where they always exist.

Instead, for $h^+_* = - 4 \pi v$ when $\bm{\psi}_{\ell=v}$ has unstable alternate boundary conditions, there are again two more fixed points: 
\begin{equation}
    c^{\pm,\text{alt.}}_* = - \pi \bigg ( \frac{N-M+2}{N+8} \epsilon \pm \sqrt{\bigg ( \frac{N-M+2}{N+8} \epsilon \bigg )^2 + \frac{4 M v(v + 1) \epsilon}{N+8}} \bigg ) = - \pi \frac{N-M+2}{N+8} \epsilon \bigg ( 1 \pm \sqrt{1 + \frac{\kappa''}{\kappa_c}} \bigg ),
\end{equation}
with $\kappa'' = v(1+v)$. The main distinguishing feature for this case is that while there is again a fixed point where $\bm{\phi}_{\ell = 0}$ has alternate boundary conditions ($c_*^{+,\text{alt}.} < 0$), there is also a fixed point where the untwisted mode has \textit{standard} boundary conditions as $(c_*^{-,\text{alt}.} > 0)$ and the defect correlation of the zero mode is suppressed. Additionally, these fixed points never annihilate and always exist for all of $0<v<1/2$ and $0<\gamma < 1$. At these fixed points, the derivative of the $\beta$-function is $\beta'(c^{\pm,\text{alt},}_*) =\frac{c^{\pm,\text{alt.}}_*}{\pi} +\frac{ (N-M+2)\lambda}{48 \pi^2}  = \mp\frac{N-M+2}{N+8}\epsilon \sqrt{1+\kappa''/\kappa_c}$. Hence note that $c_*^\text{+,\text{alt.}}$ is unstable in two directions while $c_*^{-,\text{alt.}}$ is only unstable in the $h$-direction. Also note that similarly, the subleading correction to $\frac{h_*^+}{2\pi}$, i.e. the anomalous dimension of the twisted field, is essentially Eq.~\eqref{eq:dimtwisted} with substitution $\kappa \rightarrow - \kappa''$ and total sign reversed.

\textit{Defect fixed points in the $v \lesssim \epsilon$ regime.} In this regime, we have commented on the behavior of the two fixed points when the twisted field has standard boundary conditions in the main text. For our two new fixed points when the twisted field instead has alternate boundary conditions, they of course also appear in this regime, and never annihilate for any choice of parameters $v$ and $\gamma$. In Fig.~\ref{fig:FPgrid1}, we show a schematic picture of the general classification of all four fixed points in RG that we have discussed. 

When all the fields are twisted, $M = N$, there are always two fixed points for any $v$. One of these has standard boundary conditions as mentioned in the main text, while the other has alternate. These fixed points come from solving $0 = \beta_h = \frac{h^2}{2\pi} + \frac{(N+2) \kappa \lambda}{24 \pi} + \frac{(N+2) \lambda h}{48 \pi^2} + 2vh$ and are
\begin{equation}
    h_*^\pm = - \pi \Bigg (2v + \frac{N+2}{N+8} \epsilon \pm \frac{N+2}{N+8} \eps \sqrt{1 + \frac{4(N+8)^2}{(N+2)^2\epsilon^2}\left(1+\frac{N+2}{N+8}\epsilon\right)v^2} \Bigg )
\end{equation}
where $h^-_*$ corresponds to standard boundary conditions. These expressions reduce as expected when $v \gg \epsilon$. 

\subsection{Calculation of defect scaling dimensions of higher-spin fields}

In this section, we compute the scaling dimension of $\hat{\bm{\phi}}_{\ell \neq 0}$ and $\hat{\bm{\psi}}_{\ell \neq v}$ to leading order in perturbation theory. The leading order diagrams that contribute to the defect-to-defect propagators are given by (see Fig.~\ref{fig:diagrams}): diagrams $a,b,d)$ for the untwisted field and diagrams $e,f,h)$ for the twisted field, when the external legs are brought to the defect. As before we keep twisted mode in standard boundary conditions in the free theory without loss of generality as the alternate boundary condition is perturbatively accessible in RG. 

We begin with the untwisted field. The first diagram that corrects the defect-to-defect propagator $a)$ is
\begin{align}
    \delta^1 \hat{G}_{\ell}^\phi(\vec{x}-\vec{y}) &= - \frac{2\lambda}{4!} \int \dd^d z \text{ } \langle \hat{\phi}_{i,\ell}^\dagger(\vec{x}) \bm{\phi}(z)^2 |\bm{\psi}(z)|^2 \hat{\phi}_{i,\ell}(\vec{y}) \rangle,\\
    &= - \frac{\pi M\lambda  C_\psi^+(\epsilon)}{6} \int \frac{\dd^{d-2}{\vec{z}}\dd{r_z}}{r_z^{2\Delta-1}} \text{ } K_{\ell}^\phi(\vec{x}-\vec{z},r_z) K_{\ell}^\phi(\vec{z}-\vec{y},r_z), \nonumber\\
    &= - \frac{\pi M \lambda  C_\psi^+(\epsilon)}{6} \bigg ( \frac{\Gamma(\Delta_\ell)}{4 \pi^{d/2} \Gamma(\Delta_\ell+2 - \frac{d}{2})} \bigg )^2 \int \frac{\dd^{d-2}{\vec{z}}\dd{r_z}}{r_z^{4\Delta-1}} \frac{r_z^{2\Delta_\ell}}{[|\vec{x}-\vec{z}|^2+r_z^2]^{\Delta_\ell} [|\vec{z}-\vec{y}|^2+r_z^2]^{\Delta_\ell}} \nonumber.
\end{align}
Computing this integral can be done by again using Feynman parameters. We rewrite
\begin{equation}
    \frac{1}{[|\vec{x}-\vec{z}|^2+r_z^2]^{\Delta_\ell} [|\vec{z}-\vec{y}|^2+r_z^2]^{\Delta_\ell}} = \frac{\Gamma(2\Delta_\ell)}{\Gamma(\Delta_\ell)^2} \int_0^1 \dd{u} \text{ } \frac{u^{\Delta_\ell-1}(1-u)^{\Delta_\ell-1}}{(u[|\vec{x}-\vec{z}|^2+r_z^2]+(1-u)[|\vec{z}-\vec{y}|^2+r_z^2])^{2\Delta_\ell}}.
\end{equation}
We perform the integral over the coordinates along the defect first after an appropriate shift of variables. The integrals needed are

\begin{equation*}
\begin{aligned}
    (1) \text{ } &\int \frac{\dd^{d-2}\vec{z}}{(|\vec{z}|^2+u(1-u)|\vec{s}|^2+r_z^2)^{2\Delta_\ell}} = \frac{\pi^{\Delta}\Gamma(2\Delta_\ell - \Delta)}{\Gamma(2\Delta_\ell)} \frac{1}{(u(1-u)|\vec{s}|^2+r_z^2)^{2\Delta_\ell-\Delta}},\\
    (2) \text{ } &\int \dd{r}_z \frac{r_z^{2\Delta_\ell+1-4\Delta}}{(u(1-u)|\vec{s}|^2+r_z^2)^{2\Delta_\ell-\Delta}} = \frac{\Gamma(\Delta_\ell - \frac{\epsilon}{2})\Gamma(\Delta_\ell-\Delta+\frac{\epsilon}{2})}{2\Gamma(2\Delta_\ell-\Delta)} \frac{1}{(u(1-u)|\vec{s}|^2)^{\Delta_\ell-\epsilon/2}},\\
    (3) \text{ } & \int_0^1 \dd{u} \text{ } \frac{u^{\Delta_\ell-1}(1-u)^{\Delta_\ell-1}}{[u(1-u)]^{\Delta_\ell-\epsilon/2}}= \int_0^1  \frac{\dd{u}}{[u(1-u)]^{1-\epsilon/2}} =  \frac{\Gamma(\frac{\epsilon}{2})^2}{\Gamma(\epsilon)}.
\end{aligned}
\end{equation*}

Putting everything together,
\begin{equation}
    \delta \hat{G}_\ell^\phi(\vec{x}-\vec{y}) = -\frac{\pi^{\Delta+1} M \lambda C_\psi^+(\epsilon)}{12(4 \pi^{d/2})^2}\frac{\Gamma(\Delta_\ell - \frac{\epsilon}{2})\Gamma(\Delta_\ell-\Delta+\frac{\epsilon}{2})}{\Gamma(\Delta_\ell+2-\frac{d}{2})^2} \frac{\Gamma(\frac{\epsilon}{2})^2}{\Gamma(\epsilon)} \frac{e^{\epsilon \log |\vec{s}|}}{|\vec{s}|^{2\Delta_\ell}}.
\end{equation}
Now we expand this in $\epsilon \rightarrow 0$ to find the following:
\begin{align}
    \delta^1 \hat{G}_\ell^\phi(\vec{x}-\vec{y}) &= - \frac{\pi^2 M\lambda  C_\psi^+(0)}{12(4 \pi^2)^2 |\ell|} \bigg ( \frac{4}{\epsilon} - \frac{\pi^2 \epsilon}{6}+\dots \bigg )  (1+ \epsilon \log|\vec{s}|+\dots) \frac{1}{|\vec{s}|^{2\Delta_\ell}} \nonumber,\\
    &= - \frac{ M\lambda  C_\psi^+(0)}{96 \pi^2 |\ell|} \frac{\log|\vec{s}|^2}{|\vec{s}|^{2\Delta_\ell}}  + \dots
\end{align}

The next correction to the defect-to-defect propagator from $b)$ is
\begin{align}
    \delta^2 \hat{G}^\phi_{\ell}(\vec{x}-\vec{y}) &= \frac{\lambda c}{2 \times 4!} \int \dd^d{z}_1 \dd^{d-2}{\vec{z}}_2 \text{ } \langle \hat{\phi}^\dagger_{i,\ell}(\vec{x})(\bm{\phi}(z_1)^2)^2 \hat{\bm{\phi}}_{\ell = 0}(\vec{z}_2)^2 \hat{\phi}_{i,\ell}(\vec{y}) \rangle ,\\
    &= \frac{ (N-M+2) \lambda c }{6} \int \dd^{d}{z}_1 \dd^{d-2}\vec{z}_2 \text{ } K_{\ell}^\phi(\vec{x}-\vec{z}_1,r_{z_1})K_{\ell=0}^\phi(\vec{z}_1-\vec{z}_2,r_{z_1})^2 K_{\ell}^\phi(\vec{z}_1-\vec{y},r_{z_1}) \nonumber.
\end{align}
We perform the integral over $\vec{z}_2$ as done so in an earlier section. After doing so, one arrives at
\begin{equation}
    \delta^2 \hat{G}^\phi(\vec{x}-\vec{y}) = \frac{(N-M+2) \lambda c}{768 \pi^5 |\ell|} \frac{\log |\vec{s}|^2}{|\vec{s}|^{2\Delta_\ell}} + \dots
\end{equation}

The third and final correction to the defect-to-defect propagator from $d)$ is 
\begin{align}
    \delta^3 \hat{G}_\ell^\phi(\vec{x}-\vec{y}) &= \frac{ \lambda h}{4!}\int \dd^d{z}_1 \dd^{d-2}\vec{z}_2 \text{ } \langle \hat{\phi}^\dagger_{i,\ell}(\vec{x}) \bm{\phi}(z_1)^2 |\bm{\psi}(z_1)|^2 |\hat{\bm{\psi}}_{\ell=v}(\vec{z}_2)|^2 \hat{\phi}_{i,\ell}(\vec{y}) \rangle,\\
    &= \frac{M\lambda h}{24} \int \dd^d{z}_1 \dd^{d-2}{\vec{z}}_2 \text{ } K_{\ell}^\phi(\vec{x}-\vec{z}_1,r_{z_1}) K_{\ell=v}^\psi(\vec{z}_1-\vec{z}_2,r_{z_1})^2 K_{\ell}^\phi(\vec{z}_1-\vec{y},r_{z_1}). \nonumber
\end{align}
We again perform the integral over $\vec{z}_2$ and expand to leading order in $v$, which simply results in
\begin{equation}
    \delta^3 \hat{G}^\phi(\vec{x}-\vec{y}) = \frac{M \lambda h}{768 \pi^5 |\ell|} \frac{\log |\vec{s}|^2}{|\vec{s}|^{2\Delta_\ell}} + \dots . 
\end{equation}
The free defect-to-defect propagator in $d = 4$ is $\hat{G}^\phi_{\ell}(\vec{x}-\vec{y}) = \frac{1}{4 \pi^2 |\vec{x}-\vec{y}|^{2\Delta_\ell}}$. Therefore the anomalous dimension for the untwisted fields is at a fixed point
\begin{equation}
    \alpha_\ell = \bigg ( \frac{M \lambda C_\psi^+(0)}{24|\ell|} - \frac{(N-M+2) \lambda c}{192 \pi^3 |\ell|} - \frac{M \lambda h}{192 \pi^3 | \ell|} \bigg ) = -\frac{M \epsilon}{2(N+8)|\ell|} \bigg ( \kappa  + \frac{(N-M+2)}{M} \frac{c_*}{2\pi} +  \frac{h_*}{2 \pi} \bigg ).
\end{equation}
A similar calculation shows that for the anomalous dimension of the twisted fields, one instead finds from computing diagrams $e,f,h)$ that
\begin{equation}
    \alpha_\ell = - \frac{(M+2) \epsilon}{2(N+8)|\ell|} \bigg ( \kappa + \frac{N-M}{M+2} \frac{c_*}{2 \pi} + \frac{h_*}{2 \pi} \bigg ). 
\end{equation}
Note that in the $N \rightarrow \infty$ limit, the form of these two expressions coincide and simplify dramatically. Here as we saw, $\frac{h_*^\pm}{2\pi} = - v \mp \sqrt{(c_*/2\pi)^2 + v^2}$. These expressions reduce nicely down to $$\alpha_\ell = -\frac{\epsilon}{2|\ell|} \bigg [ (1-\gamma)\frac{c_*}{2 \pi}  \pm \gamma \sqrt{(c_*/2\pi)^2+v^2}-\gamma v^2 \bigg ] = \frac{1}{2|\ell|} \bigg ( \frac{c_*}{2 \pi} \bigg )^2$$ by Eq.~\eqref{eq:rgFP-largeN}. This is consistent with results found in large $N$ where $\sigma_* = (c_*/2 \pi)^2$. Additionally, when $1 \gg v \gg \epsilon$ and all fields are twisted (i.e. $N = M$) the dimension for the twisted field reads $\alpha_\ell = - \frac{N+2}{N+8} \frac{\kappa \epsilon}{2 |\ell|}$ (for standard, $h_*^- = 0$) and $\alpha_\ell =  \frac{N+2}{N+8} \frac{\kappa'' \epsilon}{2|\ell|}$ (for alternate, $h_*^+ = - 4 \pi v$) consistent with known results \cite{Soderberg, GiombiHyperbolic}.

\section{AdS/CFT and boundary conditions}
\label{sm:adscft}

Here we will review the AdS/CFT techniques pertinent to the discussion of the large $N$ approximation in the main text, and the distinction between standard and alternate boundary conditions using a free theory as an example. 

Any DCFT with dimension $p < d$ defect can be mapped to hyperbolic or (Euclidean) Anti-de Sitter space via a Weyl transformation \cite{GiombiBCFT, NishiokaCThm}. This can be seen directly by choosing spherical coordinates for the $(d-p)$ directions perpendicular to the defect. 
\begin{equation}
    \dd{s}^2 = \dd{\vec{x}}_p^2 + \dd{r}^2 + r^2 \dd{\Omega}_{d-p-1}^2 = r^2 \bigg ( \frac{\dd{\vec{x}}_p^2 + \dd{r}^2}{r^2} + \dd{\Omega}_{d-p-1}^2 \bigg ).
\end{equation}
Hence, up to a Weyl transformation that peels off the Weyl factor $\Omega_\text{weyl} = r^{-1}$, this is the metric of $S^{d-p-1} \times \mathbb{H}^{p+1}$. For any codimension-$2$ defect, with $d = D+1$, the resulting space is $S^{1} \times \mathbb{H}^{D}$. The defect located at the origin in flat space ($r = 0$) is now mapped to the boundary in hyperbolic space. A CFT primary $\hat{O}$ with bulk scaling dimension $\Delta$ transforms accordingly: $\hat{O} \rightarrow \Omega^{-\Delta }\hat{O} = r^\Delta \hat{O}$. 

Consider the following example of a free real scalar on $\mathbb{H}^D$  (and metric $\dd{s}^2 = \frac{\dd{\vec{x}}_{D-1}^2+\dd{r}^2}{r^2}$) with mass $\ell^2$
\begin{equation}
    S = \frac{1}{2}\int \dd^D x \sqrt{g} \text{ } \bigg [(\partial \phi)^2 - \frac{(D-1)^2}{4} \phi^2 + \ell^2 \phi^2 \bigg].
    \label{eq:kkaction}
\end{equation}
When $\ell^2 = 0$, this is the free scalar at conformal coupling and is just attained from a massless free scalar in flat space after applying the above Weyl transformation. The solution to the equation of motion is, when expanded around the boundary,
\begin{equation}
    \phi(\vec{x},r\rightarrow 0) \sim r^{\frac{D-1}{2}-|\ell|} \hat{\phi}^{(0)}(\vec{x}) + r^{\frac{D-1}{2}+|\ell|} \hat{\phi}^{(1)}(\vec{x}) + \dots 
\end{equation}
There are two possible choices of consistent boundary conditions corresponding to fixing one of the two leading asymptotic solutions: $\hat{\phi}^{(0)}(\vec{x})$ or $\hat{\phi}^{(1)}(\vec{x})$. These are asymptotic analogues of usual Dirichlet and Neumann boundary conditions, and are usually referred to as ``standard'' and ``alternate'' boundary conditions. For each choice, there is a dual BCFT operator related to the unfixed field with appropriate scaling dimension \cite{KlebanovWitten}. In the quantum theory,
\begin{equation}
    \begin{aligned}
        \text{Standard boundary condition: }\phi(\vec{x},r \rightarrow 0) \sim r^{\frac{D-1}{2}+|\ell|} \hat{\phi}^{(1)}(\vec{x}) \text{ with } \hat{\Delta} = \frac{D-1}{2}+|\ell|\\
        \text{Alternate boundary condition: }\phi(\vec{x},r \rightarrow 0) \sim r^{\frac{D-1}{2}-|\ell|} \hat{\phi}^{(0)}(\vec{x}) \text{ with } \hat{\Delta} = \frac{D-1}{2}-|\ell|
    \end{aligned}
\end{equation}
With standard boundary conditions, the leading boundary operator has dimension greater than the bulk scaling dimension while for alternate boundary conditions the leading boundary operator has dimension less than the bulk scaling dimension. We use this terminology throughout the paper to distinguish different monodromy defect fixed points. Note that these boundary conditions are degenerate when $\ell = 0$, as is the case for the bare spin zero mode discussed in the manuscript. In flat space, the analogy of this boundary expansion is the bulk-to-defect or bulk-to-boundary OPE which is essentially the same expansion but with the Weyl factor $r^{-\Delta}$ restored.

\section{More details on the large $N$ limit}
\label{app:largeN}

Here we closely follow Ref.~\cite{GiombiHyperbolic}. In the following we write $d=D+1$. We map the theory to
$S^1\times\mathbb{H}^D$:
\begin{equation}
  S=\frac{1}{2}\int \dd^{D+1} x\,\sqrt{g}
  \left[
    g^{\mu\nu}\partial_\mu\Phi\partial_\nu \Phi
    -\frac{(D-1)^2}{4}\Phi^2
    +\frac{\lambda}{N}(\Phi^2)^2
  \right].
  \label{eq:on-action}
\end{equation}
Introducing a Hubbard-Stratonovich field $\sigma$, the large $N$ critical action can be written as
\begin{equation}
S=\frac{1}{2}\int \dd^{D+1} x\,\sqrt{g}
  \left[
    g^{\mu\nu}\partial_\mu\Phi\partial_\nu \Phi
    -\frac{(D-1)^2}{4}\Phi^2
    +\sigma\Phi^2
  \right].
  \label{eq:hs-action}
\end{equation}
where an IR-irrelevant $\sigma^2$ term has been dropped \cite{giombitasi}. Integrating out $\Phi$, we find the partition function
\begin{equation}
  Z=\int \mathcal{D}\sigma\,
  \exp\left[
    -\frac{N}{2}\Tr\ln\left(
      -\nabla^2+\sigma-\frac{(D-1)^2}{4}
    \right)
  \right].
  \label{eq:effective-action}
\end{equation}
At large $N$, we use a saddle-point approximation for the $\sigma$ integral and take the saddle $\sigma_*$ to be constant on the hyperbolic cylinder. Thus $\langle\sigma(x)\rangle=\sigma_{*}$ there, which Weyl transforms back to $\sigma_*/r^{\Delta_\sigma}$ in flat space ($\Delta_\sigma\!=\!2$ to leading order in large $N$). For a constant saddle, the leading large $N$ free energy is a
weighted sum of the twisted and untwisted determinants:
\begin{align}
  \frac{F}{N}\propto{}&
  (1-\gamma)\sum_{\ell\in\mathbb{Z}}
  \int_0^\infty\dd\nu\,\mu_D(\nu)
  \log\left(\nu^2+\ell^2+\sigma_*\right) +\gamma\sum_{\ell\in\mathbb{Z}+v}
  \int_0^\infty\dd\nu\,\mu_D(\nu)
  \log\left(\nu^2+\ell^2+\sigma_*\right),
  \label{eq:partial-free-energy}
\end{align}
where $\mu_D(\nu)$ is the spectral weight of the Laplacian on $\mathbb{H}^D$, given by \cite{CamporesiHiguchi1994, GiombiHyperbolic}
\begin{equation}
    \mu_D(\nu)=\frac{2{\rm Vol}(\mathbb{H}^D)}{(4\pi)^{\frac{D}{2}}\Gamma(\frac{D}{2})} \left|\frac{\Gamma(i\nu+\frac{D-1}{2})}{\Gamma(i\nu)}\right|^2 \label{eq:spectralweight}.
\end{equation}
The integral over $\nu$ can be evaluated as in Ref.~\cite{GiombiHyperbolic} to find
\begin{equation}
  (1-\gamma)T_0(\sigma_*)+\gamma T_v(\sigma_*)=0,
  \label{eq:partial-gap}
\end{equation}
where we have defined
\begin{equation}
    T_\eta(\sigma)= \sum_{\ell \in \mathbb{Z}+\eta}\frac{\Gamma(\Delta_{\ell})}{\Gamma(2-D+\Delta_{\ell})},\qquad   \Delta_{\ell}
  =\frac{D-1}{2}+ \tau_\ell\sqrt{\sigma_*+\ell^2}
    \label{eq:T}
\end{equation}
are the dimensions of defect operators of transverse spin $\ell$, and $\tau_\ell\!=\!\pm 1$ is determined by the choice of standard ($+$) or alternate ($-$) boundary conditions for the $\ell$-mode. The scaling dimensions are derived by noting that $\sigma_*$ is akin to a mass and simply shifts $\ell^2$ in Eq.~\eqref{eq:kkaction}. 

Unitarity bounds restrict the set of spin $\ell$ modes that can be assigned alternate boundary conditions. A codimension-$q$ defect breaks the ${\rm SO}(d\!+\!1,1)$ conformal symmetry of the bulk to a product ${\rm SO}(d\!-\!q+1,1)\!\times\!{\rm SO}(q)$, where the first factor consists of conformal transformations along the defect and the second of rotations around the defect. Scaling dimensions of defect scalars (with respect to ${\rm SO}(d-q)$) must thus obey the unitarity bound $\Delta_\ell\! > \! (d\!-\!q\!-\!2)/2$ of a $(d\!-\!q)$-dimensional CFT, in addition to positivity as usual. Applying these bounds to $\Delta_{\ell=0}$, we find necessarily $\sigma_*\!>\!0$, unlike the pure monodromy ($\gamma\!=\! 1$) problem with all twisted fields studied in Ref.~\cite{GiombiHyperbolic}. Moreover, positivity of $\sigma_*$ implies that only a finite number of modes can support alternate boundary conditions. For $d=3$, these are the untwisted $\ell\!=\!0$ and twisted $\ell\!=\! v$ (for $v\!\leq\!1/2$) modes.

The sum in Eq.~\eqref{eq:T} is divergent and can be regularized by defining 
\begin{align}
  R_\eta(\sigma)
  ={}
  \sum_{k=0}^{\infty}
  \left[
    f\left(\sqrt{\sigma+(k+\eta)^2}\right)
    -(k+\eta)^{D-2}
  \right]
  +\zeta(2-D,\eta),
  \label{eq:Reta-definition}
\end{align}
where $\zeta(s,\eta)$ is the Hurwitz zeta function and we have defined
\begin{equation}
  f(x)=
  \frac{\Gamma\left(\frac{D-1}{2}+x\right)}
       {\Gamma\left(\frac{3-D}{2}+x\right)}   
       =\frac{\Gamma\left(1-\epsilon/2+x\right)}
       {\Gamma\left(\epsilon/2 +x\right)}.
  \label{eq:K-definition}
\end{equation}
Then we write the series in the gap equation as
\begin{align}
    T_0(\sigma) &= f(\tau_0\sqrt{\sigma}) + 2R_1(\sigma),\nonumber\\
    T_v(\sigma) &= f(\tau_v \sqrt{v^2+\sigma}) + R_{1+v}(\sigma) + R_{1-v}(\sigma),
    \label{eq:T0Tv}
\end{align}
where the smallest $\ell$-modes have been separated out of the sum. This presentation will be the starting point of the analytical solutions discussed below. For $\tau_v=1$ we also have $T_v(\sigma)=R_v(\sigma)+R_{1-v}(\sigma)$.

This dimensional regularization scheme guarantees that $\sigma_*=0$ is a solution of the $v=0$ gap equation at any $d$. That is, $T_0(0)=0$ is satisfied. Let us check this explicitly. First we have
\begin{equation}
    f(0)=\frac{\Gamma(\frac{D-1}{2})}{\Gamma(\frac{3-D}{2})}.
\end{equation}
Recall the regularized expression for $R_\eta(0)$:
\begin{equation}
    R_\eta(0)=\sum_{n=0}^\infty \left[\frac{\Gamma(n+\eta+a)}{\Gamma(n+\eta+b)} - (n+\eta)^{D-2}\right]+\zeta(2-D,\eta).
\end{equation}
Here $a=\frac{D-1}{2}, b=1-a$. We will show below in Section \ref{sm:eval_R} that this expression evaluates to
\begin{equation}
    R_\eta(0)=\frac{\Gamma(\eta+a)}{(b-a-1)\Gamma(\eta+b-1)}=\frac{\Gamma(\eta+\frac{D-1}{2})}{(1-D)\Gamma(\eta+\frac{1-D}{2})}.
    \label{eq:R0}
\end{equation}

Therefore
\begin{equation}
    T_0(0)=f(0)+2R_1(0)=\frac{2}{(D-1)\Gamma(\frac{3-D}{2})}\left[\frac{D-1}{2}{\Gamma\Big(\frac{D-1}{2}\Big)}-{\Gamma\Big(\frac{D+1}{2}\Big)}\right]=0.
\end{equation}

In general, the regularized gap equation has to be solved numerically. One subtlety is that the  gap equation can not be directly applied to $d=3$, because as $d\rightarrow 3^+$, the left-hand side is proportional to $d-3$. However, the solutions are completely smooth near $d=3$, so in practice one can just solve it at $d=3^+$ numerically and take the limit.

Below we present several analytically solvable limits of the gap equation:
\begin{enumerate}
    \item The $\epsilon=4-d\ll 1$ limit. Here we show that the perturbative solutions ($\sigma_*\ll 1$) are exactly identical to those found in the weak-pinning RG. We also find a strong-pinning solution corresponding to the runaway FP in the RG.
    \item The $\gamma\ll 1$ limit. We find a perturbative solution in generic $d$ for both (alt, std) and (std, alt) conditions. In addition, for (std, alt) in $d=3$, we find all the solutions in the $v\rightarrow 1/2$ limit and determine $v_c^{\rm{std, alt}}(\gamma)$.
    \item The $v\ll 1$ limit. We find a perturbative solution in generic $d$ for both (alt, std) and (std, alt) boundary conditions.
\end{enumerate}

\subsection{Perturbative solutions in the $\epsilon=4-d\ll 1$ limit}

In this section, we show that perturbative solutions ($\sigma_*\!\ll\! 1$) obtained from the large $N$ gap equation are consistent with the RG fixed points obtained in the $\epsilon$-expansion. 
We start from Eq.~\eqref{eq:T0Tv} and evaluate $R_v,\,R_{1-v},$ and $R_1$ by using a small $\sigma$ (or equivalently $\ell\gg 1$) approximation, without assuming $\sigma\!\ll v$. To this end, we can expand $f(\sqrt{\sigma+t})$ for large $t$ and write
\begin{align}
    R_{1\pm v}&=\sum_{\ell=0}^\infty f(\sqrt{(\ell+1\pm v)^2+\sigma}),\nonumber\\
    &= \zeta(\epsilon-1,1\pm v) + \left(\frac{\sigma}{2}+\frac{\epsilon}{12}-\frac{\epsilon\sigma}{2}\right)\zeta(\epsilon+1,1\pm v)+...,
\end{align}
where the ellipses indicate higher order terms in $\sigma$ and $\epsilon$, both assumed to be much less than unity. A similar approximation is obtained for $R_1$ by setting $v\!=\!0$ in the above equation. Inserting in Eq.~\eqref{eq:T0Tv},
\begin{align}
    T_v(\sigma)&=f(\tau_v\sqrt{v^2+\sigma}) + \left[\zeta(\epsilon-1,1+v)+\zeta(\epsilon-1,1-v)\right] + \left(\frac{\sigma}{2}+\frac{\epsilon}{12}\right)\left[\zeta(\epsilon+1,1+v)+\zeta(\epsilon+1,1-v)\right]+\mathcal{O}(\sigma\epsilon),\nonumber\\
    &\simeq \tau_v\sqrt{v^2+\sigma} - v^2 + \frac{\sigma}{\epsilon} +\mathcal{O}(\sigma,\epsilon),
\end{align}
where the $\mathcal{O}(\epsilon,\sigma)$ terms in the second line are dropped, since they contribute $\mathcal{O}(\epsilon^2,\epsilon\sigma)$ terms to the gap equation (on multiplying the above equation by $\epsilon$). To $\mathcal{O}(\epsilon)$, the gap equation is then
\begin{equation}
    \frac{s^2}{\epsilon} + (1-\gamma)\tau_0 s + \gamma\big(\tau_v \sqrt{v^2+s^2}-v^2\big) = 0,\qquad s \equiv \sqrt{\sigma} \label{eq:ExpGapEq-LargeN}.
\end{equation}
As discussed earlier, unitarity requires $\sigma_*\!>\!0$ for all solutions for any $\gamma\!<\!1$, and this implies that admissible solutions to the above equation must have $s\geq0$. Consistency with the $\epsilon$-expansion is established by identifying $$s =\frac{|c|}{2 \pi}, \quad \tau_0=\mathrm{sgn}(c),\quad \tau_v = \mathrm{sgn}(h+2\pi v),$$ in the large $N$ limit of the RG fixed point equation Eq.~\eqref{eq:rgFP-largeN}.

\subsection{The ``strong-pinning'' (runaway) saddle points }

Here we look for a solution of $\sigma_*$ which nearly saturates the unitarity bound for the $\ell=0$ or the $\ell=v$ mode. Keeping the subleading term in the large $\ell$ expansion, the regularized expression for $R_\eta$ can be written as
\begin{align}
    R_{\eta}&=\sum_{\ell=0}^\infty\left[ f(\sqrt{(\ell+\eta)^2+\sigma}) -(\ell+\eta)^{1-\epsilon}-K(\ell+\eta)^{-1-\epsilon}\right]+\zeta(\epsilon-1,\eta) + K\zeta(\epsilon+1,\eta).
\end{align}
Here 
\begin{equation}
    K=\frac{\sigma}{2}+\frac{\epsilon}{12}-\frac{\epsilon\sigma}{2}+\mathcal{O}(\epsilon^2).
\end{equation}
The sum over $\ell$, as well as $\zeta(\epsilon-1,\eta)$, are now finite in the $\epsilon\rightarrow 0$ limit. On the other hand, the last term
\begin{equation}
    \zeta(1+\epsilon, \eta)=\frac{1}{\epsilon}-\psi(\eta)+\mathcal{O}(\epsilon),
\end{equation}
where $\psi(\eta)$ is the digamma function. We thus find
\begin{align}
    R_\eta &= 
    \sum_{\ell=0}^\infty \left[\sqrt{(\ell+\eta)^2+\sigma} - (\ell+\eta)-\frac{\sigma}{2(\ell+\eta)}\right] +\zeta(-1,\eta)+\left(\frac{\sigma}{2}+\frac{\epsilon}{12}-\frac{\epsilon\sigma}{2}\right)\left(\frac{1}{\epsilon}-\psi(\eta)\right)+\mathcal{O}(\epsilon)\\
    &= \frac{\sigma}{2\epsilon} + r_\eta(\sigma) + \mathcal{O}(\epsilon).
\end{align}
Here 
\begin{equation}
    r_\eta(\sigma)=\sum_{\ell=0}^\infty \left[\sqrt{(\ell+\eta)^2+\sigma} - (\ell+\eta)-\frac{\sigma}{2(\ell+\eta)}\right] + \zeta(-1,\eta) + \frac{1}{12}-\frac{\sigma}{2}-\frac{\sigma\psi(\eta)}{2}.
\end{equation}
Therefore we have 
\begin{align}
    T_0(\sigma)&=f(\tau_0\sqrt{\sigma})+\frac{\sigma}{\epsilon}+2r_1(\sigma)+\mathcal{O}(\epsilon)=f(\tau_0\sqrt{\sigma})+\frac{\sigma}{\epsilon}+2r_1(\sigma)+\mathcal{O}(\epsilon),\\
    T_v(\sigma)&=\frac{\sigma}{\epsilon}+f(\tau_v\sqrt{\sigma+v^2})+r_{1+v}(\sigma)+r_{1-v}(\sigma)+\mathcal{O}(\epsilon)
\end{align}
So the gap equation becomes
\begin{equation}
    (1-\gamma)f(\tau_0\sqrt{\sigma})+\gamma f(\tau_v\sqrt{\sigma+v^2})+\frac{\sigma}{\epsilon} + 2(1-\gamma)r_1(\sigma)+\gamma [r_{1+v}(\sigma)+r_{1-v}(\sigma)]=0.
\end{equation}
The last three terms are finite, so they can be neglected to the leading order in $\epsilon$.

We are interested in the solution where the unitarity bound for either the $\ell=0$ mode or the $\ell=v$ mode is nearly saturated (if $v=\mathcal{O}(\epsilon)$, both modes can become nearly saturated. Here we assume $v$ is fixed). That is, the scaling dimension $\Delta_\ell=1-\frac{\epsilon}{2}+\tau_\ell\sqrt{\sigma+\ell^2}$ is parametrically small. Thus to proceed we need to understand $f(\tau_\ell\sqrt{\sigma+\ell^2})$ when $\Delta_\ell$ is $\mathcal{O}(\epsilon)$ or smaller:
\begin{equation}
   f(\tau_\ell\sqrt{\sigma+\ell^2}) =\frac{\Gamma(\Delta_\ell)}{\Gamma(-1+\epsilon+\Delta_\ell)}=(-1+\epsilon+\Delta_\ell)\frac{\Gamma(\Delta_\ell)}{\Gamma(\Delta_\ell+\epsilon)}\approx -\frac{\Delta_\ell+\epsilon}{\Delta_\ell}.
\end{equation}

Now suppose the $\ell=0$ mode is near saturation. This is clearly only possible with $\tau_0=-1$. Because $\Delta_{\ell=0}=1-\frac{\epsilon}{2}-\sqrt{\sigma}$, we have $\sigma=1+\mathcal{O}(\epsilon)$. As a result we also need $\tau_v=1$ to satisfy the unitarity bound for the $\ell=v$ mode. The equation further simplifies to
\begin{equation}
    -(1-\gamma)\left(\frac{\epsilon}{\Delta_0}\right)+\frac{1}{\epsilon}=0.
\end{equation}
Therefore $\Delta_{\ell=0}=(1-\gamma)\epsilon^2+\mathcal{O}(\epsilon^3)$, and
\begin{equation}
    \sqrt{\sigma_*}=1-\frac{\epsilon}{2}-(1-\gamma)\epsilon^2,
\end{equation}
which gives $\sigma_*=1-\epsilon+\left(2\gamma-\frac74\right)\epsilon^2+\mathcal{O}(\epsilon^3)$.
Interestingly, even to this order $\epsilon^2$, there is no $v$ dependence.

If instead the $\ell=v$ mode is near saturation, it means $\sigma=1-v^2+\mathcal{O}(\epsilon)$. Thus
\begin{equation}
   -\gamma  \frac{\epsilon}{\Delta_v}+\frac{1-v^2}{\epsilon}=0.
\end{equation}
Therefore $\Delta_v=\frac{\gamma}{1-v^2}\epsilon^2+\mathcal{O}(\epsilon^3)$. Again, this requires $\tau_v=-1$. In this case, both $\tau_0=\pm 1$ are allowed. $\sigma_*$ is found to be
\begin{equation}
    \sigma_*=1-v^2-\epsilon+\left(\frac14-\frac{2\gamma}{1-v^2}\right)\epsilon^2+\mathcal{O}(\epsilon^3),
\end{equation}
To summarize, a strong-pinning solution exists for i) $\tau_0=-1, \tau_v=1$, when the $\ell=0$ mode is near saturation ii) $\tau_0=\pm 1, \tau_v=-1$, when the $\ell=v$ mode is near saturation. When $\tau_0=\tau_v=1$, there is no strong-pinning solution.

\subsection{Solutions in the $\gamma\ll 1$ limit}

We now consider yet another analytically solvable limit with $\gamma\ll 1$, and we expect a weak-pinning, perturbative solution $\sigma_*=\mathcal{O}(\gamma^2)$. As before we denote $s=\sqrt{\sigma_*}$. First we evaluate $T_0$.  Expand $f(\tau_0 s)$:
\begin{equation}
    f(\tau_0 s)=f(0)+\tau_0 f'(0)s+\mathcal{O}(s^2).
\end{equation}
Now for $R_1$, because it is an analytic function of $\sigma=s^2$, the Taylor expansion takes the form
\begin{equation}
    R_1(s^2)=R_1(0)+\mathcal{O}(s^2).
\end{equation}
So we have
\begin{equation}
    T_0(\sigma)=f(\tau_0 s)+2R_1(s^2)=f(0)+2R_1(0)+\tau_0f'(0)s+\mathcal{O}(s^2)=\tau_0 f'(0)s+\mathcal{O}(s^2).
\end{equation}
Here
\begin{equation}
    f'(0)=\pi\frac{\Gamma(\frac{D-1}{2})}{\Gamma(\frac{3-D}{2})}\tan\frac{\pi D}{2},
\end{equation}
which vanishes at $D=2$, and is otherwise positive in $D\in (2,3)$. Near $D=2$, we have
\begin{equation}
    f'(0)=\frac{\pi^2}{2}(D-2)+\mathcal{O}\big((D-2)^2\big).
\end{equation}

Now we evaluate $T_v(s^2)$. As before, the $R$ terms are completely regular in $s^2$ for all $v$ and we can set $s=0$ (since the leading term in $T_0(\sigma)$ is $\mathcal{O}(s)$).  Recall from Eq.~\eqref{eq:R0}:
\begin{align}
    R_\eta(0)&=\frac{\Gamma(\eta+\frac{D-1}{2})}{(1-D)\Gamma(\eta+\frac{1-D}{2})}.
\end{align}
We can write 
\begin{align}
 T_v(s^2)&=f(\tau_v\sqrt{v^2+s^2})+R_{1+v}(0)+R_{1-v}(0)+\mathcal{O}(s^2)\\  
    &=\frac{\Gamma(\tau_v\sqrt{v^2+s^2}+\frac{D-1}{2})}{\Gamma(\tau_v\sqrt{v^2+s^2}+\frac{3-D}{2})}+\frac{1}{1-D}\left(\frac{\Gamma(1+v+\frac{D-1}{2})}{\Gamma(1+v+\frac{1-D}{2})} + \frac{\Gamma(1-v+\frac{D-1}{2})}{\Gamma(1-v+\frac{1-D}{2})}\right) + \mathcal{O}(s^2).
\end{align}

The gap equation is now given by
\begin{equation}
    \tau_0(1-\gamma)f'(0)s + \gamma T_v(s^2)=0.
\end{equation}

For $D>2$, $f(\tau_v\sqrt{v^2+s^2})$ is also regular in $s^2$, so setting $s=0$ in $T_v$  gives 
\begin{equation}
    s=-\frac{\tau_0\gamma T_v(0)}{(1-\gamma)f'(0)}.
\end{equation}
This result holds if $v$ and $\epsilon=4-d$ are not small relative to $\gamma$. The $v\ll 1$ limit will be discussed in the next subsection.

Now we evaluate the expression for $D=2$. In this case, the expression is ill-defined. Thus we write $D=2+\delta$ and evaluate both the numerator and denominator to $\mathcal{O}(\delta)$. 
 We find
\begin{equation}
    T_v(0)=\delta\,\left[\pi\left(\frac{\tau_v}{2}-v\right)\tan \pi v+ \psi\Big(\frac12+\tau_v\sqrt{v^2+s^2}\Big)-\psi\Big(\frac12+\tau_vv\Big)\right]+\mathcal{O}(\delta^2).
\end{equation}
Recall $f'(0)=\frac{\pi^2}{2}\delta$.

For $\tau_v=1$, $\psi(1/2+\sqrt{v^2+s^2})$ is regular in $s$, so we can again set $s=0$. 
Therefore we find that at $d=3$
\begin{equation}
    s=-\frac{2\gamma\tau_0}{\pi}\left(\frac{\tau_v}{2}-v\right)\tan \pi v+\mathcal{O}(\gamma^2).
\end{equation}
For $\tau_0=\tau_v=\pm 1$, $s$ is negative and thus unphysical. There is no perturbatively small saddle point near $d=3$. 

For $\tau_0=-1, \tau_v=1$ we have
\begin{equation}
    s=\frac{2\gamma}{\pi}\left(\frac12-v\right)\tan \pi v+\mathcal{O}(\gamma^2).
\end{equation}
$s$ monotonically increases from $0$ at $v=0$ to $\frac{2\gamma}{\pi^2}$ at $v=\frac12$.

However, when $\tau_v=-1$, $\Delta_v^-=1/2-\sqrt{v^2+s^2}$ becomes small when $v$ is close to $1/2$, which needs to be treated separately. So first we assume $|v-1/2|$ is finite (in the $\gamma\rightarrow 0$ limit). Then we can still set $s=0$. The physical solution is obtained 
for $\tau_0=1, \tau_v=-1$:
\begin{equation}
    s=\frac{2\gamma}{\pi}\left(\frac12+v\right)\tan \pi v+\mathcal{O}(\gamma^2).
    \label{eq:s_std_alt_smallgamma}
\end{equation}
Indeed, setting $v\rightarrow 1/2$ in this expression would violate the unitarity bound for $\ell=v$. It is only applicable if $1/2-v\gg \mathcal{O}(\gamma^{2/3})$.

Now let us consider $v$ close to $1/2$. Write $v=1/2-t$, and $\Delta_v^-\approx t-s^2$. Using $\psi(x)=-1/x-\gamma_E+\mathcal{O}(x)$ for $x\rightarrow 0$, we have
\begin{equation}
    \psi(\Delta_v^-)-\psi\Big(\frac12-v\Big)\approx -\frac{1}{\Delta_v^-}+\frac{1}{t}
\end{equation}
On the other hand, the first term in $T_v(0)$ is $-\pi(1-t)\cot\pi t\approx -\frac{1}{t}+1$. So altogether, we obtain
\begin{equation}
    T'_v(0)\approx -\frac{1}{\Delta_v^-}+1+\mathcal{O}(t,s^2).
\end{equation}
The gap equation becomes
\begin{equation}
    s(t-s^2)=\frac{2 \tau_0\gamma}{\pi^2}.
    \label{eq:gap_eq_std_alt}
\end{equation}
The unitarity bound requires $\Delta_v^-\approx t-s^2>0$, so physical solutions only exist for $\tau_0=1$. 

Over the range $0<s<\sqrt{t}$, $s(t-s^2)$ is positive and has a single maximum at $s=\sqrt{t/3}$. So when $\frac{2}{\pi^2}\gamma< 2(\frac{t}{3})^{3/2}$, or equivalently $t>3(\gamma/\pi^2)^{2/3}$, there are two solutions to the gap equation. They come close and annihilate when $t=3(\gamma/\pi^2)^{2/3}$. This analysis gives 
\begin{equation}
  v_c^{\text{std},\text{alt}}(\gamma)=\frac12-3\left(\frac{\gamma}{\pi^2}\right)^{2/3},  \quad \gamma\rightarrow 0. 
\end{equation}

When $\gamma^{2/3}\ll t$, one of the solutions to Eq. \eqref{eq:gap_eq_std_alt} is $s_1\approx \frac{2\gamma}{\pi^2 t}$, which agrees with Eq. \eqref{eq:s_std_alt_smallgamma} in the $v\rightarrow 1/2$ limit. The other solution is given by $s_2\approx \sqrt{t}+\mathcal{O}(\gamma/t)$. These are precisely the two branches of solutions seen in Fig. \ref{fig:sigma-d}(a) (where $s_1$($s_2$) correspond to the lower(upper) branch).  

\subsection{The $v\ll 1$ limit}
Here we show that in the $v\ll 1$ limit, a solution with $\sigma_*\sim v^2$ can be found when $\gamma<\frac12$ is fixed. The derivation is similar to the $\gamma\ll 1$ case, and the most important difference is that the $f(\sqrt{v^2+\sigma})$ term in $T_v(\sigma)$ needs to be treated separately.

In the following we write $\sigma_*=x^2v^2$, where $x\geq 0$ is ${\cal O}(1)$ as $v\rightarrow 0$. Let us start from $T_0$. Here it is completely parallel to the $\gamma\ll 1$ case, so we have 
\begin{equation}
    T_0(\sigma)=\tau_0f'(0)xv + \mathcal{O}(v^2).
\end{equation}
A caveat here is that if one also considers small $\epsilon$, then the $\mathcal{O}(v^2)$ remainder term in $R_1$ is actually $\mathcal{O}(v^2/\epsilon)$. Thus our approximation here is only valid if $v\ll \epsilon$. When both $v, \epsilon$ are small, it is better to use the small $\epsilon$ expansion, which does not make assumptions about the relative order between $v$ and $\epsilon$.

Next is $T_v(\sigma)=f(\tau_v v\sqrt{1+x^2})+R_{1+v}(\sigma)+R_{1-v}(\sigma)$.
Observe that when Taylor expanding $R_{\eta=1\pm v}(\sigma)$ in $v$, the linear in $v$ term must come from $\eta$ since $\sigma\sim v^2$. It follows that
\begin{equation}
      R_{1+v}(\sigma)+R_{1-v}(\sigma)=2R_1(0)+\mathcal{O}(v^2).
\end{equation}
  
Therefore
\begin{equation}
\begin{split}
      T_v(\sigma)&=f(\tau_v v\sqrt{1+x^2})+R_{1+v}(0)+R_{1-v}(0)\\
      &=f(0)+2R_1(0)+\tau_vf'(0)v\sqrt{1+x^2}+\mathcal{O}(v^2)\\
      &=\tau_vf'(0)v\sqrt{1+x^2}+\mathcal{O}(v^2).
\end{split}
\end{equation}
Thus altogether to $\mathcal{O}(v)$ the gap equation becomes
\begin{equation}
    (1-\gamma)\tau_0f'(0)xv+\gamma \tau_v f'(0)v\sqrt{1+x^2}=0.
\end{equation}
Since $f'(0)\neq 0$ for $d\in (3,4)$, we get 
\begin{equation}
    (1-\gamma)\tau_0 x + \gamma \tau_v\sqrt{1+x^2}=0,
  \end{equation}
and the solution is 
\begin{equation}
    x=-\frac{\tau_0\tau_v\gamma}{\sqrt{1-2\gamma}}.
\end{equation}
When $\tau_0=\tau_v$, $x$ is negative and hence unphysical. For $\tau_0=-\tau_v$ we find
\begin{equation}
    \sigma_*=\frac{\gamma^2}{1-2\gamma}v^2+\mathcal{O}(v^3).
\end{equation}
Note that the result is independent of $d$.

\subsection{Evaluation of $R_\eta(0)$}
\label{sm:eval_R}

Recall the definition of $R_\eta(0)$:
\begin{equation}
    R_\eta(0)=\sum_{n=0}^\infty \left[\frac{\Gamma(n+\eta+a)}{\Gamma(n+\eta+b)} - (n+\eta)^{D-2}\right]+\zeta(2-D,\eta).
\end{equation}
Define $S_n(a,b)=\Gamma(n+\eta+a)/\Gamma(n+\eta+b)$. Using $\Gamma(z+1)=z\Gamma(z)$, we have
\begin{equation}
    S_{n+1}(a,b)-S_n(a,b)=\left(\frac{n+\eta+a}{n+\eta+b}-1\right)S_n(a,b)=\frac{a-b}{n+\eta+b}S_n(a,b)=(a-b)S_n(a,b+1).
\end{equation}
Hence we have the recursion relation
\begin{equation}
    S_n(a,b)=\frac{S_{n+1}(a,b-1)-S_{n}(a,b-1)}{a-b+1}.
\end{equation}
And the partial sum
\begin{equation}
    \sum_{n=0}^N S_n(a,b)=\frac{1}{a-b+1}\big[S_{N+1}(a,b-1)-S_0(a,b-1)\big].
\end{equation}
$S_{N+1}$ diverges with $N$. Using Stirling's formula with $q = N+1+\eta$, we have
\begin{align}
    S_{N+1}(a,b-1)&=\frac{\Gamma(q+a)}{\Gamma(q+b-1)}\\
    &=q^{a-b+1}\left[1+\frac{(a-b+1)(
a+b-2)}{2q}+\mathcal{O}(q^{-2})\right]\\
&= {q^{D-1}}\left[1-\frac{D-1}{2q}+\mathcal{O}(q^{-2})\right].
\end{align}

Next we can evaluate the partial sum of $(n+\eta)^{D-2}$ using the Euler-Maclaurin formula:
\begin{equation}
    \sum_{n=0}^N (n+\eta)^{D-2}=\frac{q^{D-1}}{D-1}-\frac12 q^{D-2}+\zeta(2-D,\eta)+\mathcal{O}(q^{D-3}).
\end{equation}
Since we are interested in $D<3$, $q^{D-3}$ vanishes as $N\rightarrow \infty$. Thus $R_\eta(0)$ is given by the $N\rightarrow \infty$ limit:
\begin{align}
    R_\eta(0)&=\lim_{N\rightarrow \infty} \left(\sum_{n=0}^N S_n(a,b)-\sum_{n=0}^N(n+\eta)^{D-2}\right)+\zeta(2-D,\eta)\\
    & =-\frac{S_0(a,b-1)}{a-b+1}\\
    &=\frac{\Gamma(\eta+a)}{(b-a-1)\Gamma(\eta+b-1)}.
\end{align}

\subsection{Calculation of defect scaling dimensions of $\sigma$ at $d = 3$}
\label{sm:defectsigma}

Here, we compute the defect scaling dimension for $\sigma$ at $d = 3$. Similar calculations were performed in Refs.~\cite{Carmi2019,CuomoPinning,monodromypinning}. We expand the large $N$ theory on $S^1 \times \mathbb{H}^2$ in fluctuations around the saddle point solution $\sigma = \sigma_* + \frac{\delta \sigma}{\sqrt{N}}$. The quadratic effective action for the fluctuations of the Hubbard-Stratonovich field is
\begin{equation}
    S_\text{eff} \sim  \int_{S^1 \times \mathbb{H}^2} \dd^3{\bm{x}} \dd^3{\bm{y}} \sqrt{g_x} \sqrt{g_y} \text{ }\delta \sigma(\bm{x}) \bigg [ (1-\gamma) \mathcal{K}^\phi(\bm{x},\bm{y}) + \frac{\gamma}{4} \mathcal{K}^\psi(\bm{x},\bm{y}) \bigg ] \delta \sigma(\bm{y})
\end{equation}
up to some negative constant which is unimportant for the following discussion. Here, $\bm{x} = (x,r_x,\theta_x)$ and $\bm{y} = (y,r_y,\theta_y)$. The kernels are given by $\mathcal{K}^\phi(\bm{x},\bm{y}) = G^\phi(x-y,r_x,r_y,\theta_{xy})G^\phi(y-x,r_y,r_x,\theta_{yx})$ and $ \mathcal{K}^\psi(\bm{x},\bm{y}) = G^\psi(x-y,r_x,r_y,\theta_{xy}) G^\psi(y-x,r_y,r_x,\theta_{yx})$. Upon decomposing into modes with spin $\ell \in \mathbb{Z}$ for the untwisted fields and $\ell \in \mathbb{Z}+v$ for the twisted modes around the $S^1$, this now reads
\begin{equation}
    S_\text{eff} \sim \sum_{\ell \in \mathbb{Z}} \int_{\mathbb{H}^2} \dd{x}\dd{r}_x \dd{y} \dd{r}_y \sqrt{g_x} \sqrt{g_y} \text{ } \delta\sigma_\ell(x,r_x) \bigg [(1-\gamma)\mathcal{K}^\phi_{\ell}(x-y,r_x,r_y)+\frac{\gamma}{4} \mathcal{K}^\psi_{\ell}(x-y,r_x,r_y) \bigg ] \delta \sigma_{-\ell}(y,r_y).
\end{equation}
where the new kernels for each spin are given by the following
\begin{equation}
\begin{aligned}
    \mathcal{K}^\phi_\ell(x-y,r_x,r_y) &= \sum_{\ell' \in \mathbb{Z}} G^\phi_{\ell+\ell'}(x-y,r_x,r_y) G^\phi_{\ell'}(y-x,r_y,r_x)\\
    \mathcal{K}^\psi_\ell(x-y,r_x,r_y) &= \sum_{\ell' \in \mathbb{Z}+v} G^\psi_{\ell+\ell'}(x-y,r_x,r_y) G^\psi_{\ell'}(y-x,r_y,r_x).
\end{aligned}
\end{equation}
Note that since we are in large $N$, the dimensions are $\Delta_\ell = \frac{1}{2}\pm \sqrt{\ell^2+\sigma_*}$. These are the same free propagators we discussed previously as Eq.~(\ref{eq:propagators}) and other related expressions at $d = 3$ but with the Weyl factor $(r_x r_y)^{\Delta}$ stripped off from the denominators. 

We focus on the zero mode $\delta \sigma_{\ell = 0}$ for now and look for a spectral representation of the kernel $\hat{\mathcal{K}}_{\ell = 0} = (1-\gamma) \mathcal{K}_{\ell=0}^\phi + \frac{\gamma}{4} \mathcal{K}_{\ell = 0}^\psi$. For a good spectral representation, which we will call $\hat{B}_{\ell = 0}(\nu)$, its zeroes correspond to the poles of the spectral representation of the propagator which will be located precisely at $\nu = \pm i (\Delta_{\ell = 0}^\sigma - \frac{1}{2})$ where $\Delta_{\ell = 0}^\sigma$ is the defect (or boundary) scaling dimensions of $\delta \sigma_{\ell = 0}$. We are interested in the smallest zero corresponding to the dimension of the most relevant defect singlet $\hat{\sigma}_{\ell = 0}$. Hence, our job is to compute this quantity $\hat{B}_{\ell = 0}(\nu)$ and solve for its zeroes.

Before we begin, we summarize some of the spectral properties of the propagators on $\mathbb{H}^2$. The harmonic function $\Omega^\phi_\nu(x-y,r_x,r_y)$ is defined as an eigenfunction of the $\mathbb{H}^2$ Laplacian \cite{Carmi2019, SpinningAdSProp}
\begin{equation}
    - \nabla^2_{\mathbb{H}^2} \Omega_\nu^\phi(x-y,r_x,r_y) = \bigg (\nu^2 + \frac{1}{4} \bigg ) \Omega_\nu^\phi(x-y,r_x,r_y) \label{eq:harmonic}.
\end{equation}
The propagator on $\mathbb{H}^2$ obeys the following equation
\begin{equation}
    [ - \nabla_{\mathbb{H}^2}^2 + \Delta_\ell(\Delta_\ell - 1)] G_{\Delta_\ell}^\phi(x-y,r_x,r_y) = \frac{\delta(x-y)\delta(r_x-r_y)}{2 \pi \sqrt{g}}.
\end{equation}
In this case $\Delta_\ell = \frac{1}{2} \pm \sqrt{\ell^2 + \sigma_*}$ and $G_{\Delta_\ell}^\phi(x-y,r_x,r_y)$ is just $ G^\phi_\ell(x-y,r_x,r_y)$. The harmonic function can alternatively be written as follows
\begin{equation}
    \Omega_\nu^\phi(x-y,r_x,r_y) = \frac{i \nu}{2 \pi \mu_2(\nu)}[G^\phi_{\frac{1}{2}+i \nu}(x-y,r_x,r_y) - G^\phi_{\frac{1}{2}-i \nu}(x-y,r_x,r_y)]
\end{equation}
which obeys Eq.~(\ref{eq:harmonic}). Here $\mu_2(\nu) = \frac{\nu \tanh(\pi \nu)}{4\pi}$ is the spectral density for $\nu \in \mathbb{R}$ divided by $\text{Vol}(\mathbb{H}^2)$ in order to simplify expressions. Note that $\Omega_\nu^\phi(x-y,r_x,r_y)$ does not depend on scaling dimensions, and the superscript $\phi$ is just to account for the fact that $G^\phi$ and $G^\psi$ have different normalization (one could instead use $\Omega^\psi_\nu(x-y,r_x,r_y) \equiv 2\Omega^\phi_\nu(x-y,r_x,r_y)$). One can additionally check that this indeed yields a spectral representation of the propagators whose poles correspond to $\nu = \pm i(\Delta_\ell - \frac{1}{2})$ as
\begin{equation}
\begin{aligned}
    G_{\Delta_\ell}^\phi(x-y,r_x,r_y) &= \int_{-\infty}^\infty \dd{\nu} \frac{\mu_2(\nu)}{\nu^2+(\Delta_\ell-\frac{1}{2})^2} \Omega_\nu^\phi(x-y,r_x,r_y)\\ \text{ and } G_{\Delta_\ell}^\psi(x-y,r_x,r_y) &= \int_{-\infty}^\infty \dd{\nu} \frac{2\mu_2(\nu)}{\nu^2+(\Delta_\ell - \frac{1}{2})^2} \Omega_\nu^\phi(x-y,r_x,r_y) 
\end{aligned}
\end{equation}
where the dimensions in the integrand are the appropriate $\Delta_\ell = \frac{1}{2} + \sqrt{\ell^2+\sigma_*}$ with $\ell \in \mathbb{Z}$ $ (\text{with } \ell \in \mathbb{Z}+v)$ for $G^\phi_\Delta$ (for $G^\psi_\Delta$). It follows that
\begin{equation}
    \int_{-\infty}^\infty \dd{\nu} \text{ } \mu_2(\nu) \Omega^\phi_\nu(x-y,r_x,r_y) = \frac{\delta(x-y)\delta(r_x-r_y)}{2 \pi \sqrt{g}}.
\end{equation}
Very crucially, these harmonics obey an orthogonality relation \cite{SpinningAdSProp}
\begin{equation}
    \int \dd{z} \dd{r}_z \sqrt{g_z} \text{ } \Omega_\nu^\phi(x-z,r_x,r_z) \Omega_\omega^\phi(z-y,r_z,r_y) = \frac{\delta(\nu-\omega)+\delta(\nu+\omega)}{4 \pi \mu_2(\nu)} \Omega_\nu^\phi(x-y,r_x,r_y)
\end{equation}
and are normalized such that $\Omega_\nu^\phi(0,r_x,r_x) = \frac{1}{2\pi}$.

The upshot of all these properties is that these harmonics can diagonalize the kernel $\hat{\mathcal{K}}_{\ell = 0}(x-y,r_x,r_y)$. That is,
\begin{equation}
   \langle \delta \sigma_{\ell = 0}(x,r_x) \delta \sigma_{\ell = 0}(y,r_y) \rangle \sim \hat{\mathcal{K}}_{\ell = 0}^{-1}(x-y,r_x,r_y) = 4 \pi^2 \int_{-\infty}^\infty \dd{\nu} \text{ } \mu_2(\nu) \hat{B}_{\ell = 0}^{-1}(\nu) \Omega_\nu^\phi(x-y,r_x,r_y).
\end{equation}
The inverse kernel is defined as per usual by $\int \dd{z} \dd{r}_z\sqrt{g_z} \text{ } \hat{\mathcal{K}}_{\ell =0}(x-z,r_x,r_z) \hat{\mathcal{K}}^{-1}_{\ell = 0}(z-y,r_z,r_y) = \delta(x-y)\delta(r_x-r_y)/\sqrt{g}$. The spectral kernel is given by 
\begin{equation}
\begin{aligned}
    \frac{\hat{B}_{\ell = 0}(\nu)}{4 \pi^2} &=  \int \dd{x} \dd{r}_x\sqrt{g_x} \text{ } \hat{\mathcal{K}}_{\ell = 0}(x-y,r_x,r_y) \Omega_\nu^\phi(x-y,r_x,r_y),\\
    &= \int \dd{x}\dd{r}_x \sqrt{g_x} \bigg[ (1-\gamma) \sum_{\ell \in \mathbb{Z}}G_{\Delta_\ell}^\phi(x-y,r_x,r_y) G_{\Delta_\ell}^\phi(y-x,r_y,r_x) \\&+ \frac{\gamma}{4} \sum_{\ell \in \mathbb{Z}+v} G_{\Delta_\ell}^\psi(x-y,r_x,r_y) G_{\Delta_\ell}^\psi(y-x,r_y,r_x) \bigg ] \Omega_\nu^\phi(x-y,r_x,r_y).
\end{aligned}
\end{equation}
To proceed, we need to rely on an identity of propagators on $\mathbb{H}^2$, namely that they exhibit closure under multiplication \cite{Fitzpatrick2011, Fitzpatrick2012}
\begin{equation}
\begin{aligned}
    G^\phi_{\Delta_\ell}(x-y,r_x,r_y) G^\phi_{\Delta_\ell'}(y-x,r_y,r_x) &= \sum_{n=0}^\infty a^\phi_{2n}(\Delta_\ell,\Delta_\ell') G^\phi_{\Delta_\ell+\Delta_\ell'+2n}(x-y,r_x,r_y), \\\text{ with } a^\phi_{2n}(\Delta_\ell,\Delta_\ell') &= \frac{(\frac{1}{2})_n}{4\pi^{3/2}n!} \frac{(\Delta_\ell+\Delta_\ell'+2n)_{1/2}(\Delta_\ell+\Delta_\ell'+n)_n}{(\Delta_\ell+n)_{1/2}(\Delta_\ell'+n)_{1/2}(\Delta_\ell+\Delta_\ell'+n-\frac{1}{2})_n}
\end{aligned}
\end{equation}
where $(a)_n = \Gamma(a+n)/\Gamma(a)$ is the Pochhammer symbol. Likewise, of course the same identity holds for the twisted fields but with $a^\psi_{2n}(\Delta_\ell,\Delta_\ell') = 2a^\phi_{2n}(\Delta_\ell,\Delta_\ell')$. This can be derived by verifying these propagators obey an orthogonality relation under a contour integral over the cross-ratio $\zeta$ extended to complex plane \cite{Fitzpatrick2011, Fitzpatrick2012}. The contour can be taken as a very large circle.

With this identity, we just have
\begin{equation}
\begin{aligned}
    \frac{\hat{B}_{\ell = 0}(\nu)}{4 \pi^2} &=  (1-\gamma) \sum_{\ell \in \mathbb{Z}} \sum_{n=0}^\infty  a_{2n}^\phi(\Delta_\ell,\Delta_\ell)\int \dd{x}\dd{r}_x \sqrt{g_x} \text{ } G^\phi_{2\Delta_\ell+2n}(x-y,r_x,r_y) \Omega_\nu^\phi(x-y,r_x,r_y) \\&+ \frac{\gamma}{4} \sum_{\ell \in \mathbb{Z}+v} \sum_{n=0}^\infty  a^\psi_{2n}(\Delta_\ell,\Delta_\ell)\int \dd{x}\dd{r}_x \sqrt{g_x} \text{ }G_{2\Delta_\ell+2n}^\psi(x-y,r_x,r_y) \Omega_\nu^\phi(x-y,r_x,r_y).
\end{aligned}
\end{equation}
By the orthogonality relation we discussed before, we can perform the integral to get
\begin{equation}
    \hat{B}_{\ell = 0}(\nu)= (1-\gamma) \sum_{\ell \in \mathbb{Z}} \sum_{n=0}^\infty \frac{ a^\phi_{2n}(\Delta_\ell,\Delta_\ell)}{\nu^2+(2\Delta_\ell+2n- \frac{1}{2})^2}+ \frac{\gamma}{2} \sum_{\ell \in \mathbb{Z}+v} \sum_{n=0}^\infty \frac{a^\psi_{2n}(\Delta_\ell,\Delta_\ell)}{\nu^2+(2\Delta_\ell+2n- \frac{1}{2})^2} \label{eq:B0}.
\end{equation}
Now lastly, to compute scaling dimensions we need to numerically solve  for the zeroes of this spectral kernel as
\begin{equation}
    \hat{B}_{\ell = 0}\bigg (\nu = \pm i \bigg (\Delta^\sigma_{\ell = 0}-\frac{1}{2}\bigg ) \bigg ) = 0.
\end{equation}
The smallest physical solution, $\Delta^\sigma_{\ell = 0}$, to this equation is the dimension of the defect singlet $\hat{\sigma}_{\ell = 0}$. In this spectral representation, the alternate boundary condition for the lowest mode(s) is chosen through analytic continuation of the scaling dimension in the above expressions. 

The numerical solution to this equation for the case of (alt, std) -- when the untwisted zero mode has alternate boundary conditions and the lowest twisted mode has standard boundary conditions -- is shown in the main text as Fig.~\ref{fig:sigma_dimension_alt_std}. From looking at Eq.~(\ref{eq:B0}), one can see that $\hat{B}_{\ell = 0}( \pm i (\Delta - \frac{1}{2}))$ has poles at $\Delta = 1 - 2 \sqrt{\sigma^*}$ and $\Delta = 1 + 2 \sqrt{\sigma^*+v^2}$ and must cross a zero between them which is roughly around $1$ for small $v$ and $\sigma_*$. When $\gamma < \frac{1}{2}$ and $v \rightarrow 0$ such that $\sigma_* \rightarrow 0$ these poles overlap, and the first zero has to now be between $\Delta = 1$ and $\Delta = 3$, which ends up being the bulk value $\Delta_{\ell=0}^\sigma = 2$. 
\begin{figure}
    \centering
    \includegraphics[width=0.6\linewidth]{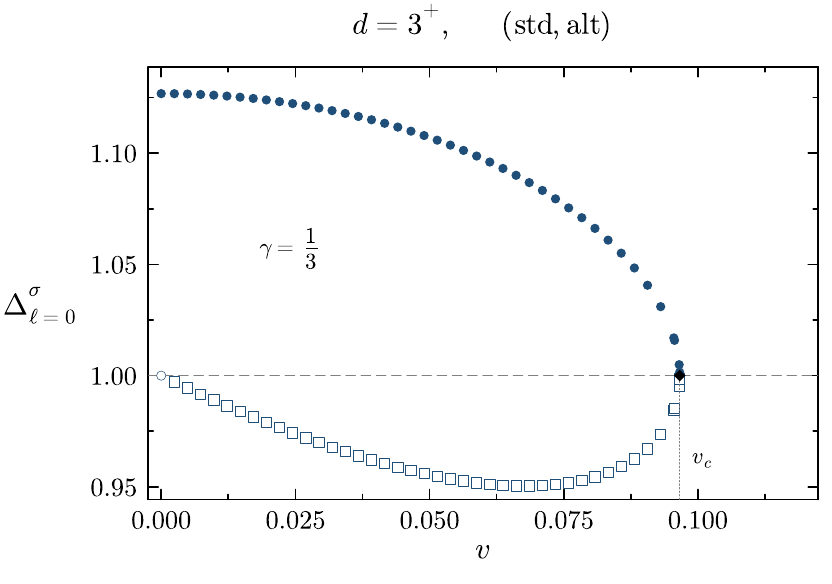}
    \caption{The defect scaling dimension $\Delta^\sigma_{\ell=0}$ of the leading singlet $\hat{\sigma}_{\ell=0}$ as a function of $v$ for $\gamma=1/3$, with (std, alt) boundary condition. The large $N$ gap equations are solved at $d=3.0001$. The top (bottom) branch corresponds to the stable (unstable) saddle-point solutions we have already found. Note that the top branch still has a nontrivial solution for $v \rightarrow 0$. This is because this limit is the same as the $v \rightarrow 0$ limit of the (alt, std) solution when $\gamma = 2/3$.}
    \label{fig:sigma_dimension_std_alt}
\end{figure}
The solution for the case of (std, alt) is shown in Fig.~\ref{fig:sigma_dimension_std_alt}. In this case, $\Delta^\sigma_{\ell=0}$ for the stable (unstable) solution is indeed greater (smaller) than $1$, as expected, and they both reach marginality at $v=v_c$.

We can also perform a similar calculation for the spin $\ell = \pm 1$ case. For $\ell = 1$, we get the spectral kernel
\begin{equation}
    \hat{B}_{\ell = 1}(\nu) = (1-\gamma) \sum_{\ell \in \mathbb{Z}} \sum_{n=0}^\infty \frac{ a^\phi_{2n}(\Delta_\ell,\Delta_{\ell+1})}{\nu^2+(\Delta_\ell+\Delta_{\ell+1}+2n- \frac{1}{2})^2}+ \frac{\gamma}{2} \sum_{\ell \in \mathbb{Z}+v} \sum_{n=0}^\infty \frac{a^\psi_{2n}(\Delta_\ell,\Delta_{\ell+1})}{\nu^2+(\Delta_\ell+\Delta_{\ell+1}+2n- \frac{1}{2})^2},
\end{equation}
and an identical expression for $\ell = -1$. From solving $\hat{B}_{\ell = 1}(\nu = \pm i(\Delta_{\ell=1}^\sigma - 1/2)) = 0$ numerically, we find the spectrum of defect operators at $\ell = \pm 1$. There is always an operator with dimension $\Delta = 2$ for any $\gamma$ and $v$, which should be a healthy displacement operator, when a DCFT is expected to exist. This includes the case for $\gamma > \frac{1}{2}$ and $v \rightarrow 0$, which is another signature that this limit corresponds to a nontrivial DCFT. The displacement operator is defined through the divergence of the stress tensor at the defect, and is a spin $\ell = \pm 1$ operator in $d = 3$.

\begin{figure}
    \centering
    \includegraphics[width=0.6\linewidth]{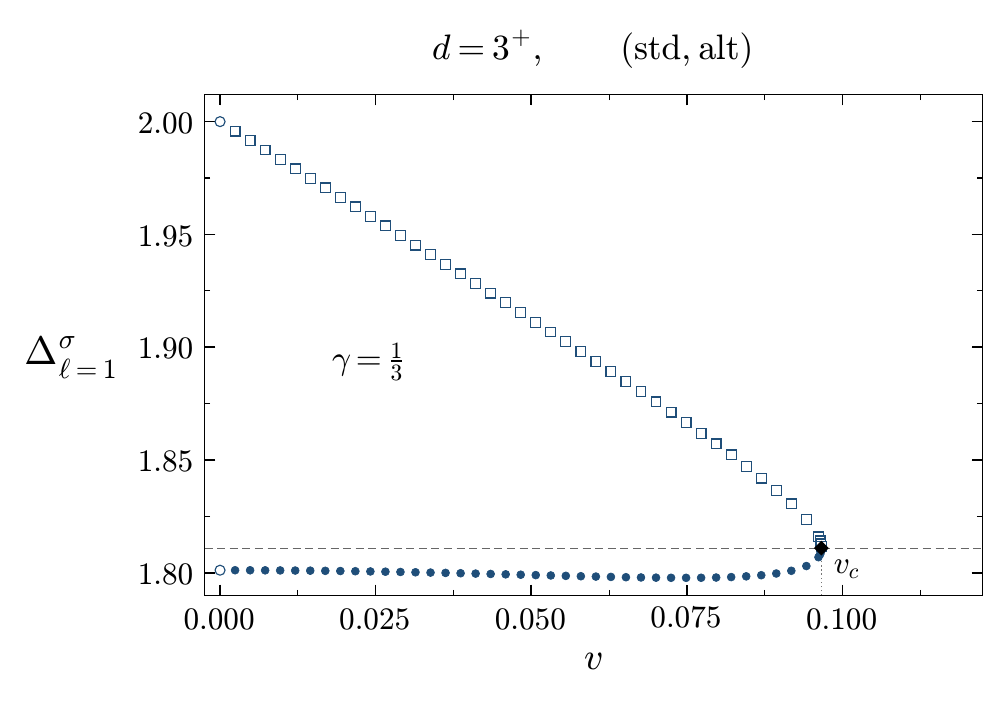}
    \caption{The defect scaling dimension $\Delta^\sigma_{\ell=1}$ of the leading higher-spin singlet $\hat{\sigma}_{\ell=1}$ as a function of $v$ for $\gamma=1/3$, with (std, alt) boundary condition. The large $N$ gap equations are solved at $d=3.0001$. The bottom (top) branch corresponds to the stable (unstable) saddle-point solutions we have already found. In this plot the two displacement operators at $\Delta_{\hat{D}} = 2$ are explicitly not shown. }
    \label{fig:higherspin_sigma_dimension_std_alt}
\end{figure}

Interestingly, for each (std, alt) solution, we find a spin $\ell = \pm 1$ defect operator $\hat{\sigma}_{\ell = 1}$ that has smaller dimension than the displacement operator as shown in Fig.~\ref{fig:higherspin_sigma_dimension_std_alt}. This is not unnatural as even for the all twisted free theory in $d = 3$ with alternate boundary conditions on the lowest mode, there is $\hat{\bm{\psi}}_{\ell = v-1}^\dagger \hat{\bm{\psi}}_{\ell =v}$ that has spin $\ell = 1$ and dimension $2-2v<2$, whereas the displacement operator corresponds to $\hat{D} \sim \hat{\bm{\psi}}_{\ell=v}^\dagger \hat{\bm{\psi}}_{\ell = 1+v}$  \cite{GiombiHyperbolic}.

\subsection{Additional numerical results}

\begin{figure}[t]
  \centering
  \begin{minipage}[t]{0.49\textwidth}
    \centering
    \includegraphics[width=\linewidth]{
      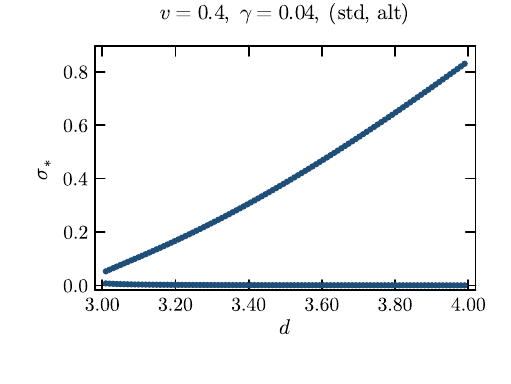
    }
  \end{minipage}
  \begin{minipage}[t]{0.49\textwidth}
    \centering
    \includegraphics[width=\linewidth]{
      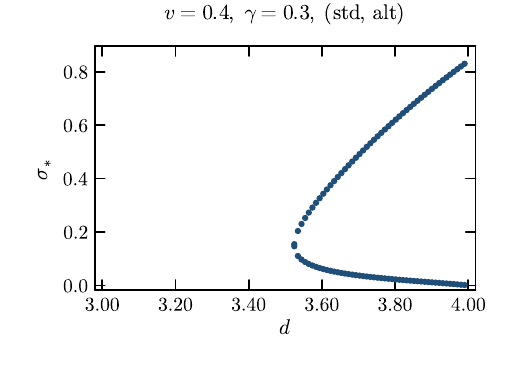
    }
  \end{minipage}
   \begin{minipage}[t]{0.49\textwidth}
    \centering
    \includegraphics[width=\linewidth]{
      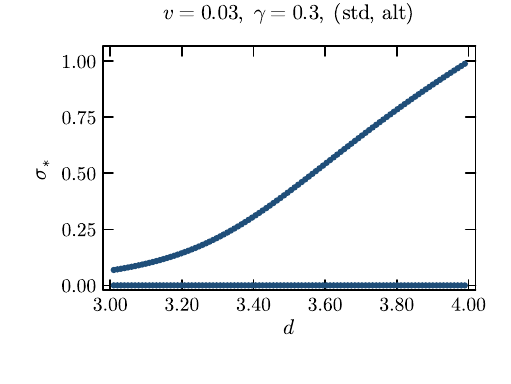
    }
    \end{minipage}
    \begin{minipage}[t]{0.49\textwidth}
    \centering
    \includegraphics[width=\linewidth]{
      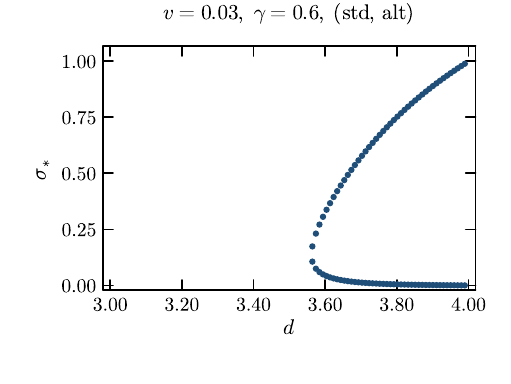
    }
  \end{minipage}
  \caption{
    Solutions $\sigma_*$ of the gap equation as a function of
    $d$ with alternate boundary condition for the $\ell=v$ mode and standard boundary condition for the $\ell=0$ mode. 
  }
  \label{fig:sigma-d}
\end{figure}

Fig. \ref{fig:sigma-d} illustrates the evolutions of the saddle points with $\tau_0=1,\tau_v=-1$ boundary conditions with $d$. We see that if $v<v_c^{\text{std},\text{alt}}(\gamma)$, the two branches, corresponding to the perturbative and the strong-pinning fixed points near $d=4$, continue all the way to $d=3$, consistent with the analytical result. However, when $v>v_c^{\text{std},\text{alt}}(\gamma)$, the two branches annihilate at $d_c>3$, and no solutions exist for $d$ close to 3. 

\begin{figure}
    \centering
    \includegraphics[width=0.5\linewidth]{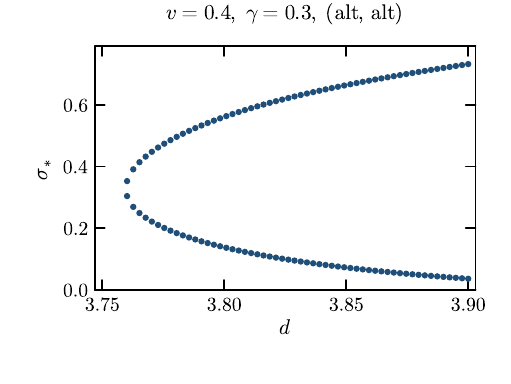}
    \caption{$\sigma_*$ as a function of $d$ for $v=0.4, \gamma=0.3$ when both $\ell=0,v$ modes have alternate boundary conditions.}
    \label{fig:sigma_alt_alt}
\end{figure}

In Fig. \ref{fig:sigma_alt_alt} we show a similar plot for $v=0.4, \gamma=0.3$, with $\tau_0=\tau_v=-1$. There are no solutions near $d=3$. The perturbative solution annihilates with the strong-pinning one as $d$ reaches the critical dimension $d_c\approx 3.76$. The same pattern is observed numerically for other values of $v, \gamma$, and is consistent with the analytical solutions in the small $\gamma$ or $v$ limit.

\section{More data on Monte Carlo simulations}
\label{sm:MCmore}

\subsection{Details on the update algorithm and the pinning strength selection}

Here we describe the Monte Carlo update algorithm in our simulation.
Configurations are generated using a local hybrid update consisting of one Metropolis sweep followed by overrelaxation sweeps with deterministic reflection proposals.
For a site $\mathbf{x}$, define the local molecular field
\begin{equation}
    \boldsymbol{B}_\mathbf{x} = \sum_{\mu=1}^{3} \left( U_{\mathbf{x},\mu} \, \boldsymbol{\Phi}_{\mathbf{x}+\hat{\mu}} + U_{\mathbf{x}-\hat{\mu},\mu}^{T} \, \boldsymbol{\Phi}_{\mathbf{x}-\hat{\mu}} \right) .
\end{equation}
We also define
\begin{equation}
    w_{\mathbf{x}} = \frac{1}{4} \sum_{a=1}^2 \mathbf{1}_{\mathbf{x}_\perp \in \mathcal{S}_a}, \quad V(\boldsymbol{\Phi}) = \boldsymbol{\Phi}^2 + \lambda \left( \boldsymbol{\Phi}^2 - 1 \right)^2 .
\end{equation}
So that $H_{\rm pin} = \Tilde{c} \sum_{\mathbf{x}} w_{\mathbf{x}} |P_{\rm u} \boldsymbol{\Phi}_{\mathbf{x}}|^2$.
A Metropolis proposal is generated according to
\begin{equation}
    \boldsymbol{\Phi}_{\mathbf{x}}^{\prime} = \boldsymbol{\Phi}_{\mathbf{x}} + \epsilon \boldsymbol{\eta}, \ \eta_\mu \sim \mathcal{N}(0,1) ,
\end{equation}
with $\epsilon=0.55$.
The local Gibbs weight change used in the simulation is
\begin{equation}
    \Delta \mathcal{H} = V(\boldsymbol{\Phi}_{\mathbf{x}}^{\prime}) - V(\boldsymbol{\Phi}_{\mathbf{x}}) - \beta \left( \boldsymbol{\Phi}_{\mathbf{x}}^{\prime} - \boldsymbol{\Phi}_{\mathbf{x}} \right) \cdot \boldsymbol{B}_\mathbf{x}  + \Tilde{c} w_{\mathbf{x}}\left[  |P_{\rm u} \boldsymbol{\Phi}_{\mathbf{x}}^{\prime}|^2 - |P_{\rm u} \boldsymbol{\Phi}_{\mathbf{x}}|^2 \right].
\end{equation}
The proposal is accepted with probability $\min (1,e^{-\Delta \mathcal{H}})$, which satisfies detailed balance with respect to Eq.~\eqref{eq:lattice_model}.
Each Metropolis sweep is followed by reflections of the field about its local molecular field,
\begin{equation}
    \boldsymbol{\Phi}_{\mathbf{x}}^{\prime}  = 2 \, \frac{\boldsymbol{\Phi}_{\mathbf{x}} \cdot \boldsymbol{B}_\mathbf{x}}{\boldsymbol{B}_\mathbf{x}^2} \, \boldsymbol{B}_\mathbf{x} - \boldsymbol{\Phi}_{\mathbf{x}},
    \label{eq:overrelaxation}
\end{equation}
provided $\boldsymbol{B}_\mathbf{x}^2$ is nonzero.
When $\Tilde{c}=0$, this transformation preserves both $\boldsymbol{\Phi}_{\mathbf{x}}^2$ and $\boldsymbol{\Phi}_{\mathbf{x}} \cdot \boldsymbol{B}_\mathbf{x}$ and is accepted with unit probability.
When $\Tilde{c} \neq 0$, the proposal Eq.~\eqref{eq:overrelaxation} is however accepted with probability $\min \left( 1, \exp \left[ -\Tilde{c} w_{\mathbf{x}}\left(  |P_{\rm u} \boldsymbol{\Phi}_{\mathbf{x}}^{\prime}|^2 - |P_{\rm u} \boldsymbol{\Phi}_{\mathbf{x}}|^2 \right) \right] \right)$.
Finally, each run is initialized with independent Gaussian components and thermalized before measurements begin.

Finally we elaborate how we choose the nonzero pinning strengths $\tilde c_*$ used for the finite-size scaling collapse.
For each monodromy, we performed preliminary scans of the lattice pinning strength $\tilde c$ on systems with $L\leq64$, comparing the finite-size collapse of the defect correlation functions. 
Among the scanned values with no observed signatures of SSB, we selected pinning strengths yielding comparatively good collapse. 
The resulting choices, $\tilde c_*=-1.8,-1.5,-1.75$ for the $v=1/2$ rotational, $v=1/4$ rotational, and reflection monodromies, respectively, were then held fixed for the subsequent larger-size simulations.

\subsection{Additional data for $v=1/2$ rotational monodromy in $N=3$}

We explain the finite-size collapse method that was used to estimate the scaling dimensions of defect primaries in Figs.~\ref{Fig:twisted_collapse} and \ref{Fig:untwisted_collapse}.
Denoting $u\equiv r_z/L$, we introduce $x_i = \ln u_i$ and $y_i=\ln G_i +2 \hat{\Delta} \ln L_i$ where $\hat{\Delta}$ is a trial value for the scaling dimension,
According to Eq.~\eqref{eq:scaling}, when correct scaling dimension is used, all transformed data should lie on a single master curve 
\begin{equation}
    y_i (\hat{\Delta}) =  \ln \mathcal{F}(u_i) \equiv Y(x_i) .
\end{equation}
For each $x_i$, we estimate $\hat{Y}(x_i)$ by a weighted linear regression through the nearby points.
We evaluate the collapse quality by minimizing the cost function which is defined as
\begin{equation}
\label{eq:fitting-cost-function}
    S(\hat{\Delta}) = \frac{1}{N} \sum_{i=1}^{N} \frac{\left[ y_i(\hat{\Delta}) - \hat{Y}\left(x_i\right) \right]^2}{\sigma_{y_i}^2 + \sigma_{\widehat{Y}_i}^2}.
\end{equation}
$S \simeq 1$ corresponds to collapse within statistical errors.
$\Delta$ is determined by minimizing $S$.
The uncertainty of $\Delta$ is reported as the halfwidth value satisfying $S(\hat{\Delta}) \le S_{\rm min}+1$.
These values of $S$ indicate that the collapse is generally within statistical errors.

\begin{figure}[h]
    \centering
    \includegraphics[width=0.55\linewidth]{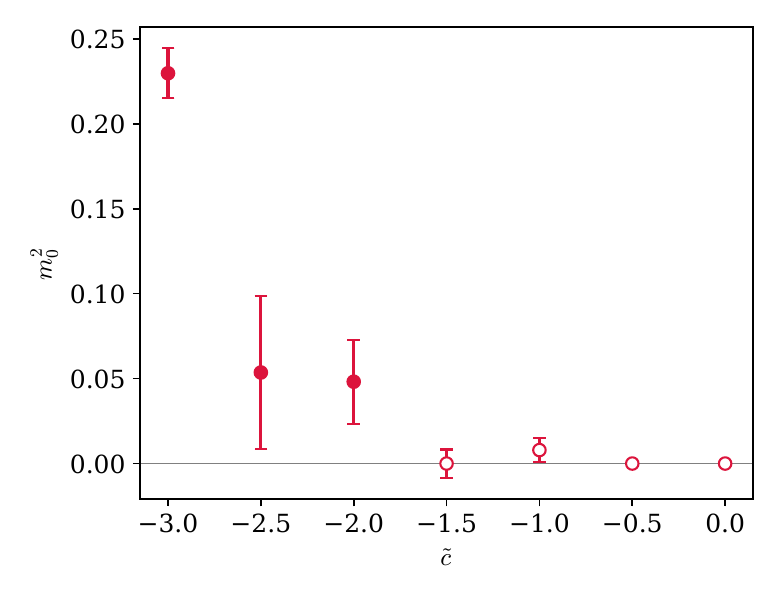}
    \caption{Evidence of spontaneously symmetry breaking on the remaining $\mathbb{Z}_2$ symmetry for $v=1/2$ rotational monodromy in $N=3$. Here $m_0^2$ denotes the squared spontaneous line magnetization, and $\tilde{c}$ denotes the lattice pinning strength. Nonzero $m_0^2$ signals $\mathbb{Z}_2$ SSB. When $|\tilde{c}|\lesssim 2 $ the $\mathbb{Z}_2$ symmetry is unbroken (hollow symbols), while for $|\tilde{c}|\gtrsim 2 $ the $\mathbb{Z}_2$ symmetry is spontaneously broken (filled symbols).} 
    \label{fig:Z2_SSB}
\end{figure}

Next we briefly discuss the spontaneous breaking of the remaining $\mathbb{Z}_2$ symmetry.
We denote the line magnetization by $\mathcal{M}_\phi = \frac{1}{L} \sum_{z=0}^{L-1} \phi^{\mathrm{lat}}(z)$ where $\phi^{\mathrm{lat}}(z)$ is defined in Eq.~\eqref{eq:untwisted}. Our ansatz for squared line magnetization reads
\begin{equation}
  \left\langle \mathcal{M}_\phi^2 \right\rangle_L
    = m_{0}^2 + A_\phi L^{-p_\phi} + \cdots ,
\end{equation}
where $m_{0}^2 = \lim_{L \to \infty} \langle \mathcal{M}_\phi^2 \rangle_L$ is nonzero only if the $\mathbb{Z}_2$ symmetry sending $\hat{\phi} \to -\hat{\phi}$ spontaneously breaks.
Figure~\ref{fig:Z2_SSB} shows $m_{0}^2$ for pinning strength $\Tilde{c}$ between $0$ and $-3$.
We find that when $|\tilde{c}|\lesssim 2 $ the $\mathbb{Z}_2$ symmetry is unbroken, while for $|\tilde{c}|\gtrsim 2 $ the $\mathbb{Z}_2$ symmetry spontaneously breaks.
We leave a systematic study for the spontaneous symmetry breaking induced by pinning-field defect in critical O($N$) model in future work.

\subsection{$v=1/4$ rotational monodromy in $N=3$}

\begin{figure}[h]
    \centering    \includegraphics[width=0.65\columnwidth]{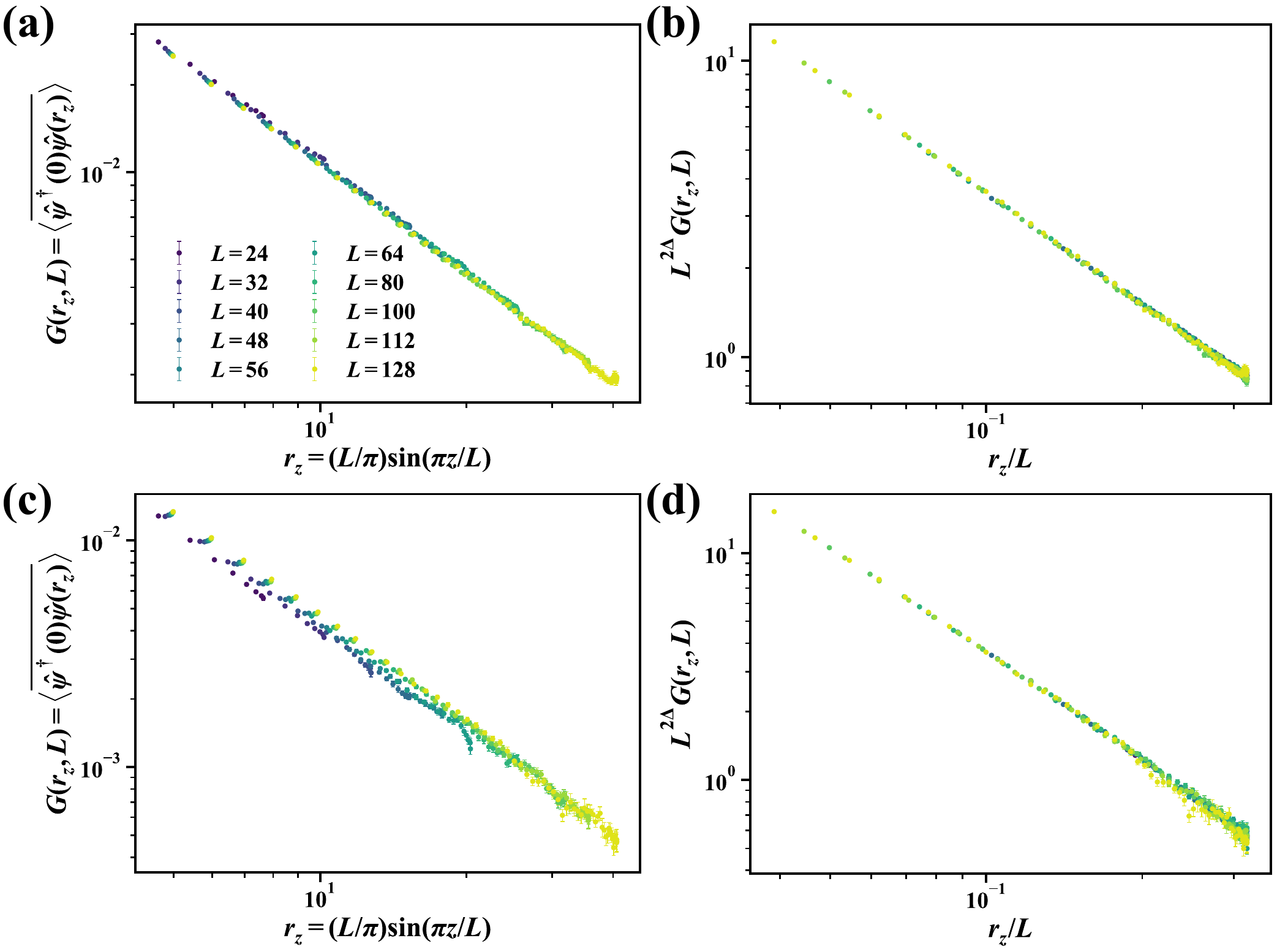}
    \caption{Determination of the scaling dimensions of the twisted defect primary $\hat{\psi}$ for $v=1/4$ rotational monodromy in $N=3$ (a)(b) without explicit lattice pinning ($\Tilde{c}=0$) and (c)(d) with explicit lattice pinning ($\Tilde{c}_*=-1.5$) using finite-size scaling collapse. We get $\Delta_{\psi}=0.632 \pm 0.011$ when $\Tilde{c}=0$ and $\Delta_{\psi}=0.725 \pm 0.016$ when $\Tilde{c}_*=-1.5$.} 
    \label{Fig:rotation_1_4_collapse_twisted}
\end{figure}

\begin{figure}[h]
    \centering    \includegraphics[width=0.65\columnwidth]{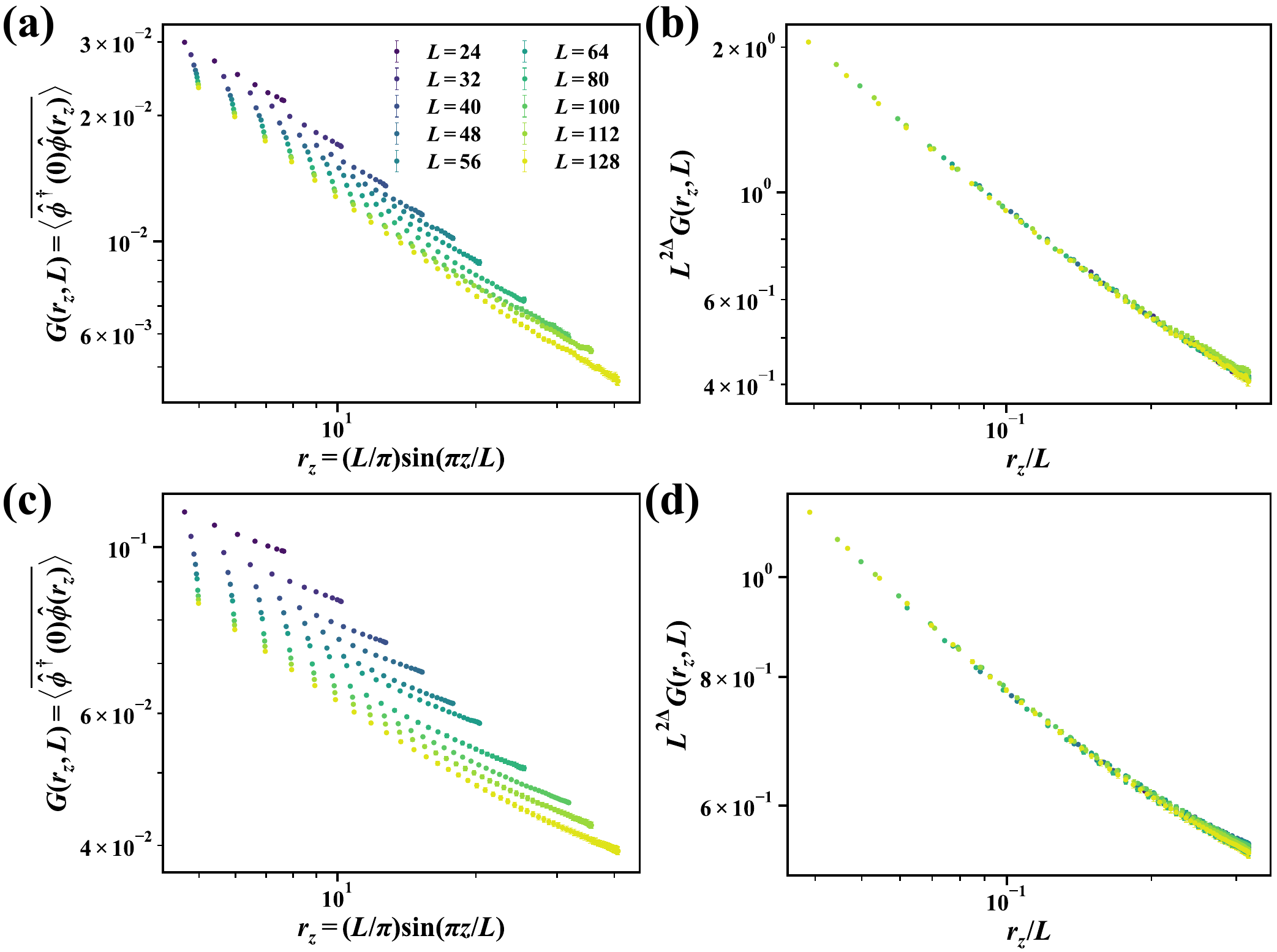}
    \caption{Determination of the scaling dimensions of the untwisted defect primary $\hat{\phi}$ for $v=1/4$ rotational monodromy in $N=3$ (a)(b) without explicit lattice pinning ($\Tilde{c}=0$) and (c)(d) with explicit lattice pinning ($\Tilde{c}_*=-1.5$) using finite-size scaling collapse. We get $\Delta_{\phi}=0.461 \pm 0.011$ when $\Tilde{c}=0$ and $\Delta_{\phi}=0.270 \pm 0.006$ when $\Tilde{c}_*=-1.5$.} 
    \label{Fig:rotation_1_4_collapse_untwisted}
\end{figure}

Figures~\ref{Fig:rotation_1_4_collapse_twisted} and \ref{Fig:rotation_1_4_collapse_untwisted} show the results for the $v=1/4$ rotational monodromy.
The operators we measure are given in Eqs.~\eqref{eq:twisted} and \eqref{eq:untwisted}.

Without explicit pinning term ($\Tilde{c}=0$) we get $\Delta_{\psi}=0.632 \pm 0.011$ for twisted field, and $\Delta_{\phi}=0.461 \pm 0.011$ for untwisted field.
With explicit lattice pinning ($\Tilde{c}_*=-1.5$), we get $\Delta_{\psi}=0.725 \pm 0.016$ for twisted field, and $\Delta_{\phi}=0.270 \pm 0.006$ for untwisted field.
The cost functions are: $S=1.55$ for Fig.~\ref{Fig:rotation_1_4_collapse_twisted}(b), $S=1.20$ for Fig.~\ref{Fig:rotation_1_4_collapse_twisted}(d), $S=1.40$ for Fig.~\ref{Fig:rotation_1_4_collapse_untwisted}(b), and $S=1.72$ for Fig.~\ref{Fig:rotation_1_4_collapse_untwisted}(d), indicating that the collapse is generally within statistical errors.
We confirm that with $\Tilde{c}_*=-1.5$, the spontaneous symmetry breaking on the remaining $\mathbb{Z}_2$ does not happen yet. For comparison, the large $N$ solution for $\gamma=\frac23, v=\frac14$ gives $\Delta_{\ell=0}\approx 0.243, \Delta_{\ell=\frac14}\approx 0.860$.

\end{document}